\documentclass[12pt]{article}

\usepackage[active]{srcltx}
\usepackage{amsmath,amssymb,epsfig,yfonts,esint}
\usepackage{wrapfig}
\usepackage{floatflt}
\usepackage{mathrsfs}
\usepackage{mathabx}
\usepackage{dsfont}
\newcommand{\be}{\begin{equation}}
\newcommand{\ba}{\begin{eqnarray}}
\newcommand{\ea}{\end{eqnarray}}
\newcommand{\nn}{\nonumber}

\def\a{\alpha}
\def\b{\beta}

\def\d{\delta}
\def\e{\epsilon}

\def\l{\lambda}
\def\m{\mu}
\def\n{\nu}
\def\oo{\omega}
\def\p{\pi}

\def\s{\sigma}
\def\t{\tau}

\def\G{\Gamma}

\def\OO{\Omega}
\def\P{\Pi}

\def\S{\Sigma}

\def\ca{{\cal A}}
\def\cb{{\cal B}}
\def\cc{{\cal C}}

\def\ce{{\cal E}}

\def\cg{{\cal G}}
\def\ch{{\cal H}}

\def\ck{{\cal K}}
\def\cl{{\cal L}}
\def\cm{{\cal M}}

\def\co{{\cal O}}
\def\cp{{\cal P}}

\def\cs{{\cal S}}
\def\ct{{\cal T}}

\newcommand{\pa}{\partial}

\newtheorem{thm}{Theorem}[subsection]

\newtheorem{definition}[thm]{Definition}

\newcommand{\One}{{\boldmath 1}}

\newcommand{\Hom}{\operatorname{Hom}}

\newcommand{\Tr}{\operatorname{Tr}}

\fontfamily{yfrak}
\newcommand{\levi}{\nabla^{\mbox{\tiny\bf lc}}}
\newcommand{\nablato}{\nabla^{\mbox{\tiny\bf 2}}}

\begin{document}

\vskip 15mm

\begin{center}

{\Large\bfseries On Spectral Triples in
%\\[2mm]
 Quantum Gravity I\\[2mm]  
%Quantum Gravity via a Projective System of\\[2mm] Spectral Triples. draft %Diffeomorphism Invariant Spectral Triples via Projective Systems of Triangulations %\\[2mm]
}

\vskip 4ex

Johannes \textsc{Aastrup}$\,^{a}$\footnote{email: \texttt{johannes.aastrup@uni-muenster.de}},
Jesper M\o ller \textsc{Grimstrup}\,$^{b}$\footnote{email: \texttt{grimstrup@nbi.dk}}\\ \& Ryszard \textsc{Nest}\,$^{c}$\footnote{email: \texttt{rnest@math.ku.dk}}

\vskip 3ex  

$^{a}\,$\textit{SFB 478 "Geometrische Strukturen in der Mathematik"\\
  Hittorfstr. 27, D-48149 M\"unster, Germany}
\\[3ex]
$^{b}\,$\textit{The Niels Bohr Institute \\Blegdamsvej 17, DK-2100 Copenhagen, Denmark}
\\[3ex]
$^{c}$ \textit{Matematisk Institut\\ Universitetsparken 5, DK-2100 Copenhagen, Denmark}
\end{center}

\vskip 5ex

\begin{abstract}

This paper establishes a link between Noncommutative Geometry and canonical quantum gravity.
A semi-finite spectral triple over a space of connections is presented. The triple involves an algebra of holonomy loops and a Dirac type operator which resembles a global functional derivation operator. The interaction between the Dirac operator and the algebra reproduces the Poisson structure of General Relativity. Moreover, the associated Hilbert space corresponds, up to a discrete symmetry group, to the Hilbert space of diffeomorphism invariant states known from Loop Quantum Gravity. Correspondingly, the square of the Dirac operator has, in terms of canonical quantum gravity, the form of a global area-squared operator. Furthermore, the spectral action resembles a partition function of Quantum Gravity. The construction is background independent and is based on an inductive system of triangulations. This paper is the first of two papers on the subject.

\end{abstract}

\newpage
\tableofcontents

%\newpage
\section{Introduction}

Ever since the discovery of the Standard Model of Particle Physics physicists have worked to understand the apparently arbitrary structure of this theory. Natures choice of gauge group, the Higgs sector, the 20-30 apparently unrelated parameters etc. almost begs for a deeper explanation.

With the pioneering work of Alain Connes and co-workers on the Standard Model \cite{ConnesBook}-\cite{Chamseddine:2007ia} such an explanation now appears to emerge. In Connes' work the Standard Model coupled to General Relativity is expressed as a single gravitational theory. 
The language used for this unification is Noncommutative Geometry \cite{ConnesBook}.
%and the geometry behind it is characterised by an almost commutative algebra. 
%Within the framework of Noncommutative Geometry, and u
%Under
%Under a few mathematical assumptions, the Standard Model coupled to gravity can be shown to be almost unique .
Within this framework, and under a few mathematical assumptions, the Standard Model coupled to gravity can be shown to be almost unique \cite{Chamseddine:2007hz,Chamseddine:2007ia}.

Noncommutative Geometry is based on the result \cite{ConnesBook,Rennie:2006pi} that Riemannian spin geometry has an equivalent formulation in terms of commutative $\ast$-algebras and Dirac operators. In this formulation it is the Dirac operator that carries metric information of the underlying manifold which is now written as the spectrum of the $\ast$-algebra. In total, a Riemannian spin geometry can be described in terms of a {\it spectral triple} which is the collection $(B,H,D)$ of the algebra $B$, the Dirac operator $D$ and the Hilbert space $H$ which carries a representation and action of $B$ and $D$. To obtain equivalence the triple is required to satisfy a set of axioms of Noncommutative Geometry.

The language of spectral triples has a natural generalisation which includes also noncommutative $\ast$-algebras and corresponding Dirac type operators. It is this generalisation which leads to the aforementioned formulation of the Standard Model. It turns out that the classical action of the Standard Model coupled to the Einstein-Hilbert action emerges from a spectral action principle applied to a specific spectral triple \cite{Chamseddine:1991qh,Chamseddine:1996rw,Chamseddine:1996zu}. This triple involves an almost commutative $\ast$-algebra which is an algebra that factorises into a commutative part times a matrix factor. This means that the classical action describing all fundamental physics\footnote{Here we give the bosonic part only.} emerges from an asymptotic expansion of the spectral action
\[
\mbox{Tr}\phi(\tilde{D}/\lambda)= \sum \lambda^n c_n \;.
\]
Here $\tilde{D}$ denotes the Dirac type operator $D$ subjected to certain inner fluctuations stemming from the noncommutativity of the algebra.
Without the matrix factor, that is, for a commutative $\ast$-algebra, the same expression leads to the Einstein-Hilbert action alone. Thus, it is the inclusion of noncommutative $\ast$-algebras in the language of spectral triples that permits the formulation of all fundamental forces and particles in terms of pure gravity.

This success of Noncommutative Geometry as a framework to describe fundamental physics in a unified manner raises, however, a fundamental question regarding quantization. The Standard Model by itself is a quantum field theory. However, in its noncommutative formulation it arises as an integrated part of a purely gravitational theory. This theory is essentially classical.  Quantum Field Theory enters the construction in a secondary step where the spectral action has been expanded in a gravitational sector involving only the metric field and a matter sector including all the fermionic and bosonic fields of the Standard Model. Quantization is applied to the latter sector only.

So this is the question: how does the quantization procedure of Quantum Field Theory fit into the language of Noncommutative Geometry? Since the structure of the Standard Model is so readily translated into the language of Noncommutative Geometry one would expect also Quantum Field Theory to have a corresponding translation. This question is further complicated by the intrinsic gravitational nature of Connes' formulation of the Standard Model. If a notion of quantization does exist within this framework one would expect this to involve, at some level, Quantum Gravity.

%And one would expect this translation to involve, at some level, Quantum Gravity since the Standard Model formulated within Noncommutative Geometry is essentially General Relativity on a noncommutative space.

% it is so far unclear how the quantization procedure itself fits into the framework of Noncommutative Geometry. This issue is further complicated by the intrinsic gravitational nature of Connes' formulation of the Standard Model. If a notion of quantization does exist within this framework one would expect this to involve, at some level, Quantum Gravity.

This paper is motivated by these considerations. We wish to address the question regarding the inclusion of the quantization procedure in the general framework of Noncommutative Geometry and, in particular, in the noncommutative formulation of the Standard Model. We start the investigation with the assumption that the answer will involve some notion of Quantum Gravity. Thus, to guide our intuition we consult first canonical quantum gravity and one of its modern trends known as Loop Quantum Gravity \cite{Rovelli:1987df}-\cite{Ashtekar:2004eh}. 

Loop Quantum Gravity is an approach to non-perturbative Quantum Gravity which is based on a formulation of General Relativity in terms of Ashtekar variables \cite{Ashtekar:1986yd,Ashtekar:1987gu}. These variables include the Ashtekar connection and its conjugate variable, the inverse densitised dreibein. In Loop Quantum Gravity one takes as the classical phase space variables the holonomy loop of the Ashtekar connection and its conjugate, a certain flux vector.
These variables are then used in a Dirac-type quantization procedure. This involves the representation of the corresponding Poisson structure on a kinematical Hilbert space as well as the formulation of constraints encoding the symmetries of the classical theory. Here, the holonomy loops are represented as multiplication operators. This means that the kinematical Hilbert space involves functions over a space of connections.

On a technical level, Loop Quantum Gravity exploits the holonomy-formulation of gravity to recast the problem of quantizing gravity in terms of a inductive system of graphs. It turns out that the space of connections, when seen in terms of holonomies restricted to a specific graph, is a manifold. This manifold is related to the gauge group which, in the case of Loop Quantum Gravity, is $SU(2)$. This means that the space of connections itself is a pro-manifold, the projective limit of manifolds. This, in turn, permits a formulation of the quantization procedure on the level of finite graphs whose complexity are subsequently increased infinitely. This construction is due to Ashtekar and Lewandowski \cite{Ashtekar:1993wf,Ashtekar:1994wa}.

In this paper we aim to construct a model which involves elements of both Noncommutative Geometry and Quantum Gravity. We use two central elements of Loop Quantum Gravity to obtain such a model. First, the fact that gravity can be formulated in terms of Wilson loops leads us to consider a spectral triple which involves an algebra of holonomy loops. These loop variables serve as functions on an underlying space of connections. Second, we wish to exploit the pro-manifold structure of the space of connections to describe the algebra of loops and to construct a Dirac type operator on the space of connections. This means that we aim to construct a spectral triple over each manifold associated to graphs in the projective system. These spectral triples are required to be compatible with all embeddings between graphs. This requirement will ensure that the limit where the complexity of graphs is increased infinitely gives rise to a limit spectral triple.

This program was initiated first in \cite{Aastrup:2005yk} (see also \cite{Aastrup:2006ib}) where the authors attempted to construct such a spectral triple. There the authors found that the inductive system of graphs used in Loop Quantum Gravity, which is the system of all piecewise analytic graphs, is too large to permit a Dirac type operator on the space of connections. Technically, the multitude of possible embedding of different graphs was found to be too large for a Dirac type operator compatible with all embeddings to exist.

In the present paper we return to this problem to consider now different systems of embedded graphs. In particular, we study the countable system of embedded graphs given by a triangulation and its barycentric subdivisions. It turns out that this restricted system of graphs does permit a Dirac type operator on the associated projective system of manifolds. Furthermore, we find that the limit of infinitely many barycentric subdivisions gives us an accurate description of the full space of connections as well as the associated algebra of holonomy loops. The construction is general and only assumes the gauge group $G$ to be compact.

What we obtain is the following. Given a triangulation $\ct$ and a compact Lie group $G$, we construct a spectral triple 
\ba 
(\cb_{\smalltriangleup},D_{\smalltriangleup},\ch_\smalltriangleup)\;,
\label{ONee}
\ea
where $\cb_{\smalltriangleup}$ is the $\ast$-algebra of holonomy loops obtained via the inductive system of triangulations. The algebra is represented on the separable Hilbert space $\ch_\smalltriangleup$ which carries an action of the Dirac type operator $D_{\smalltriangleup}$. 
If we denote by $\ca$ the space of smooth connections in a trivial principal bundle $\cm\times G$, where $\cm$ is a manifold which corresponds to the triangulation $\ct$, then we find that $\ca$ is densely contained in the pro-manifold associated to the algebra $\cb_{\smalltriangleup}$. This means that the spectral triple (\ref{ONee}) is a geometrical construction over the space of connections.

The construction of the spectral triple (\ref{ONee}) depends, as mentioned, crucially on the choice of graphs. This choice is closely related to the group of diffeomorphisms acting on the Hilbert space $\ch_\smalltriangleup$. We find that the choice of a restricted system of graphs amounts to a type of gauge fixing of the diffeomorphism group. Thus, the Hilbert space $\ch_\smalltriangleup$ does not carry an action of any smooth diffeomorphisms. Rather, it carries an action of a discrete group of diffeomorphisms associated to the inductive system of graphs. This means that the construction reduces the diffeomorphism group to a countable group.

The spectral triple (\ref{ONee}) has a clear interpretation in terms of a non-perturbative, background independent Quantum Field Theory related to gravity. First of all, since the triple exists over a space of connections the Dirac type operator $D_\smalltriangleup$ should be interpreted as a global functional derivation operator. Also, the Hilbert space $\ch_\smalltriangleup$ has an inner product which involves a functional integral.
Next, we find that the interaction between the Dirac type operator $D_\smalltriangleup$ and the loop algebra $\cb_\smalltriangleup$ reproduces the Poisson structure of General Relativity. Furthermore, the Hilbert space $\ch_\smalltriangleup$ is found to be directly related to the Hilbert space of (spatial) diffeomorphism invariant states known from Loop Quantum Gravity \cite{Fairbairn:2004qe}. The difference between the two is given by the group of discrete diffeomorphisms acting on $\ch_\smalltriangleup$. Thus, we interpret the Hilbert space $\ch_\smalltriangleup$ in terms of a partial solution to the (spatial) diffeomorphism constraint.

The square of the Dirac type operator $D_\smalltriangleup$ has the form of an integral over the underlying manifold $
\cm$. The integrand is a quantity which, in terms of canonical quantum gravity, has an interpretation as an area-squared density operator. This operator resembles the area operators known in Loop Quantum Gravity \cite{Ashtekar:1996eg}. Furthermore, we suggest that $(D_\smalltriangleup)^2$ should be interpreted in terms of an action. This, in turn, gives the spectral action of $D_\smalltriangleup$ the form of a Feynman integral. Thus, at the core of the construction we find an object which resembles a partition function related to Quantum Gravity.

It is important to realize that the construction works in any dimension and does not require a foliation of the underlying manifold.

The construction of the Dirac type operator $D_{\smalltriangleup}$ involves an infinite dimensional Clifford bundle. This structure entails the {\it canonical anticommutation relations} (CAR) algebra, which appears as a tensor factor acting on the Hilbert space $\ch_\smalltriangleup$.

Technically, the triple (\ref{ONee}) satisfies the requirements of a semi-finite spectral triple. This is due to the fact that the infinite dimensional Clifford bundle entails a large degeneracy of the spectrum of $D_{\smalltriangleup}$ which, naively, fails to have compact resolvent. The solution to this problem is, in short, to integrate out the symmetry group related to this degeneracy. This process leads to a semi-finite spectral triple.

The Dirac type operator $D_\smalltriangleup$ is gauge invariant. It is, however, not invariant under the group of discrete diffeomorphisms acting on $\ch_\smalltriangleup$. This means that $D_\smalltriangleup$ is, by itself, not an observable in a physical sense. We suggest to apply a standard trick of Noncommutative Geometry: One obtains invariant quantities by multiplying the algebra of observables with the relevant symmetry group. A prime example of this method is the identification of two points: By applying the noncommutative trick one obtains, besides invariance under exchange of the points, additional degrees of freedom which, ultimately, leads to the Higgs mechanism. In the present case we expect this general mechanism to entail further degrees of freedom.

Let us note in passing, that the algebra of holonomy loops that we work with in this paper is smaller than the correponding algebra of Wilson loops. However, the missing information is recovered by the Hopf algebra structure of the algebra of holonomy loops. The missing information is coded in the extra structure coming from the fact that it is given as a (norm closure of a) group algebra
\[
\mathbb{C}[f_L,\;L \;\mbox{loops on } \cs_\smalltriangleup]\;,
\]
where $\cs_\smalltriangleup$ denotes the entire inductive system of triangulations.
As such it has a Hopf algebra structure and for example the connections on $\cm$ can be recovered as Hopf algebra homomorphisms into the group algebra of $G$.

The construction of the Dirac type operator $D_\smalltriangleup$ is not unique. In fact, we find a large class of Dirac type operators labelled by infinite sequences $\{a_n\}$ of real parameters. The operator $D_\smalltriangleup$ has compact resolvent whenever the sequence diverges sufficiently fast. These parameters are related to the scaling behaviour of the operator and are clearly of metric origin. Thus, in order to obtain a Dirac type operator $D_\smalltriangleup$ we are forced to choose a certain scaling behaviour.

It is clear that a correct interpretation of the sequence $\{a_n\}$ is imperative since the existence of the operator $D_\smalltriangleup$ depends crucially hereon. 
In this paper we present one possible solution as to how these free parameters should be dealt with. We propose an extension of the spectral triple (\ref{ONee}) to include the sequence $\{a_n\}$ as dynamical degrees of freedom. The result is a new triple $(\cb_t,D_t,\ch_t)$ which is a fibration of spectral triples $(\cb_{\smalltriangleup},D_{\smalltriangleup},\ch_\smalltriangleup)$. We emphasise, however, that the question as to how the sequence $\{a_n\}$ should be understood and dealt with remains open.

This paper is this first of two papers concerned with the spectral triple (\ref{ONee}). This paper is primarily concerned with the general construction and physical interpretation of the triple. The second paper \cite{Aastrup} deals with the concise mathematical construction.

\subsection{Outline of the construction}

Before we go into details we first give a brief outline of the construction. The first step is concerned with the formulation of a semi-finite spectral triple over a space of connections. The triple is, as mentioned, based on a $\ast$-algebra of loops which we denote by $\cb_{\smalltriangleup}$. A smooth loop $l$ gives a map from the space of smooth connections, denoted $\ca$, into the structure group $G$
\[
l:\nabla\rightarrow Hol(\nabla,l)\in G\;,
\]
\begin{figure}[t]
\begin{center}
 \input{figure_4a.pstex_t}
\label{figfigfig}
\caption{A graph $\G$ with edges $\{\e_1,\e_2,\e_3,\e_4\}$.}
\end{center}
\end{figure}
\noindent where $Hol(\nabla,l)$ is the holonomy of the connection $\nabla\in\ca$ along $l$ and $G$ is a compact connected Lie group. 
%In fact, one can, up to gauge transformations, completely determine the space of connections in terms of holonomy loops (see for example \cite{Barrett:1991aj} or references therein). 
In order to describe the algebra of holonomy loops we first restrict the loop algebra to a finite graph $\G$ with edges $\e_i$ (see figure \ref{figfigfig}). Seen from $\G$ the connection $\nabla$ can be seen as point $\nabla$ in the space $G^n$
\[
\nabla = (g_1,\ldots,g_n)\in G^n =\ca_\G\;,
\]
where $n$ is the number of edges in $\G$ and where $g_i = Hol(\nabla,\e_i)$ is the holonomy transform along the $i$'th edge. That is, a connection is given by its holonomy transforms along edges $\e_i$. Clearly, $\ca_\G$ is a highly inaccurate picture of the full space $\ca$ of connections. However, one can show \cite{Aastrup} that for a suitable choice of embedded graphs
\ba 
\G_1\subset \G_2 \subset \G_3 \ldots \subset \G_n \subset \ldots
\label{PROJEKTIVT}
\ea
the space $\ca$ is densely contained in the limit space
\[
\overline{\ca}=\lim_{n\rightarrow\infty} \ca_{\G_n}\;.
\]
This provides us with the strategy to construct a spectral triple over $\ca$. With the system of graphs given by nested triangulations we construct, at the level of each graph $\G_n$, a spectral triple
\[
(\cb_n,D_n,\ch_n)_{\G_n}\;,
\]
where the algebra $\cb_n$ is generated by loops in $\G_n$, the operator $D_n$ is some Dirac operator on $\ca_{\G_n}=G^n$ and where $\ch_n=L^2(G^n,Cl(T^\ast G^n))$. This triple is almost canonical and the little choice one has is mostly eliminated by the requirement that the construction of the triple should be compatible with structure maps
\[
P_{nm}: \ca_{\G_n}\rightarrow \ca_{\G_m}\;,\quad n>m
\]
between the different coarse-grained spaces of field configurations. 

The Dirac type operator obtained in the limit is an operator on the space $G^\infty$ and is of the general form
\[
a_1 D_1 + a_2 D_2 + \ldots + a_k D_k + \ldots
\]
where $D_k$ is an operator corresponding to a certain level in the projective system (\ref{PROJEKTIVT}) and where the infinite sequence $\{a_k\}$ determines the weight assigned to each operator $D_k$. We find that for sequences satisfying
\ba 
\lim_{k\rightarrow\infty}a_k=\infty \quad \mbox{sufficiently fast}
\label{CONDICONDI}
\ea
the limit spectral triple
\ba 
(\cb_{\smalltriangleup},D_{\smalltriangleup},\ch_\smalltriangleup) := \lim_{n\rightarrow\infty}(\cb_n,D_n,\ch_n)_{\G_n}
\label{TRIPLETWO}
\ea
satisfies the requirements of a semi-finite spectral triple.

The second part of the construction is more tentative. This part is concerned with the infinite sequence $\{a_k\}$ of free parameters which enters the construction of the semi-finite spectral triple $(\cb_{\smalltriangleup},D_{\smalltriangleup},\ch_\smalltriangleup)$. 
The sequence $\{a_k\}$ is readily seen to carry metric data since it determines the scaling behaviour of the operator $D_{\smalltriangleup}$. Also, the sequence determines the measure that arises in the square of $D_{\smalltriangleup}$. Furthermore, the choice of parameters $\{a_k\}$ is related to the invariance properties of $D_{\smalltriangleup}$. 
%This means that the diffeomorphisms acting on the triple $(\cb_{\smalltriangleup},D_{\smalltriangleup},\ch_\smalltriangleup)$ consist of rearrangements of the edges in graphs and therefore of rearrangements of parameters $a_k$ which are assigned to edges. 
Based on these observations we choose to include the sequence $\{a_k\}$ as dynamical parameters in the construction. This means that we first construct a spectral triple
\ba 
(A_a,D_a,\ch_a)\;,
\label{TRIPLEONE}
\ea
where elements in the algebra $A_a$ are functions $f(x_1,x_2,\ldots)$ on the moduli space of sequences $\{a_k\}$ satisfying condition (\ref{CONDICONDI}). The construction of the triple (\ref{TRIPLEONE}) is inspired by Higson and Kasparov \cite{Higson}. The Hilbert space $\ch_a$ is a $L^2$ space of functions over this moduli space and $D_a$ is a Dirac operator hereon. Next, we merge the triple (\ref{TRIPLEONE}) with the triple (\ref{TRIPLETWO}) to obtain a total spectral triple
\ba 
(\cb_t,D_t,\ch_t)\;,
\label{thetriple}
\ea
where the Dirac type operator $D_t$ combines the operators
$D_{\smalltriangleup}$ and $D_a $.
It is important to see that the first part of the Dirac operator, the operator $D_{\smalltriangleup}$, depends on the parameters $\{a_k\}$ which the second part, the operator $D_a$, probes. The exact form of this interdependency, which makes the operator $D_t$ highly nontrivial, is governed by the requirement that $\ch_a$ is a Hilbert space over an infinite dimensional space corresponding exactly to those sequences $\{a_k\}$ which leaves the operator $D_{\smalltriangleup}$ $\theta$-summable. \\

The paper is organised as follows. Sections \ref{section1} - \ref{projlim} give a self contained presentation of the mathematical construction of the spectral triple $(\cb_{\smalltriangleup},D_{\smalltriangleup},\ch_\smalltriangleup)$. Here, section \ref{section1} introduces the basic machinery used in this paper: First the loop algebra associated to an abstract simplicial complex and the concept of abstract connections associated to a simplicial complex. Thereafter projective systems of simplicial complexes. In section \ref{sec14} we construct a spectral triple on the level of a simplicial complex and in section \ref{projlim} we obtain the triple $(\cb_{\smalltriangleup},D_{\smalltriangleup},\ch_\smalltriangleup)$ via an inductive limit of simplicial complexes.

Next, sections \ref{cangrav} - \ref{EINstein} are concerned with the physical significance of the spectral triple $(\cb_{\smalltriangleup},D_{\smalltriangleup},\ch_\smalltriangleup)$. In section \ref{cangrav} we show that the spectral triple involves a representation of the Poisson algebra of General Relativity. Then in section \ref{section2} we give a detailed comparison between the setup presented in this paper and the setup used in Loop Quantum Gravity and find that the two constructions, on a technical level, differ primarily in the way the diffeomorphism group is treated. 
%Subsequently, section \ref{link2} provides two possible interpretations of the spectral triple, both in terms of quantization schemes of General Relativity on Hilbert spaces which corresponds to partial solutions of the diffeomorphism constraint. 
Section \ref{sectionarea} briefly reviews area operators in Loop Quantum Gravity and finds that the square of the Dirac operator $D_{\smalltriangleup}$ has a natural interpretation as a kind of global area operator. This leads naturally, in section \ref{EINstein}, to an interpretation of the operator $D_{\smalltriangleup}^2$ in terms of an action. This, in turn, gives the spectral action of $D_{\smalltriangleup}$ a strong resemblance to a partition function related to gravity.

Then, in section \ref{includea}, we describe the construction of the spectral triple $(\cb_t,D_t,\ch_t)$ which includes the sequence $\{a_k\}$ as dynamical variables. 
%Section \ref{background} gives a general discussion of background dependence of the construction and
Finally, in section \ref{DISTa} we mention that the Dirac type operator $D_{\smalltriangleup}$ defines a distance on the space $\overline{\ca}^{\smalltriangleup}$. Sections 
\ref{DISCu} and \ref{section4} contain a discussion and conclusion. Three appendices are added to outline first an extended setup which avoids the choice of a basepoint and next to discuss again a notion of diffeomorphism invariance. The last appendix is concerned with certain symmetric states.

\section{Projective systems of simplicial complexes}
\label{section1}

As explained in the introduction, the aim is to study geometrical structures over a space of connections. To do this we apply a dual picture of the space of connections. This means that, rather than studying the space of connections itself, we will work with an algebra of functions on the space. This algebra is an algebra of loops and has a natural interpretation in terms of holonomy loops.

The purpose of this section is to introduce the machinery needed to describe this algebra of loops. The strategy is to break up the algebra into finite parts, introduce various geometrical structures on each finite part and finally let the complexity of the construction increase infinitely to obtain the full algebra of loops. 

To emphasise the purely combinatorial nature of the construction we adopt a formalism which, for the main part of the analysis, avoids any reference to an underlying manifold. This means that we will work with abstract graphs and their loop algebras. An alternative approach is to work directly with graphs on a manifold and their loop algebras. This approach, which may be intuitively clearer, is applied in \cite{Aastrup}.

%In this section we introduce first an abstract graph given by a simplicial complex and an algebra of abstract loops associated to this graph. In a second step we prescribe a systematic refinement procedure given by repeated barycentric subdivisions of the simplicial complex. At the level of each simplicial complex we introduce the space of homomorphisms from the loop algebra to a compact Lie group. This space has a natural interpretation as a coarse-grained functional space of connections.

\subsection{Simplicial complexes and their loop algebras}
\label{sec11}
\begin{figure}[t]
\begin{center}
 \input{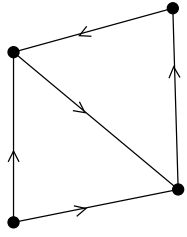}
\caption{An abstract simplicial complex with directed edges.}
\end{center}
%\label{abstractgraph}
\end{figure}

We first introduce the notion of an abstract graph and its associated loop algebra. 
The abstract graphs we consider are given by simplicial complexes.
Consider therefore first an abstract, finite, $d$-dimensional simplicial complex $\ck$ with vertexes $ v_i$ and directed edges
\[
\e_j:\{0,1\}\rightarrow \{ v_i \}\;,\quad \e_j(0)\not=\e_j(1)\;,
\]
connecting the vertexes. We shall refer to the two elements of the set $\{0,1\}$ as start and endpoint of the edge. 
The construction which we present works for a large class of simplicial complexes. We will, however, restrict ourselves to simplicial complexes which corresponds to triangulations of $d$-dimensional manifolds.
%The complex is build entirely of $d$-dimensional simplices with the restriction that, if $\ck$ involves more than one simplex, then an intersection between two simplices in $\ck$ will be a $(d-1)$-face or empty. Furthermore, a $(d-1)$-face in $\ck$ can belong to at most two $d$-simplices in $\ck$. Denote by $V_\ck$ the set of vertices in $\ck$ and by $E_\ck$ the set of edges in $\ck$.

%Given a suitable $d$-dimensional manifold $\cm$ we shall later also consider triangulations $\ct$ of $\cm$ associated to the complex $\ck$. We denote by $\phi$ the corresponding embedding
%\[
%\phi:\ck\rightarrow\ct\;.
%\]

We will consider based simplicial complexes which means that the complex has a preferred vertex $v_0$.

Given a simplicial complex $\ck$ we wish to describe an algebra of based loops living on $\ck$. First, a path is a finite sequence $L=\{\e_{i_1},\e_{i_2},\ldots,\e_{i_n}\}$ of edges in $E_\ck$ with the property that 
\[
\e_{i_k}(1)=\e_{i_{k+1}}(0)\;.
\]
Next, a loop is a path satisfying
\[
\quad\e_{i_1}(0)=\e_{i_n}(1) \;,
\]
and a based loop is a loop satisfying the additional requirement
\[
\quad\e_{i_1}(0)=\e_{i_n}(1)=v_0 \;.
\]
By $\e_j^*$ we denote the edge $\e_j$ with reversed direction
\[
\e_j^*(\t) = \e_j(1-\t)\;,\quad \t\in\{0,1\}\,.
\]
A path may contain both edges and their reverse. We discard trivial backtracking by which we mean sequences that contain successions of edges $\e_i$ and their reverse $\e_i^*$. Thus, we introduce the equivalence relation
\[
\{\ldots,\e_j,\e_k,\e_k^*,\e_l,\ldots\}\sim \{\ldots,\e_j,\e_l,\ldots\}\;,
\] 
and regard a path as an equivalence class with respect to this relation.

We define a product between two based loops $L_i=\{\e_{j_i}\}$, $i\in\{1,2\}$, simply by gluing
\[
L_1\circ L_2 = \{\{\e_{j_1}\},\{\e_{j_2}\}\}\;.
\]
Notice that this product is noncommutative. 
%A product between paths and loops which are not based is defined likewise whenever start or end points  of the paths or loops coincide. 

The inversion of a based loop $L=\{\e_{j_1},\ldots,\e_{j_i},\ldots,\e_{j_n}\}$ defined by
\[
L^*=\{\e_{j_n}^*,\ldots,\e_{j_i}^*,\ldots,\e_{j_1}^*\}\;,
\]
is again a based loop and satisfy the requirements of an involution
\[
(L^*)^*=L\;,\quad (L_1\circ L_2)^* = L_2^*\circ L_1^*\;.
\]
Furthermore, we define the based identity loop $L_0$ as the equivalence class that includes the empty loop
\[
L_0 = \{\O\}\;.
\]
The based identity loop clearly satisfies
\[
L_0\circ L^\prime = L^\prime\;\quad\forall\;\; \mbox{based loops }\; L^\prime\;.
\]
This, together with the observation that
\[
L^\ast\circ L = L\circ L^\ast = L_0\;,
\]
implies that, for based loops, the involution equals an inverse. This provides the set of based loops with a group structure. We call the group of based loops associated to $\ck$ for the hoop group ({\it holonomy loops}, see \cite{Ashtekar:1993wf}), denoted $\ch\cg_\ck$. 
%Equally, the sets of paths and un-based loops has a groupoid structure. In this case there is an empty loop for each vertex in $V_\ck$, see \cite{Johannes}. 

For the remaining part of this paper we shall consider only based loops and will therefore drop the prefix 'based'.

We finally consider formal, finite series of loops living on a complex $\ck$
\ba
a =\sum_i a_i L_i\;,\quad L_i\in \ch\cg_\ck\;,\quad a_i\in\mathbb{C}\;.
\label{element}
\ea
The product between two elements $a$ and $b$ is defined by
\[
a\circ b = \sum_{i,j}(a_i\cdot b_j) L_i\circ L_j\;,
\]
and the involution of $a$ is defined by
\[
a^* =\sum_i \bar{a}_i L^*_i\;.
\]
The set of elements of the form (\ref{element}) is a $\star$-algebra. We denote this algebra with $\cb_\ck$.

\subsection{Loop group homomorphisms and connections}
\label{sec-loop}

We now introduce the notion of an abstract connection. Let $G$ be a compact, connected Lie-group and, for later reference, fix a matrix representation of $G$. 

Next we introduce a $G$-bundle over a smooth manifold $\cm$ and a loop $l$ in $\cm$. A smooth connection can be understood as a map
\ba 
\nabla:l\rightarrow \nabla(l)\in G\;,
\label{angelaW}
\ea
which satisfies the condition
\ba 
\nabla(l_1\circ l_2)= \nabla(l_1)\cdot \nabla(l_2)\;,
\label{toronto}
\ea
where $l_1$ and $l_2$ are loops in $\cm$. Here, the map (\ref{angelaW}) is the holonomy transform of $\nabla$ along $l$,
$
\nabla(l)=Hol(\nabla,l)
$. 

This motivates the following definition of an abstract connection as a map
\ba
\nabla:\{\e_j\}\rightarrow G\;,\quad 
\label{identification2}
\ea
that associates to each edge $\e_j\in E_\ck$ a point $g_j\in G$. The map is required to satisfy
\[
\nabla(\e_j)=(\nabla(\e_j^\ast))^{-1}\;.
\]
Denote by $\ca_\ck$ the space of all abstract connections associated to $\ck$. The action of $\nabla$ is extended to a path $L=\{\e_{i_1},\e_{i_2},\ldots,\e_{i_n}\}$ simply by
\ba
\nabla(L)=\nabla(\e_{i_1})\cdot\nabla(\e_{i_2})\cdot\ldots\cdot\nabla(\e_{i_n})\;,
\label{actionon}
\ea
where the product on the rhs is matrix multiplication. This makes $\nabla$ a group homomorphism from the hoop group $\ch\cg_\ck$ into $G$ which means that it satisfies
\[
\nabla(L_1\circ L_2) = \nabla(L_1)\cdot\nabla(L_2)\;.
\]
This corresponds to condition (\ref{toronto}) and justifies the terminology abstract connection.

Via the space $\ca_\ck$ we can equip the $\star$-algebra formed by elements of the form (\ref{element}) with a natural norm given by
\ba
\| a \|= \sup_{\nabla\in\ca_\ck}\|\sum a_i \nabla(L_i)  \|_G\;,
\label{norm}
\ea
where the norm on the rhs of (\ref{norm}) is the matrix norm given by the representation of $G$. The closure of the $\star$-algebra $\cb_\ck$ of loops with respect to this norm is a $C^\star$-algebra. We denote this loop algebra by $B_\ck$.

In fact, the algebra $\cb_\ck$ is a function algebra over the space $\ca_\ck$ with values in the matrix representation of the group $G$. A loop $L$ gives rise to a function $f_L$ via
\[
f_L(\nabla):= \nabla(L)\;,\quad L\in\ch\cg_\ck\;,\quad \nabla\in\ca_\ck\;.
\]
Notice that the algebra of functions $f_L$ with the natural product
\[
f_{L_1}\cdot f_{L_2}= f_{L_1\circ L_2}\;,
\] 
is noncommutative whenever $G$ is non-Abelian.

The space $\ca_\ck$ is identified as a manifold
\ba
\ca_\ck\simeq G^{n(\ck)}\;,
\label{identification1}
\nonumber
\ea
via the bijection
\ba
\ca_\ck\ni\nabla\rightarrow (\nabla(\e_1),\ldots,\nabla(\e_{n(\ck)}))\in G^{n(\ck)} \;.
\ea
where $n(\ck)$ denotes the number of edges in $V_\ck$. This identification gives rise to various structures on $\ca_\ck$. For example the topological structure is given by the topological structure of $G^{n(\ck)}$.

A loop $L=\{\e_{i_1},\e_{i_2},\ldots,\e_{i_k} \}$ gives, according to (\ref{identification1}), rise a function on $G^{n(\ck)}$ given by
\[
f_L:(g_1,g_2,\ldots,g_{n(\ck)})\rightarrow g_{i_1}\cdot g_{i_2}\cdot\ldots\cdot g_{i_k}\in G\;.
\]

We think of the space $\ca_\ck$ as a "coarse grained version of a space of (smooth) connections". To clarify this interpretation consider an embedding of the simplicial complex into a triangulation
\ba 
\phi:\ck\rightarrow\ct
\label{EmBeD}
\ea 
and a trivial principal bundle $P=\cm\times G$. Denote by $\ca$ the space of smooth connections in $P$. There is a natural map 
\ba
\chi_\ck:\ca\rightarrow\ca_\ck\;,\quad\chi_\ck(\nabla)(\e_i)=Hol(\nabla,\phi(\e_i))\;,
\label{firstchi}
\ea
where $Hol(\nabla,\phi(\e_i))$ denotes the holonomy of $\nabla$ along the edge $\e_i$ which now lives in $\cm$ via the embedding $\phi$. This means that points in $\ca_\ck$ can be understood in terms of connections. Clearly, the map (\ref{firstchi}) is not injective which is exactly what is meant by 'coarse grained'. The idea is to gradually increase the complexity of the simplicial complex and thereby to turn the map (\ref{firstchi}) into an injection. Therefore, the next step is to  introduce a refinement procedure for the simplicial complex $\ck$.

%Let us end this section with a few remarks. The identification (\ref{identification1}) shows that the space $\ca_\ck$ of abstract connections is a manifold. This gives us immediate access to various geometrical structures. For example, it is easy to construct a Dirac or Laplace operator on $G^{n(\ck)}$. Such operators should then, according to (\ref{firstchi}), be interpreted as coarse grained functional derivations. Also, a Hilbert space structure over $G^{n(\ck)}$ is easily obtained and will involve an inner product which should be interpreted as a coarse grained functional integral. As the complexity of the underlying simplicial complex is increased, these structures will approach real functional derivations and functional integrals. 

\subsection{The projective system}
\label{sec-proje}

\begin{figure}[t]
\begin{center}
 \input{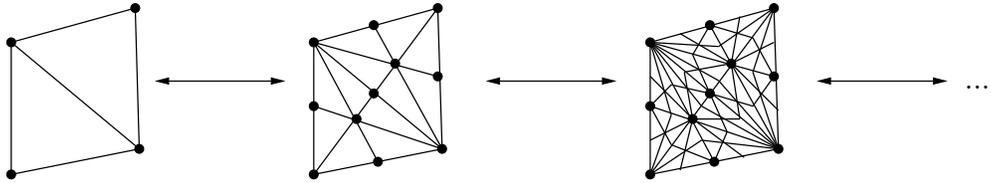}
\caption{The lhs shows a triangulation (dimension $=2$) and the rhs its two first barycentric subdivisions.}
\end{center}
\end{figure}
The key tool to refine the simplicial complex $\ck$ is the barycentric subdivision of simplexes. We will consider repeated barycentric subdivisions of $\ck$. Thus, the basic element in our analysis of geometrical structures over spaces of connections is an inductive systems of simplicial complexes. We start with the following
\begin{definition}
An inductive system of simplicial complexes is a countable set $(\{\ck_i\},\{I_{jk}\})$ of nested abstract simplicial complexes $\ck_i$ and embeddings
\ba
I_{jk}:\ck_j\rightarrow \ck_k\;\;,\quad j<k
\label{proj}
\ea
so that $I_{jk}$ is either a barycentric subdivision or an inversion of edges. The set is required to satisfy
\[
\lim_i n(\ck_i)=\infty\;.
\]
A barycentric subdivision of a simplicial complex $\ck$ is here understood as the simultaneous subdivision of all simplices in $\ck$.
\end{definition}
The simplest simplicial complex in the inductive system is called the initial complex and is denoted $\ck_o$. The orientation of the initial complex is not important and one simply chooses one.

The embedding (\ref{proj}) gives rise to a projection between spaces of abstract connections
\[
P_{n_2,n_1}:\ca_{\ck_{n_2}}\rightarrow \ca_{\ck_{n_1}}\;,\quad n_2\geq n_1\;,
\]
which, via (\ref{identification1}), is identified as a projection between manifolds
\ba 
P_{n_2,n_1}:  G^{n_2}\rightarrow G^{n_1}\;,\quad n_2\geq n_1\;.
\label{prO}
\ea
The projection $P_{n_2,n_1}$ is given by composition of one or more of the following operations: 
\begin{itemize}
\item 
multiplying $g_{i_1}$ with $g_{i_2}$.
\item 
inverting $g_i$.
\item 
leaving out some $g_i$ in $(g_1,\ldots,g_{n_2})\in G^{n_2}$.
\end{itemize}
%It is important that the inductive system of simplicial complexes involve a rather limited number of embeddings. Thus, at the level of associated manifolds, a given copy of $G$ will either be removed by a projection or merged with another copy of $G$ corresponding to a neighbouring edge. Or it will be inverted. The same copy of $G$ can never be removed by one projection and merged with another copy of $G$ by a second projection.

The embedding (\ref{proj}) commute by construction with the identification $\ca_\ck\simeq G^{n(\ck)}$, i.e. the diagram:
\begin{equation}
\begin{array}{lcl}
\ca_{\ck_{i+1}} & \stackrel{\s_{i+1}}{\longrightarrow} & G^{n(\ck_{i+1})}\nn\\
\;\downarrow &&\;\downarrow\nn\\
\ca_{\ck_{i}} & \stackrel{\s_{i}}{\longrightarrow} & G^{n(\ck_i)}\nn\\
\end{array}
\end{equation}
where $\s_i$ is given by (\ref{identification1}), commutes. This means that the limit space
\[
\overline{\ca}^{\smalltriangleup}:=\lim_{\stackrel{\ck}{\longleftarrow}}\ca_\ck
\]
is a pro-manifold. That is, it is the projective limit of manifolds. This gives us immediate access to various structures on $\overline{\ca}^{\smalltriangleup}$.
For instance, since the projections $P$ in (\ref{prO}) are continuous they give a topological structure on $\overline{\ca}^{\smalltriangleup}$. In general, the structure of a pro-manifold is a powerful tool that leads us to both Hilbert space and metric structures on the limit space  $\overline{\ca}^{\smalltriangleup}$. 

In section \ref{funcspace} we show that the space $\ca$ of smooth connections is densely embedded in $\overline{\ca}^{\smalltriangleup}$ (once a suitable embedding of the simplicial complex has been applied, see (\ref{EmBeD})). This fact is the very reason for studying the space $\overline{\ca}^{\smalltriangleup}$. It means that structures on $\overline{\ca}^{\smalltriangleup}$ can equally be understood as structures on $\ca$.

\section{A spectral triple on a simplicial complex}
\label{sec14}

The aim is to construct a spectral triple at the level of each simplicial complex in an inductive system of simplicial complexes. The triples are required to involve the loop algebras $\cb_{\ck_i}$ as function algebras over the associated spaces $\ca_{\ck_i}$ of abstract connections. To ensure that the limit of increased complexity is well defined we require the spectral triples to be compatible with all projections induced by the inductive system of complexes. This means that geometrical structures over $\ca_{\ck_i}$ will converge to geometrical structures over $\overline{\ca}^{\smalltriangleup}$.

\subsection{Outline of the construction.}

Before we go into details let us outline the construction of the spectral triple at the level of a simplicial complex $\ck$. The starting point is the manifold $\ca_{\ck}\simeq G^{n(\ck)}$. It is natural to consider first the Hilbert space
\[
\ch_\ck= L^2(G^{n(\ck)})\;,
\]
where $L^2$ is with respect to the Haar measure on $G^{n(\ck)}$. Since we wish to construct both a Dirac operator acting on $\ch_\ck$ and to have a representation of the algebra $\cb_\ck$ on $\ch_\ck$ we need to equip the Hilbert space with additional structure. Consider therefore instead the Hilbert space
\ba
\ch_\ck= L^2(G^{n(\ck)},Cl(T^*G^{n(\ck)})\otimes M_l(\mathbb{C}))\;,
\label{Hilbert}
\ea
where $l$ is the size of the representation of $G$ and $Cl(T^*G^{n(\ck)})$ is the Clifford bundle involving the cotangent bundle over $G^{n(\ck)}$. If we recall that points in $G^{n(\ck)}$ represents homomorphisms $\nabla$ from the hoop group into $G$ we immediately have a representation of the loop algebra $\cb_\ck$ on $\ch_\ck$
\ba
f_L\cdot \psi(\nabla)= (\One\otimes\nabla(L))\cdot\psi(\nabla)\;,\quad \psi\in\ch_\ck\;,
\label{rep}
\ea
where the first factor acts on the Clifford part of the Hilbert space and the second factor acts by matrix multiplication on the matrix part of the Hilbert space. Finally, we choose some Dirac operator $D_\ck$ on $G^{n(\ck)}$ and obtain the triple
\ba
(\cb_\ck,D_\ck,\ch_\ck)\;,
\label{suus}
\ea
on the level of the simplicial complex $\ck$.

It is not difficult to construct the candidate (\ref{suus}) for a spectral triple on the space $\ca_{\ck}$. The crucial point, however, is to ensure that the construction is compatible with the induced projections between the simplicial complexes (\ref{prO}). This requirement turns out to restrict the choice of Dirac operator on $\ca_\ck$ considerably.

To ease the notation we shall from now on write $\ca_i$ for $\ca_{\ck_i}$, $\cb_i$ for $\cb_{\ck_i}$, $\ch_i$ for $\ch_{\ck_i}$, $D_i$ for $D_{\ck_i}$ and $n_i$ for $n(\ck_i)$.

\subsection{Notation and basic setup}

Before we proceed with the construction we need some preparations. A point
 $(g_1,\ldots,g_n)\in G^n$ is denoted $\bar{g}$. Let $R_g$
 denote right translation on the group $G$, i.e. $R_g(h)=hg$. Accordingly,
 $L_g$ denotes left translation. We also denote by $R_g$ the corresponding
 differential, $R_g: T_hG\rightarrow T_{hg}G$. $L_g$ likewise. Given a cotangent vector $\phi$
 at the identity we define the right translated cotangent vector field $R\phi$
 by:
\[
R\phi(g)(v) = \phi(R_{g^{-1}}(v))\;,\quad v\in T_g G\;.
\]
The left translated cotangent vector fields are defined equivalently.
Given a projection $P:G^n\rightarrow G^{m}$ we denote by $P_\ast$ the
corresponding differential $$P_\ast: T_{\bar{g}}G^n\rightarrow
T_{P(\bar{g})}G^m $$ and by $P^\ast$ the induced map on cotangent spaces
$$P^\ast: T^\ast_{P(\bar{g})}G^m\rightarrow
T^\ast_{\bar{g}}G^n\;. $$
Consider the particular projection
\[
P: G^2\rightarrow G\;;\quad (g_1,g_2)\rightarrow g_1\cdot g_2\;.
\]
An element in $(v_1,v_2)\in T_{(g_1,g_2)}G^2$ transforms according to
\[
P_\ast (v_1,v_2)\rightarrow (R_{g_2}v_1+L_{g_1}v_2)\;,
\]
which is best seen by writing $v_1=\dot{\gamma}_1(\t)|_{\t=0}$ where $$\gamma_1(\t)=(\gamma_{11}(\t),g_2)\in G^2$$ is
the one parameter subgroup in $G^2$ generated by $v_1$. $P$ clearly maps
$\gamma_1(\t)$ into $\gamma_{11}(\t)\cdot g_2= R_{g_2}\gamma_{11}(\t)\in G$. The same
argument applies to $v_2$. Correspondingly, given an element of the cotangent
bundle $\phi\in T^\ast_{g_1\cdot g_2}G$ the dual maps yields
\ba
P^\ast \phi = (R_{g_2^{-1}}\phi,L_{g_1^{-1}}\phi)\in T^\ast_{(g_1,g_2)}G^2\;,
\label{cota}
\ea
where
\ba 
R_{g_2^{-1}}\phi(v)&=&\phi(R_{g_2}v)\;,\quad  v\in T_{g_1}G\nn\\
L_{g_1^{-1}}\phi(v)&=&\phi(L_{g_1}v)\;,\quad v\in T_{g_2}G\;.\nn
\ea

\subsection{The Hilbert space}
To construct the Hilbert space related to a simplicial complex
$\ck_i$ we first
 construct an inner product on  $T^\ast G^{n_i}$. We choose a left and right invariant
metric on $G$.
The edges in $\ck$ are numbered with $1,\ldots , n_i$. In the following we shall occasionally write $n$ instead of $n_i=n(\ck_i)$. Consider two
elements $\phi_1,\phi_2\in T^\ast_{\bar{g}}G^{n_i}$. We write
$\phi_k=(\phi_{k1},\ldots,\phi_{kn})$ and define the
inner product by
\ba
\langle \phi_1 |\phi_2 \rangle_{\bar{g}} = \frac{1}{2^N}\sum_{j=1}^{n} \langle \phi_{1j} |\phi_{2j} \rangle^G_{g_j}\;,
\label{innerproduct}
\ea
where $\langle \cdot|\cdot\rangle^G_{g_j}$ is the inner product on $T^\ast_{g_j}G$. In (\ref{innerproduct}) $N$ is the number\footnote{There is some freedom in the choice of $N$. We could also introduce a factor $N_j$ associated to each individual edge $\e_j$. Here, $N_j$ counts the number of barycentric subdivisions that lies between the edge $\e_j$ and the simplest complex $\ck_k$ ($k<j$) that has the edge $\e_j$ as a segment of an edge in $\ck_k$. This choice therefore associated different weights to edges in a simplicial complex according to their position in the complex.} of barycentric subdivision between the simplicial complex $\ck_i$ and the initial complex $\ck_o$. 
In \cite{Aastrup} we prove that the inner product (\ref{innerproduct}) is compatible with projections induced by embeddings (\ref{proj}). 

With the inner product (\ref{innerproduct}) on $T^\ast G^n$ we construct the Clifford bundle $Cl(T^\ast G^n)$ and define an inner product on the Hilbert space (\ref{Hilbert})
\ba
\langle\cdot|\cdot\rangle =\int d\m \cdot \Tr\cdot \langle \cdot|\cdot\rangle_{Cl} \;,
\label{ip2}
\ea
where $d\m$ is the Haar measure on $G^n$, $\Tr$ is the trace on $M_l$ and $\langle \cdot|\cdot\rangle_{Cl}$ is the inner product of the Clifford bundle $Cl(T^\ast G^n)$. The inner product (\ref{ip2}) is compatible with projections (\ref{proj}). This gives us the Hilbert space (\ref{Hilbert}).

\subsection{The Dirac operator}
\label{sectiondirac}

\begin{figure} [t]
\begin{center}
 \input{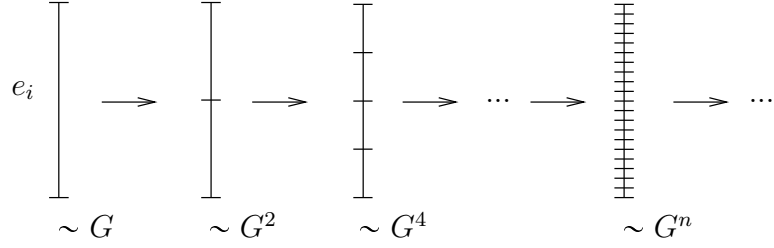}
\caption{an edge $e_i$ and its repeated barycentric subdivisions.}
\label{FigContBary}
\end{center}
\end{figure}

The Dirac-like operator on each space $\ca_{i}$ is required to be compatible with the projections in the inductive system of complexes. Thus, the induced maps
\ba 
P^*: T^*G^m\rightarrow T^*G^n\;,\quad \quad m<n\;,
\label{PrOj}
\ea
give rise to the compatibility conditions 
\ba
P^* (D_m v)(g_1,\ldots,g_n) = D_n (P^*v)(g_1,\ldots,g_n) \;,
\label{kom}
\ea
where $v\in L^2(G^m,Cl(T^*G^m))$. These conditions largely restrict the Dirac-like operator $D_i$ on $\ck_i$. On the initial complex $\ck_o$ with the corresponding manifold $G^{n_o}$ the Dirac operator is chosen to have the canonical form\footnote{There is some freedom of choice here since the Dirac operator on $\ck_o$ need not involve the Levi-Civita connection. However, to obtain a self-adjoint operator certain invariance properties must be satisfied \cite{Aastrup}.}
\ba
D_o = \sum_{i,j} e^i_j\cdot \levi_{\hat{e}^i_j}\;,
\label{dirac1}
\ea
where $\{e_j^i\}$ is a global basis on $T^*G$ corresponding to the $i$'th copy of $G$. $\{\hat{e}^i_j\}$ is the corresponding basis on $TG$, obtained from $\{e_j^i\}$ via the inner product (\ref{innerproduct}). $\levi$ is the Levi-Civita connection on $G$. 

The problem is therefore to find a Dirac operator on $G^{n_i}$ compatible with projections (\ref{proj}). Here, it is sufficient to deal with an edge $\e$ and its partitions generated by barycentric subdivisions of the complex to which $\e$ belongs, see figure \ref{FigContBary}. Once the operator is constructed here we obtain an operator on the entire system of simplicial complexes by gluing the individual operators in an obvious manner.

The generic problem is the projection 
\ba
P:G^2\rightarrow G\;;\quad (g_1,g_2)\rightarrow g_1\cdot g_2\;,
\label{proje}
\ea
which we now consider together with the induced map between cotangent bundles (\ref{cota}).
Let $\{e_i\}$ be an orthonormal basis of $T^*_{id}G$ and denote by
\[
e_i(g) = L_g e_i (id)\equiv e_i
\]
the left-translated basis covectors at $g\in G$.
The push-forward of the basis covector $e_i(g)$ in $T^*_{(g_1,g_2)} G$ by $P^*$ gives
\[
P^*e_i(g_1\cdot g_2)=(R_{g_2^{-1}}e_i(g),L_{g_1^{-1}}e_i(g))\;,
\]
where $g=g_1\cdot g_2$. This suggests a natural orthonormal basis of $T^*_{(g_1,g_2)}G^2$ given by
\ba 
\ce^{2,s}_i=(\ce_i^1,\pm\ce_i^2)\;,
\label{firstE}
\ea
where
\ba
\ce_i^1(g_1)=L_{g_1\cdot g_2}R_{g_2^{-1}}e_i(id)\;,\quad \ce_i^2(g_2)=L_{g_2}e_i(id)\;,
\label{ba1}
\ea
and where $s$ in (\ref{firstE}) represents the appropriate sign combinations
\[
s\in\{ (+,+), (+,-) \}
\]
characterising the two orthonormal covectors. Denote by $\hat{\ce}^j_i$ and $\hat{\ce}^{2,s}_i$ the corresponding sections in $TG$ and $TG^2$ respectively, defined via the inner product (\ref{innerproduct}).

We consider a Dirac operator corresponding to $G^2$ of the form
\ba
D_2=\frac{1}{2}\sum_{s,i}\ce^{2,s}_i\cdot\nablato_{\hat{\ce}^{2,s}_i}\;,
\label{dirac2}
\ea
where $\nablato$ is a connection on $T^*G^2$. It turns out that the operator (\ref{dirac2}) satisfies the compatibility condition (\ref{kom}) if the connection $\nablato$ satisfies
\ba
\nablato_{(\hat{\ce}^1_i,0)}(\ce^1_i,0)&\equiv&(\levi_{\hat{\ce}^1_i}\ce^1_i,0)\;,
\nn\\
\nablato_{(0,\hat{\ce}^2_i)}(0,\ce^2_i)&\equiv&(0,\levi_{\hat{\ce}^2_i}\ce^2_i)\;.
\nn
\ea
%Next, we find
%\ba
%\lefteqn{D (P^* e_j) (g_1,g_2)}\nn\\&=& \frac{1}{2}\sum_i\Big( (\ce^1_i,\ce^2_i)\cdot\nablato_{(\ce^1_i,\ce^2_i)} (R_{g_2^{-1}}e_j(g_1\cdot g_2),L_{g_1^{-1}}e_j(g_1\cdot g_2))
%\nn\\
%&&+ (\ce^1_i,-\ce^2_i)\cdot\nablato_{(\ce^1_i,-\ce^2_i)} (R_{g_2^{-1}}e_j(g_1\cdot g_2),L_{g_1^{-1}}e_j(g_1\cdot g_2))\Big)
%\nn\\
%&=&\frac{1}{2}\sum_i\Big((\ce^1_i,\ce^2_i)\cdot
%\nablato_{(\ce^1_i,\ce^2_i)} (\ce^1_j(g_1),\ce^2_j(g_2))
%\nn\\&&
%+(\ce^1_i,-\ce^2_i)\cdot
%\nablato_{(\ce^1_i,-\ce^2_i)} (\ce^1_j(g_1),\ce^2_j(g_2))
%\label{no2}\ea
%The equality between (\ref{no1}) and (\ref{no2}) holds if we introduce the following conditions
%\ba
%\nablato_{(\ce^1_i,\ce^2_i)}(\ce^1_j,\ce^2_j)&\equiv&(\levi_{\ce^1_i}\ce^1_j,\levi_{\ce^2_i}\ce^2_j)\;,
%\nn\\
%\nablato_{(\ce^1_i,-\ce^2_i)}(\ce^1_i,\ce^2_i)&\equiv&0\;,
%\label{condition1}
%\ea
%which are sufficient to define a connection leading to a Dirac operator compatible with (\ref{proje}).

%We fix the remaining freedom in $\nablato$ by the following condition
%\ba
%\nablato_{(\ce^1_i,\pm\ce^2_i)}(\ce^1_j,-\ce^2_j)&\equiv&0\;.
%\label{conditionextra}
%\ea
%The additional condition (\ref{conditionextra}) permits the construction of the trace, see section \ref{sectiontrace}.

To obtain the general form of the Dirac like operator on $G^n$ we first define the following twisted covectors on $G$
\ba
\ce^1_i(g_1)&=&L_{g_1\cdot\ldots\cdot g_n}R_{(g_2\cdot\ldots\cdot g_n)^{-1}}e_i(id)
\nn\\
&\vdots&
\nn\\
\ce^j_i(g_j)&=& L_{g_j\cdot\ldots\cdot g_n}R_{(g_{j+1}\cdot\ldots\cdot g_n)^{-1}}e_i(id)
\nn\\
&\vdots&
\nn\\
\ce^n_i(g_n)&=& L_{g_n}e_i(id)\;.
\ea
Next, we write 
$$
\ce^{n,s}_i=(\ce^1_i,\pm\ce^2_i,\ldots,\pm\ce^n_i)\;,
$$
where $s=(+,\pm,\ldots,\pm)$ is the sequence of signs which characterises the covector. Again, we denote by $\hat{\ce}^{n,s}_i$ the corresponding sections in $TG^n$. 
The global frames $\ce^{n,s}_i$ are found by repeated lifts of the covector $e_i(g)$ on $G$ to $G^n$. Therefore, they are constructed to satisfy 
\[
\ce^{n,s}_i= P^*(\ce^{n/2,s'}_i)\;,
\]
where the sequence $s$ is obtained from the sequence $s'$ by replacing each sign in $s'$ with the same sign twice, and where
\ba
P:G^n\rightarrow G^{n/2}\;;\quad(g_1,\ldots,g_n)\rightarrow (g_1\cdot g_2,\ldots,g_{n-1}\cdot g_n)\;.
\label{projek}
\ea
The Dirac like operator on $G^n$ has the form
\ba
D_n = \frac{1}{n}\sum_{s,i}\ce^{n,s}_i\cdot \nabla^{\mbox{\tiny\bf n}}_{\hat{\ce}^{n,s}_i}\;,
\label{DIRAC}
\ea
where the sum runs over $i$ as well as all appropriate sign sequences $s$. 

We define the connections $\nabla^{\mbox{\tiny\bf n}}$ recursively. That is, the action of $\nabla^{\mbox{\tiny\bf n}}$ on basis covectors $\ce^{n,s}_i$ are given recursively and thereafter extended via linearity and the requirements of a derivation to the entire Clifford bundle. Thus, we require
%\ba
%\nabla^{\mbox{\tiny\bf n}}_{(\bar{\ce}^1,0)}(\bar{\ce}^1,0)&=& 
%(\nabla^{\mbox{\tiny\bf n/2}}_{\bar{\ce}^1}\bar{\ce}^1,0)
%\nn\\
%\nabla^{\mbox{\tiny\bf n}}_{(\bar{\ce}^1,0)}(0,\bar{\ce}^2)&=& 
%(0,\nabla^{\mbox{\tiny\bf n/2}}_{\bar{\ce}^2}\bar{\ce}^2)
%\nn\\
%\nabla^{\mbox{\tiny\bf n}}_{(0,\bar{\ce}^2)}(\bar{\ce}^1,0)&=& 
%(\nabla^{\mbox{\tiny\bf n/2}}_{\bar{\ce}^1}\bar{\ce}^1,0)
%\nn\\
%\nabla^{\mbox{\tiny\bf n}}_{(0,\bar{\ce}^2)}(0,\bar{\ce}^2)&=& 
%(0,\nabla^{\mbox{\tiny\bf n/2}}_{\bar{\ce}^2}\bar{\ce}^2)
%\nn
%\ea
\ba
\nabla^{\mbox{\tiny\bf n}}_{P^*\hat{\ce}}(P^*\ce)&=&P^*(\nabla^{\mbox{\tiny\bf n/2}}_{\hat{\ce}} \ce)\;,
\nn\\
\nabla^{\mbox{\tiny\bf n}}_{(P^*\hat{\ce})_{\perp}}(P^*\ce)&=&0\;,
\label{CONDI}
\ea
where $\ce$ and $\hat{\ce}$ are basis covectors and vectors, respectively, of the type $(\ce^1_i,\pm\ce^2_i,\ldots,\pm\ce^{n/2}_i)$ at the level $\frac{n}{2}$. In equation (\ref{CONDI}) and in the following we denote by $(P^*\hat{\ce})_{\perp}$ general elements in the orthogonal complement to vectors (and covectors) of the form $P^*\hat{\ce}$.
We fix the remaining freedom in $\nabla^{\mbox{\tiny\bf n}}$ with the additional condition
\[
\nabla^{\mbox{\tiny\bf n}}_{(P^*\hat{\ce})_{\perp}}(P^*\bar{\ce})_{\perp} = \nabla^{\mbox{\tiny\bf n}}_{(P^*\hat{\ce})}(P^*\bar{\ce})_{\perp}=0\;,
\]
which is required for the construction of the trace, see section \ref{sectiontrace} and 
\cite{Aastrup}. The properties given in (\ref{CONDI}) are again dictated by the requirement that the Dirac operator (\ref{DIRAC}) is compatible with the projections.

The proof that the operator (\ref{DIRAC}) satisfies the compatibility condition (\ref{kom}) is given in \cite{Aastrup}.

\subsection{The general Dirac operator associated to $\ck_i$}

The Dirac type operator (\ref{DIRAC}) is not the most general operator satisfying the requirements of compatibility with the structure maps (\ref{kom}). In fact, these requirements renders substantial parts of the operator (\ref{DIRAC}) free to modifications. This observation is closely related to the fact that the Dirac type operator (\ref{DIRAC}) will, at it stands, not descend to an operator with a compact resolvent in the inductive limit of repeated barycentric subdivisions. It turns out that the additional degrees of freedom are exactly the necessary leverage needed to obtain a well-behaved Dirac type operator in the limit.

Let us start with the case $n=2$. We observe that a rescaling of parts of the operator (\ref{dirac2}) according to
\ba
D_2=\frac{1}{2}\sum_i\Big( (\ce^1_i,\ce^2_i)\cdot\nablato_{(\ce^1_i,\ce^2_i)} + a_1 (\ce^1_i,-\ce^2_i)\cdot\nablato_{(\ce^1_i,-\ce^2_i)}\Big)\;.
\label{dirac3}
\ea
where $a_1\in\mathbb{R}$, does not affect the compatibility with the embedding (\ref{cota}). In the general case, the modification of (\ref{DIRAC}) is as follows: Let $\{a_k\}$ be an infinite sequence of real numbers and put $a_0=1$. Consider a single edge $\e$ and its subdivisions. Consider the $n$'th subdivision. Let $s$ be a finite sequence of $n$ signs and denote by $\#(s)$ the number of '+'s in the beginning of the sequence. Define the number
\[
m(s)=\log_2\Big(\frac{2^N}{\#(s)} \Big)\;.
\]
The modified Dirac type operator now has the form
\ba
D_n = \frac{1}{n}\sum_{s,i}a_{m(s)}\ce^{n,s}_i\cdot \nabla^{\mbox{\tiny\bf n}}_{\hat{\ce}^{n,s}_i}\;.
\label{DIRACmod}
\ea
Let us consider the eigenvalues of this operator. For simplicity, consider the Abelian case $G=U(1)$. If we define the product between a sequence $s=(+,\pm,\pm,\ldots)$ of signs with a sequence of real numbers\footnote{or, to be general, any sequence of objects.} $(n_1,n_1,\ldots)$ by
\ba
(n_1,n_2,\ldots)\cdot s =n_1\pm n_2\pm \ldots
\label{cdot}\ea
where the signs on the rhs are read of the sequence $s$, then we can write the spectrum of the Dirac type operator (\ref{DIRACmod}) as
\[
\mbox{spec}(D) =\left\{\pm \frac{1}{2^N}\sqrt{\sum_{k}a^2_{m(s_k)}((n_1,n_2,\dots,n_{2^N})\cdot s_k)^2}\right\}\;,
\]
where $s_k$ is the sequence of signs corresponding to the $k$'th orthonormal vector $\ce^{n,s_k}$. In this case, the number of eigenvalues of the limit operator $D_{\smalltriangleup}=\lim_n D_n$ in a range $[0,\Lambda],\Lambda<\infty$, is finite whenever
\ba 
a_n = 2^n b_n\;\,\quad \mbox{where} \;\;\; \lim_{n\rightarrow\infty} b_n=\infty\;.
\label{CONDITION}
\ea
The general case where $G$ is a non-Abelian Lie-group is more complicated. The full analysis is given in \cite{Aastrup} where we prove that for any compact Lie-group $G$ there exist a sequence $\{a_i\}$ so that the number of eigenvalues within a finite range is finite up to a controllable degeneracy. This result will be clarified in the next section.

\section{The limit of the triple $(\cb_n,D_n,\ch_n)$}
\label{projlim}

Up till now we have considered a system of embedded, abstract simplicial complexes $\{\ck_n\}$. To each simplicial complex $\ck_n$ we introduced a space $\ca_{n}$ which we interpreted as a coarse-grained version of a space of connections. 
The next step was to construct a spectral triple $(\cb_n,D_n,\ch_n)$ on each of the spaces $\ca_{n}$. 
The spectral triple satisfies requirements of compatibility with the operation of barycentric subdivision. This means that we can take the limit of infinitely many barycentric subdivisions. The resulting triple, which is denoted by $(\cb_{\smalltriangleup},D_{\smalltriangleup},\ch_\smalltriangleup)$, has the following elements:

 First, the Hilbert space $\ch_\smalltriangleup$ is constructed by adding all Hilbert spaces $\ch_{n}$
\ba
\ch'=\oplus_{\ck}L^2 (G^{n(\ck)},Cl(T^*G^{n(\ck)})\otimes M_l(\mathbb{C}))/N\;, 
\label{alg}
\ea 
where $N$ is the subspace generated by elements of the form 
\[
(\ldots , v, \ldots , -P^*_{ij}(v),\ldots )\;,
\label{N}
\] 
where $P^*_{ij}$ are maps between Hilbert spaces $\ch_n$ induced by (\ref{PrOj}). The Hilbert space $\ch_\smalltriangleup$ is the completion of $\ch'$. The inner product on $\ch_{\smalltriangleup}$ is the inductive limit inner product. This Hilbert space is manifestly separable. 
Next, the algebra
\ba
\cb_{\smalltriangleup}:= \lim_{\stackrel{\ck}{\longrightarrow}}\cb_{\ck}\;.
\label{cb-tri}
\ea
contains loops defined on a simplicial complex $\ck_n$ in $\{\ck_n\}$ as well as their closure. Again, the algebra $\cb_{\smalltriangleup}$ is separable.
Finally, the Dirac-like operator $D_n$ descends to a densely defined operator on the limit space $\ch_{\smalltriangleup}$
\ba
D_{\smalltriangleup} = \lim_{\stackrel{\ck}{\longrightarrow}}D_n\;.
\label{didi}
\ea

\subsection{A semi-finite spectral triple}
\label{sectiontrace}

The question is whether the triple $(\cb_{\smalltriangleup},D_{\smalltriangleup},\ch_\smalltriangleup)$ satisfies the requirements of a spectral triple. Clearly, to have a true spectral triple is highly desirable, first of all to ensure that we are operating on mathematically safe ground. Also, the toolbox of Noncommutative Geometry is a rich one. It provides not only metric structures to the underlying spaces involved, it also involves aspects which does not exist in ordinary Riemannian geometry.

To have a spectral triple $(B,D,H)$ where $B$ is a $C^\star$-algebra represented on a Hilbert space $H$ on which an unbounded, selfadjoint operator $D$ acts, means that the two requirements
\begin{enumerate}
\item
$(\l - D)^{-1}$, where $\l\not\in\mathbb{R}$, is a compact operator
\item
$[b,D]$, where $b\in \cb$, is a bounded operator and where $\cb$ is a dense $\star$-subalgebra of $B$
\end{enumerate}
are satisfied. However, the triple $(\cb_{\smalltriangleup},D_{\smalltriangleup},\ch_\smalltriangleup)$ is not spectral in this sense. The spectrum of $D_{\smalltriangleup}$ involves a large degeneracy due to the infinite dimensional Clifford bundle. This means that the resolvent of $D$ will not be compact.

There exist, however, another sense in which a triple $(B,D,H)$ can be called spectral. If there exist a trace on the algebra containing $B$ and the spectral projections of $D$, then $(B,D,H)$ is called a semi-finite spectral triple if $(\l-D)^{-1}$ is compact with respect to this trace.

It turns out that the triple $(\cb_{\smalltriangleup},D_{\smalltriangleup},\ch_\smalltriangleup)$ does form a semi-finite spectral triple.  
The sufficient requirement is that the sequence $\{a_k\}$ satisfies
\ba
\lim_k a_k\rightarrow \infty\;,\quad \mbox{sufficiently fast}\;.
\label{suff}
\ea

A semi-finite spectral triple can be thought of as a spectral triple which involves a redundant symmetry group. This symmetry group resembles a gauge group and one must integrate out the extra degrees of freedom. In the present case the symmetry group is identified as the endomorphisms of the infinite dimensional Clifford bundle. To deal with this redundancy we first define a trace on the algebra containing $\cb_\smalltriangleup$ as well as the spectral projections of $D_{\smalltriangleup}$.

We will construct the trace at each level in the inductive system of simplicial complexes and subsequently take the limit of repeated barycentric subdivisions. For the construction we factorize the Hilbert space. Choose a global orthonormal frame in $T^*_eG$. This choice gives rise to a decomposition 
\ba 
L^2(G^n, M_l\otimes Cl(T^* G^n ))=L^2(G^n )\otimes M_l(\mathbb{C}) \otimes Cl(T^*_e G^n).
\label{CliFF}
\ea
We will construct the trace on the algebra 
$$C_n=\ck (L^2(G^n ))\otimes End (M_l) \otimes End (Cl(T^*_e G^n)).$$
where $\ck$ here denotes the compact operators. 
%Note that $\cb_n$ and $\ck (L^2(G^n ))\otimes M_N \otimes End(Cl(T^*_e G^n))$ have zero intersection and that 
%$$\cb_n \cdot \ck (L^2(G^n ))\otimes M_N \otimes End(Cl(T^*_e G^n))\subset \ck (L^2(G^n ))\otimes M_N \otimes End(Cl(T^*_e G^n)).$$

Let $Tr_{op}$ denote the operator trace on $\ck (L^2(G^n ))$. For each $n$ we have the normalised trace $tr$ on $End (Cl(T^*_e G^n)).$ 
We define the trace as
\ba 
Tr = Tr_{op}\otimes Tr_l\otimes tr.
\label{TraCE}
\ea
Note that the trace is independent of the choice of global orthonormal frame  in $T^*G$, since a different choice of basis is given by a unitary transformation.

In \cite{Aastrup} we prove that the limit $$C_{\smalltriangleup}:=\lim C_{n}$$ will be a $C^*$-algebra and that the trace $Tr$ gives a trace on $C_{\smalltriangleup}$. Since $C_{\smalltriangleup}$ is contained in the weak closure of $\cb_{\smalltriangleup}$ the trace extends to a trace on $\cb_{\smalltriangleup}$ as well.

The important point in this construction is the normalisation of the trace $tr$ on $End (Cl(T^*_e G^n))$. To explain this consider the step going from $n$ to $n+m$. 
Let $P_\lambda$ be the spectral projection of $D_n$ for the eigenvalue $\lambda$. Going from $n$ to $n+m$ the spectral projection will roughly speaking be mapped to $P_\lambda\otimes \bf{1}$. Thus the size of the eigenspace $\lambda$ grows in the same rate as the dimension of the Clifford bundle. To remedy this defect we must ensure that $$tr ({\bf 1})=1\;,$$ 
which is what the normalised trace does.

The proof that $(\cb_{\smalltriangleup},D_{\smalltriangleup},\ch_\smalltriangleup)$ form a semi-finite spectral triple with respect to the trace (\ref{TraCE})
is given in \cite{Aastrup}. 
 It turns out that in general the Dirac-type operator (\ref{didi}) may require a perturbation at each level in the inductive system. This perturbation deals with the fact that the operators $D_n$ may, in general, have nontrivial kernels which obstructs the control of the eigenvalues of the limit operator $D_{\smalltriangleup}$. These perturbations lift the entailed degeneracy, which would otherwise destroy the spectral properties of the triple. 
The required perturbation is, at each level in the projective system, bounded and does not affect the commutator between $D_{\smalltriangleup}$ and the algebra $\cb_{\smalltriangleup}$ significantly. For the special case $G=SU(2)$ we find that the operators $D_n$ have trivial kernels and therefore that no perturbations are needed.\\

To recapitulate, we have successfully constructed a large class of semi-finite spectral triples $(\cb_{\smalltriangleup},D_{\smalltriangleup},\ch_\smalltriangleup)$. The triples are labelled by infinite sequences $\{a_k\}$ of real numbers satisfying $\lim_k a_k\rightarrow\infty$ sufficiently fast.\\

Let us end this section with a short discussion of the structure of the Hilbert space $\ch_\smalltriangleup$ the role of the infinite dimensional Clifford bundle. Consider first the decomposition in (\ref{CliFF}) and rewrite it in the suggestive form
\[
L^2(G^n )\otimes M_l(\mathbb{C}) \otimes Cl(T^*_e G^n)= \ch_{n,b}\otimes \ch_{n,f}\;,
\]
where $\ch_{n,b}=L^2(G^n )\otimes M_l(\mathbb{C})$. In the inductive limit this leads to the decomposition
\[
\ch_\smalltriangleup = \ch_{\smalltriangleup,b}\otimes \ch_{\smalltriangleup,f}\;.
\]
Next we factorize the algebra $\ch_\smalltriangleup$ according to the above factorisation of the Hilbert space. The result is 
$$\cc_\smalltriangleup=\ck ( \ch_{\smalltriangleup,b})\otimes C , $$
where 
$$C=\lim_n End( Cl(T^*_eG^n)),$$
where the morphisms in this inductive system are the unital ones. In particular $C$ is a UHF-algebra and since $Cl(T^*_eG^n)$ has dimension $2^{n\cdot dim (G)}$, the $C^*$-algebra $C$ is isomorphic to the CAR algebra.

The CAR algebra is an integral element of fermionic Quantum Field Theory. We find it interesting that this algebra naturally emerges from the triple $(\cb_{\smalltriangleup},D_{\smalltriangleup},\ch_\smalltriangleup)$ which, a priori, is an entirely 'bosonic' construction. In section \ref{DISCu} we shall comment further on the appearance of the CAR algebra.

\subsection{The space $\overline{\ca}^{\smalltriangleup}$ and generalised connections}
\label{funcspace}

\begin{figure} [t]
\begin{center}
 \input{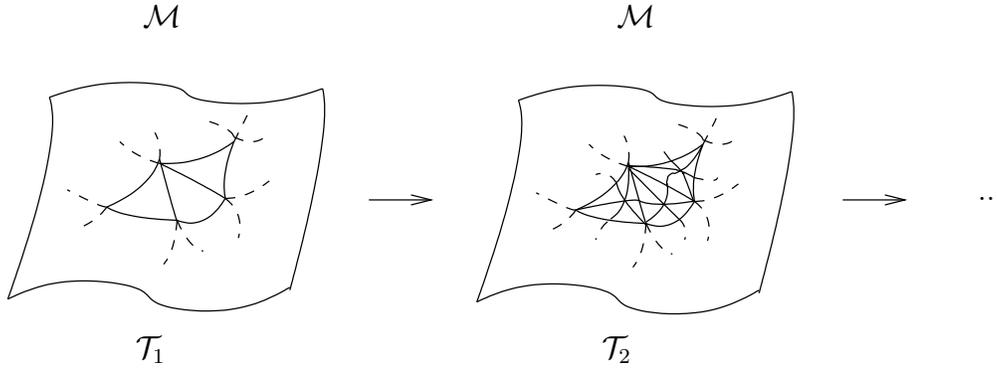}
\caption{The embedding of the system $\{\ck_i\}$ into $\cm$ gives rise to an inductive system of triangulations $\{\ct_i\}$.}
\label{JohannesFig}
\end{center}
\end{figure}
We have already indicated that the space $\overline{\ca}^{\smalltriangleup}$ is a space of generalised connections. This means that it is a closure of the space $\ca$ of smooth $G$-connections. To see this consider the embedding $\phi$ of the simplicial complex $\ck_i$
\[
\ct_i :=\phi(\ck_i)\;,
\]
where $\ct_i$ is a triangulation of the manifold $\cm$. There is a natural map 
\ba
\chi_{\smalltriangleup}:\ca\rightarrow\overline{\ca}^{\smalltriangleup}\;,\quad\chi_{\smalltriangleup}(\nabla)(\phi(\e_i))=Hol(\nabla,\phi(\e_i))\;,
\label{mAp}
\ea
where $Hol(\nabla,\phi(\e_i))$ is the holonomy of $\nabla$ along the edge $\phi(\e_i)$ which now has a location in $\cm$ via the embedding $\phi$. Further, if we are given two different connections $\nabla_1,\nabla_2\in\ca$ they will differ in a point, say $m\in\cm$, and hence in a neighbourhood of $m$. We can therefore choose a small, directed edge $\phi(\e_i)$ in a triangulation $\ct_i$ that is sufficiently refined, so that $\phi(\e_i)$ lies in the neighbourhood of $m$ where the connections $\nabla_1$ and $\nabla_2$ differ. Furthermore, because the system of triangulations contain edges in all directions in $\cm$ we can choose $\phi(\e_i)$ so that $$Hol(\nabla_1,\phi(\e_i))\not=Hol(\nabla_2,\phi(\e_i))\;.$$ In other words, $\chi_{\smalltriangleup}$ is an embedding, and hence the terminology generalised connection is justified.

This means that we have successfully turned the map (\ref{firstchi}) into an injection by repeating the barycentric subdivisions of the simplicial complexes. 

The identification of $\overline{\ca}^{\smalltriangleup}$ as a space of connections provides us with a new understanding of the spectral triple $(\cb_{\smalltriangleup},D_{\smalltriangleup},\ch_\smalltriangleup)$. First, as already mentioned, according to (\ref{mAp}) the algebra of loops should be interpreted in terms of holonomy loops
\[
f_L(\nabla)\sim Hol(L,\nabla)\;.
\]
It is a well known result (see \cite{Barrett:1991aj} and \cite{Loll:1993yz} and references therein) that the complete set of holonomies, as well as their associated Wilson loops, contain the full information, up to gauge transformations, about the underlying space of smooth connections. 
The fact that the map (\ref{mAp}) is an injection means that the algebra $\cb_{\smalltriangleup}$, which involves a highly restricted set of loops, does in fact contain the same information about the underlying space of connections as does the full set of smooth or piece-wise analytic loops.

Second, since the Dirac-type operator $D_{\smalltriangleup}$ is a derivation on the space $\overline{\ca}^{\smalltriangleup}$ it should be understood in terms of a functional derivation operator
\[
D_{\smalltriangleup} \sim \frac{\d}{\d \nabla}
\]
in some integrated sense which shall become clearer soon. Finally, elements in the Hilbert space $\ch_\smalltriangleup$ are functions over field configurations of connections and the inner product in $\ch_\smalltriangleup$ comes in the form of a functional integral
\ba
\langle\Psi(\nabla)|\ldots |\Psi(\nabla)\rangle\sim\int_{\overline{\ca}^{\smalltriangleup}}[d\nabla]\ldots\;,\quad\Psi\in\ch_{\smalltriangleup}\;.
\label{funcint}
\ea
That is, an integral\footnote{This integral resembles functional integrals found elsewhere in physics. First, it is similar to the inner product on the Hilbert space $\ch_{\small diff}$ of diffeomorphism invariant states in Loop Quantum Gravity, see below. Second, it also resembles functional integrals in lattice gauge theories. Here, the main difference is the 'lattice spacing' $a$ in lattice gauge theories which gives the continuum limit $a\rightarrow 0$.  } over $\overline{\ca}^{\smalltriangleup}$.

 All together it is clear that the construction should be interpreted in terms of Quantum Field Theory. \\

Notice that the diffeomorphism group diff($\cm$) has no natural action on $\overline{\ca}^{\smalltriangleup}$ or the algebra $\cb_{\smalltriangleup}$, except for a few, discrete diffeomorphisms. We shall comment on this fact in section \ref{link}.

\subsection{Gauge transformations}

It remains to clarify whether the Dirac type operator $D_{\smalltriangleup}$ is gauge invariant. To this end
let $U$ be an element of the gauge group $\cg$ of $M\times P$, i.e. $U:M\rightarrow G$ is a smooth function. Given a connection $\nabla\in\ca$, $U$ induces a gauge transformed connection $\tilde{\nabla}$. Given a path $L$ with startpoint $x_1$ and endpoint $x_2$ the holonomy along $L$ transforms according to
\[
Hol(\nabla,L)\rightarrow U(x_1)Hol(\nabla,L)U^\ast(x_2)\;.
\]
To determine the properties of the Dirac type operator $D_\smalltriangleup$ when subjected to a gauge transformation we consider first the case $\ca_n=G^n$ that corresponds to $n$ divisions of a single edge. Consider the general transformation
\[
U_n: G^n\rightarrow G^n\;,\quad (g_1,g_2,\ldots,g_n)\rightarrow (u_0 g_1 u_1^{-1},u_1 g_2 u_2^{-1},\ldots,u_{n-1} g_n u_{n}^{-1} )\;,
\]
where $u_0,u_1,\ldots,u_n$ are unitary group elements in $G$. This transformation generates a map
\[
U_n:L^2(G^n,Cl(T^\ast G^n))\rightarrow L^2(G^n,Cl(T^\ast G^n))
\]
between Hilbert spaces. We need to check whether 
\ba 
D_n\xi = U_n D_n U_n^\ast \xi\;,\quad \xi\in L^2(G^n,Cl(T^\ast G^n))\;.
\label{COMPa}
\ea
In \cite{Aastrup} we find that (\ref{COMPa}) holds whenever the connections used to construct $D_n$ satisfy a certain gauge compatibility condition. In particular, we find that the special flat connections entering the construction of $D_n$ do satisfy this condition. This, in turn, implies that the full Dirac type operator $D_{\smalltriangleup}$ is gauge invariant. For the full analysis we refer the reader to \cite{Aastrup}.

\subsection{The commutator between $D_{\smalltriangleup}$ and the algebra $\cb_{\smalltriangleup}$}
\label{commutatorsection}

Section \ref{cangrav} is concerned with the relation between the Poisson algebra of General Relativity and the algebra $\cb_{\smalltriangleup}$. The point is that the interaction between the Dirac type operator $D_{\smalltriangleup}$ and the algebra $\cb_{\smalltriangleup}$ equals the interaction between conjugate variables of gravity. Before we show this we need to calculate the commutator between the operator (\ref{DIRACmod}) and a loop operator. 

First of all, the commutator is non vanishing
\[
[D_{\smalltriangleup},b]\not = 0\;,\quad b\in\cb_{\smalltriangleup}
\]
since, on the level of refinement corresponding to $n$ edges, $b$ is a non-trivial matrix valued function on $G^n$ and $D_{\smalltriangleup}$ is the Dirac-type operator on $G^n$. 

Consider first the simple case where a loop $L_j$ corresponds to the function on $G^n$
\[
f_{L_j}:(g_1,\ldots,g_n)\rightarrow g_j\;.
\]
Thus, the loop $L_{j}$ is really just the $j$'th edge. We wish to calculate the commutator
\[
[d_{\hat{\ce}^j_i},f_{L_j}]\xi(g_1,\ldots,g_n)\;,
\]
where $\xi\in L^2(G^n)$. If we denote by $\mathfrak{e}_i$ the generators of the Lie algebra $\mathfrak{g}$, then we introduce the twisted generators of $\mathfrak{g}$
\[
\mathfrak{E}_i^j= g_{j+1}g_{j+2}\ldots g_n\mathfrak{e}_i g^{-1}_{n}\ldots g^{-1}_{j+1}
\]
corresponding to $\hat{\ce}^j_i$. We now calculate
\ba 
[d_{\hat{\ce}^j_i},f_{L_j}]\xi(g_1,\ldots,g_n)&=& \left(d_{\hat{\ce}^j_i}f_L\right)  \xi(g_1,\ldots,g_n)
\nn\\
&=&\frac{d}{dt}\left( g_j\ldots g_n \exp{(t e_i)} g_n^{-1}\ldots g^{-1}_{j+1} \right)\xi(g_1,\ldots,g_n)
\nn\\
&=&g_j\mathfrak{E}_i^j\xi(g_1,\ldots,g_n)\;.
\label{cccom}
\ea
It is important to notice that product in the last line of (\ref{cccom}) is a matrix multiplication.

 Equation (\ref{cccom}) implies that\footnote{we here consider the subdivision of an edge into $n$ segments. Therefore, $D_n$ is the operator given by (\ref{DIRACmod}).}
\ba
[D_n,f_{L_j}]=\frac{1}{n}\sum_{s,i}\pm a_{m(s)}\ce^{n,s}_i \cdot g_j \mathfrak{E}^j_i  \;.
\label{comg}
\ea
The sign on the rhs of (\ref{comg}) correspond to the $j$'th sign in the sequence $s$. Thus, for $i=1$ all the signs are positive and else the total number of $+$'s and $-$'s
are equal.

Next, the commutator between $D_{\smalltriangleup}$ and a general loop $L$ 
\[
f_L:(g_1,\ldots,g_n)\rightarrow g_{i_1} g_{i_2}\ldots g_{i_k}\;.
\]
simply consist of repeated applications of $(\ref{comg})$ according to
\ba
[D_{\smalltriangleup},f_L]= [D_{\smalltriangleup},f_{L_{i_1}}]g_{i_2}\ldots g_{i_k} + g_{i_1}[D_{\smalltriangleup},f_L{_{i_2}}]\ldots g_{i_k} + \ldots
\label{coml}
\ea
where each commutator on the rhs is calculated by inserting the appropriate operator $D_n$ corresponding to the level of refinement given by the loop $L$.

Thus, the action of $D_{\smalltriangleup}$ on a single loop is to insert the 'twisted' generators $\mathfrak{E}^j_i$ of the Lie-algebra into the loop at each vertex the loop passes through and to multiply with an appropriate element in the Clifford bundle. 
%In the next section we show that this action of $D_{\smalltriangleup}$ can be understood as a functional derivation on an underlying space of connections and that the commutators (\ref{comg}) and (\ref{coml}) are related to the Poisson brackets of General Relativity.

%Let us end this section with the observation that a loop operator $f_L$ does not map the subspace of $\ch^+$ of states in $\ch_{\ck_i}$ with positive eigenvalues of $D_{\smalltriangleup}$ into itself. Thus, if we denote by $P^\pm$ projections into positive/negative eigenspaces, then it is easy to see that
%\[
%P^{+} f_L P^{-}\Psi\not=0\;,
%\]
%where $\Psi\in\ch_{\ck_i}$.

\subsection{The role of the sequence $\{a_n\}$}
\label{rolE}

In section \ref{includea} we will include the sequence $\{a_n\}$ as dynamical variables in the construction. To motivate this step we need a better understanding of the sequence $\{a_n\}$ and the role it plays in the spectral triple $(\cb_{\smalltriangleup},D_{\smalltriangleup},\ch_\smalltriangleup)$.

Primarily, the role of the sequence $\{a_n\}$ is to shift an otherwise impossible degeneracy in the spectrum of the operator $D_{\smalltriangleup}$ via the condition (\ref{suff}).
The parameters $a_i$ introduces a hierarchy between the eigenvalues of $D_{\smalltriangleup}$ on the infinitely many copies of $G$ in $\overline{\ca}^{\smalltriangleup}$. This hierarchy is closely related to the scaling behaviour of the construction. 

Consider first a line segment divided in two, corresponding to the projection 
$$P:G^2\rightarrow G\;.$$ 
Functions on $G^2$ naturally fall into two classes: push-forward of functions on $G$ 
$$
P^\ast: L^2(G)\rightarrow L^2(G^2)\;,\quad (P^\ast f)(g_1\cdot g_2)\;,
$$
and the orthogonal complement hereof. We denote the former $V$ and the latter $V_\perp\equiv(P^\ast L^2(G))_\perp$. Functions in $V$ correspond to information which is also contained in the less refined picture which involves only one copy of $G$. This is the simplest and most coarse-grained description of parallel transports along the line segment available. Function in $V_\perp$, on the other hand, contain additional information which cannot be traced back to the simpler picture. Each additional division of the line segment refines the picture further.

%The functional space $\overline{\ca}^{\smalltriangleup}$ is constructed using line segments and their subdivisions. It corresponds to the infinite sequence
%\[
%\ca_{\ck_0}\leftarrow \ca_{\ck_1}\leftarrow \ca_{\ck_2}\leftarrow \ldots \overline{\ca}^{\smalltriangleup}
%\]
%of coarse-grained versions of the functional space. Each subdivision of line segments and simplicial complexes refines the picture further, moving us from $\ca_{\ck_i}$ to $\ca_{\ck_{i+1}}$ towards the complete picture.

\begin{figure} [t]
\begin{center}
 \input{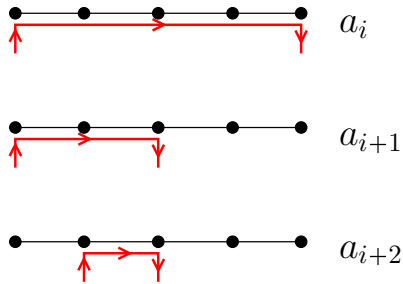}
\caption{The role of the parameters $\{a_i \}$ is to weight the different segments of a given loop according to the refinement of the segments.}
\label{Fig...}
\end{center}
\end{figure}
 It is clear that any division of a line segment into two will involve new segments which are shorter - independently of any choice of metric - than the original segment. Therefore, functions in $V_\perp$ correspond to information about $\overline{\ca}^{\smalltriangleup}$ which is {\it deeper}, i.e. at a shorter distance, compared to information carried by functions in $V$.  

The role of the sequence $\{a_i\}$ is to take this scaling behaviour into account. The operator $D_{\smalltriangleup}$ weights functions on $\overline{\ca}^{\smalltriangleup}$ according to where in the projective system of spaces $\ca_{i}$ the information originates (see figure \ref{Fig...}). Therefore, if a function $\psi\in L^2(\overline{\ca}^{\smalltriangleup})$ is the push-forward of functions on $L^2(\ca_{i})$ then the corresponding eigenvalues of $\psi$ are weighted with the appropriate parameter $a_i$.

Thus, if we wish to probe $\overline{\ca}^{\smalltriangleup}$ at a very short distance, this information will come with a very high weight factor $a_i$ corresponding to high energy. On the other hand, if $\overline{\ca}^{\smalltriangleup}$ is probed at a more coarse-grained level, then the corresponding eigenvalues are weighted with smaller weights $a_i$.

Let us end this subsection with a curious observation. If we start with a meter, then we find that it takes about 116 subdivisions of the meter to reach the Planck length of
$
1.6 \times 10^{-35}
$ meter. This corresponds then to 116 of the parameters $\{a_i\}$. Thus, although the sequence $\{a_i\}$ is infinite, the number of parameters involved when probing physical scales is certainly finite.

\section{Link to canonical quantum gravity I. Operator brackets.}
\label{cangrav}

In the next two sections we relate the construction of the spectral triple $(\cb_{\smalltriangleup},D_{\smalltriangleup},\ch_\smalltriangleup)$ to canonical quantization of field theories and in particular to canonical quantum gravity. 

The first section is concerned with the Poisson brackets of General Relativity. 
We first introduce the formulation of General Relativity based on Ashtekar variables. The fact that this formulation has a gauge connection as a primary variable is the Raison D'\'{e}tre for the study of spaces of connections in this paper.
Next, we show that the interaction between the Dirac type operator $D_{\smalltriangleup}$ with the algebra $\cb_{\smalltriangleup}$ reproduces the structure of the Poisson brackets of General Relativity when these are formulated in terms of loop variables. This means that the triple $(\cb_{\smalltriangleup},D_{\smalltriangleup},\ch_\smalltriangleup)$ includes a representation of the Poisson algebra of General Relativity. 
%Finally, we discuss briefly the area operators known from Loop Quantum Gravity and show that the square of the operator $D_{\smalltriangleup}$ has, in terms of canonical quantum gravity, a natural interpretation as a global area operator. This observation leads us to an interpretation of the spectral action of $D_{\smalltriangleup}$ as a kind of partition function of gravity. 

Section \ref{section2} is primarily concerned with a more detailed comparison between the construction presented in this paper (the choice of graphs, the Hilbert space) and the study of spaces of connections within Loop Quantum Gravity. It turns out that the key difference between the two lies in the treatment of the diffeomorphism group. We find that the Hilbert space $\ch_\smalltriangleup$ is directly related to the Hilbert space of diffeomorphism invariant states known from Loop Quantum Gravity. The difference between the two is a group of discrete diffeomorphisms. This means that the representation of the Poisson brackets given by the triple $(\cb_{\smalltriangleup},D_{\smalltriangleup},\ch_\smalltriangleup)$ includes partial diffeomorphism invariance as a first principle. 

 For the remaining part of this section we will, unless otherwise stated, restrict ourselves to $G=SU(2)$ and dimension three.

\subsection{Canonical gravity}

For an introduction to canonical gravity see for example \cite{Baez:1992tj,Nicolai:1992xx}. Let us first fix some notation. We follow \cite{Nicolai:1992xx} and \cite{Nicolai:2005mc}. Consider the vierbein formulation of General Relativity where $E^A_\m$ is the vierbein and $g_{\m\n}=E^A_\m E^B_\n \eta_{AB}$ is the corresponding space-time metric where $\eta_{AB}=\{-1,1,1,1\}$. Here $\m,\n,\ldots$ and $A,B,...$ denote curved and flat space-time indices respectively. We assume that space-time can be foliated according to $\cm=\Sigma\times \mathbb{R}$ where $\Sigma$ is a spatial manifold. Let $m,n,\ldots$ and $a,b,\ldots$ denote curved and flat spatial indices, respectively. Denote by $e^a_m$ the spatial dreibein. The spin connection $\oo_{\m AB}$ is given by
\[
\oo_{\m AB} = E_{A}^{\n}\nabla_{\m} E_{B\n}\;,
\]
where $\nabla_\m$ is the covariant derivative which involves the Christoffel connection.

The canonical momenta $\P_a^m$ corresponding to the dreibein are obtained from the Einstein-Hilbert Lagrangian $\cl$ \cite{Nicolai:1992xx}
\[
\P_a^m:= \frac{\d\cl}{\d (\pa_t e^a_m)} = \frac{1}{2} e e^m_b(K_{ab}-\d_{ab}K)\;,
\]
where $e=\det(e^a_m)$ and $K_{ab}=\oo_{ab0}$ is the extrinsic curvature. We write $K=K_{aa}$. The non-vanishing canonical Poisson brackets read
\ba
\{ e_{ma}(x),\P^n_b(y) \} &=& \d_{ab}\d^n_m \d^{(3)}(x,y)\;.
\label{equaltime0}
\ea
The Hamiltonian formulation of gravity involves a set of 3 constraints\footnote{The Gauss constraint corresponds to the connection-formalism, see below.} related to the symmetries of the theory: 
 the Gauss constraint corresponding to gauge invariance; the diffeomorphism constraint corresponding to spatial diffeomorphisms within $\Sigma$; the Hamiltonian constraint encoding the full 4-dimensional diffeomorphism invariance and thus containing the dynamics of General Relativity. The diffeomorphism and Hamiltonian constraints corresponds to 4 of the 10 Einstein field equations. The Hamiltonian itself is a linear combination of the two  constraints. In the quantum theory, this leads to the famous Wheeler-DeWitt equation.

A change of variables from the spatial spin connection and dreibein field to the connection 
\ba
A_{ma}:= -\frac{1}{2}\e_{abc}\oo_{mbc} +\gamma K_{ma}\;,
\label{AshCon}
\ea
where $\e_{abc}$ is the totally antisymmetric symbol, and the inverse densitised spatial dreibein $\tilde{E}^m_a=e e^m_a$ leads to the Poisson brackets (all other vanish)
\ba
\{ A^a_m(x),\tilde{E}^n_b(y) \}=\gamma \d^a_b\d^n_m \d^{(3)}(x,y)\;.
\label{equaltime}
\ea
Often the variables $\{\tilde{E},A\}$ are contracted with Pauli matrices $\t^a$ according to
\[
\tilde{E}^m_{\a\b}:= \tilde{E}^m_a \t_{a\a\b}\;,\quad A_{m\a\b}:= A_{ma}\t_{a\a\b}
\]
in order to replace Lorentz indices by spinorial $SU(2)$ indices.

The variables $\{ \tilde{E},A\}$ are the well-known Ashtekar variables \cite{Ashtekar:1986yd,Ashtekar:1987gu} and the parameter $\gamma\not=0$ in (\ref{AshCon}) is known as the 'Barbero-Immirzi parameter' \cite{Barbero,Immirzi}. 

The identification of a gauge connection $A_{ma}$ as a primary variable of General Relativity permits applications of techniques from Yang-Mills theory. In particular, one might shift focus from connections to their holonomies \cite{Rovelli:1987df}
\[
Hol_{C}(A)=\cp \exp\int_C A\;,
\]
where $C$ is a curve in $\Sigma$, and express the Poisson brackets in terms of holonomies and a set of conjugate variables. To find a suitable choice of conjugate variables pick any two-dimensional surface $S$ in $\Sigma$ and define the flux vector
\[
F^a_S(\tilde{E}):= \int_S dF^a\;.
\]
where the area element $dF^a$ is given by
\ba 
dF^a=\e_{mnp}\tilde{E}^{ma} dx^n\wedge dx^p\;.
\label{areaelement}
\ea
To obtain the Poisson brackets pick a curve $C=C_1\cdot C_2$ which intersects $S$ at the single point $C_1\cap C_2$. Then the Poisson bracket between the new variables reads \cite{Ashtekar:1998ak}
\ba
\{ F^a_S(\tilde{ E}),Hol_C(A)  \}= - \iota(C,S) \gamma Hol_{C_1}(A)\t^a Hol_{C_2}(A)\;,
\label{com-hol}
\ea
where $\iota(C,S)=\pm 1$ or $0$ encodes information about the intersection of $S$ and $C$ ($\iota(C,S)$ vanishes when $C$ and $S$ do not intersect).

Notice that the structure of the bracket (\ref{com-hol}) is identical to structure of the commutators (\ref{comg}) and (\ref{coml}) between the loop algebra $\cb_{\smalltriangleup}$ and the Dirac type operator $D_{\smalltriangleup}$. Both set of  commutators prescribe an insertion of an element in the Lie-algebra into the loop or curve at an "intersection point": either at the intersection between the surface $S$ and the curve $C$, or at the vertex between neighbouring edges in a given simplicial complex. To investigate this correspondence we first need to consider the canonical quantization approach to Quantum Gravity.

\subsection{Canonical Quantum Gravity}
\label{canqua}

When the canonical quantization procedure is applied to gravity the philosophy is to rewrite Poisson brackets like (\ref{equaltime0}), (\ref{equaltime}) or (\ref{com-hol}) as operator brackets and represent the canonical variables on a suitable Hilbert space as multiplication and derivation operators. Clearly, there is a freedom in choice as to which variables should be represented as multiplication operators and which should be represented as differential operators. 
%The first attempts, known as geometrodynamics \cite{}, involved a representation of either the metric or the dreibein as a multiplication operator on some (ill defined) Hilbert space over the configuration space of dreibeins or metrics. 
Using Ashtekars variables it is possible to represent the holonomies $Hol_{C}(A)$ as multiplication operators $$Hol_C(A)\rightarrow{\bf C}$$ and the corresponding triad and flux variables $\tilde{E}$ and $F$ as differential operators
\ba 
\tilde{E}^m_a(x)\rightarrow {\bf E}^m_a(x)=\frac{\hbar}{\rm{i}}\frac{\d}{\d A^a_m(x)}\;,\quad F^a _S\rightarrow {\bf F^a_S}=\int_S \e_{mnp}{\bf E}^{ma} dx^n\wedge dx^p\;.
\label{mmm}
\ea
on a suitable Hilbert space corresponding to a configuration space of connections. In the following we refer to the setup used in Loop Quantum Gravity. We postpone to section \ref{section2} the construction of the Hilbert space on which the operators ${\bf C}$ and ${\bf F}$ are represented within Loop Quantum Gravity. For now it suffices to state that the Hilbert space is based on a projective system of piecewise analytic graphs. 
Therefore, we assume that the bracket
\ba
[{\bf F}^a_S,{\bf C}] = \pm\gamma {\bf C}_1\t^a {\bf C}_2\;,
\label{opbra}
\ea
where $C=C_1\cdot C_2$ is piecewise analytic, is defined as an operator bracket acting on a Hilbert space of functions over a configuration space of connections. 
\begin{figure} [t]
\begin{center}
 \input{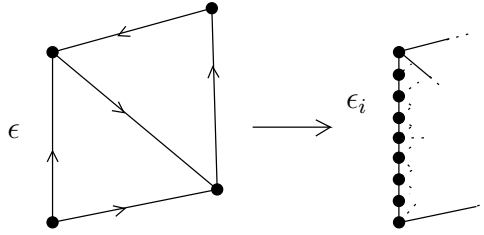}
\caption{Repeated division of an edge $\e$ in the initial simplicial complex.}
\label{simpII}
\end{center}
\end{figure}

We wish to show that the spectral triple $(\cb_{\smalltriangleup},D_{\smalltriangleup},\ch_\smalltriangleup)$ reproduces the operator bracket (\ref{opbra}). To do this we will use the operators ${\bf F}^a_S$ to construct a new operator $\textfrak{D}$. The commutator between this operator and an operator ${\bf C}$ is then shown to be identical to the commutator between the Dirac type operator $D_{\smalltriangleup}$ and an element of the algebra $\cb_{\smalltriangleup}$.

\begin{figure} [t]
\begin{center}
 \input{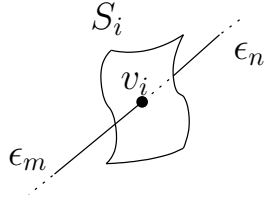}
\caption{The surface $S_i$ is chosen to intersect only the vertex $v_i$.}
\label{IntersectionS}
\end{center}
\end{figure}

To proceed we restrict the algebra (\ref{opbra}) to an inductive system of simplicial complexes $\{\ck_i\}$ and an embedding hereof $\phi(\ck_i)=\ct_i\in\cm$. Thus, we consider only curves ${\bf C}_j$ which coincide with edges\footnote{To be exact, we here assume that the triangulation $\ct_i$ is piecewise analytic.} $\e_j$ in the some triangulation $\ct_i$. To simplify matters further we first consider a single edge $\e$ in the initial complex $\ck_0$. This edge corresponds to a sequence of edges $\{\e_1,\e_2,\ldots,\e_{2^N} \}$ in the $i$'th triangulation $\ct_i$ arising through $N$ barycentric subdivisions, see figure \ref{simpII}. For simplicity we set $\gamma=1$. For each edge $\e_j$ there exist a set of sections $\hat{\ce}_i^j(g_1,\ldots,g_n)$ in the tangent bundle of the $j$'th copy of $G$, see section \ref{sectiondirac}. Expand the corresponding generators $\mathfrak{E}^j_i$ in terms of Pauli matrices
\[
\mathfrak{E}_i^j= b^{j}_{i,a}\t^a\;,
\]
and define the new operators
\[
{\bf F}^j_i=\sum_{a} b^j_{i,a}{\bf F}^a_{S_j}\;,
\]
where we introduce a set of surfaces $\{S_k\}$ chosen so that $S_k$ intersects the triangulation $\ct_i$ at its vertices only. Choose the numbering of the surfaces so that $S_k$ intersects the edge $\e_k$ at its endpoint, see figure \ref{IntersectionS}. Then we obtain (no summation over repeated indices)
\ba
[{\bf F}^j_i,{\bf C}_j]= {\bf C}_j\mathfrak{E}^j_i\;.
\label{opbra1}
\ea
We chose $S_k$ in a way so that the sign in (\ref{opbra1}) is always positive when the orientation of the curve $C_k$ coincide with the orientation of the triangulation. 

The commutator (\ref{opbra1}) has the same structure as the commutator (\ref{cccom}) between the vector field $\hat{\ce}^j_i$ and the group element corresponding to the $j$'th copy of $G$ in $G^n$. This suggest that the vector field $\hat{\ce}^j_i$ corresponds to the flux operator ${\bf F}^j_i$
\ba 
{\bf F}^j_i\leftrightarrow \hat{\ce}^j_i
\label{coRRe}
\ea
when ${\bf F}^j$ is restricted to the triangulation $\ct_i$.

Next, we define the operator
\ba
\textfrak{D}_n=\frac{1}{n}\sum_{s,i} a_{m(s)}\ce^{n,s}_i\cdot\big(({\bf F}^1_i,{\bf F}^2_i,{\bf F}^3_i,\ldots)\cdot s\big)\;,
\label{Dfrak}
\ea
where $s$ is the sequence of signs $(+,\pm,\pm,\ldots)$ corresponding to the vector $\ce^{n,s}_i$ and the '$\cdot$' in the bracket to the right is defined in (\ref{cdot}). Then, the final commutator reads
\ba 
[{\bf\textfrak{D}}_n,{\bf C}_j]=\frac{1}{n}\sum_{s,j} a_{m(s)} \big(\pm {\bf C}_j \mathfrak{E}^j_i\big)\cdot \ce^{n,s}_i\;.
\label{OPBRII}
\ea
In (\ref{OPBRII}) the sign on the rhs is again given by the $i$'th entries in the sequences $s$. The commutator (\ref{OPBRII}) has precisely the same form as the commutator (\ref{comg}).  This shows that the operator ${\textfrak{D}}_n$ corresponds exactly to the Dirac operator $D_n$ defined in equation (\ref{DIRACmod}):
\[
{\textfrak{D}}_n\leftrightarrow D_n\;.
\]

Notice also that the choice of surfaces $S_k$ has no importance for the definition of (\ref{Dfrak}). Only the intersection points, the vertices, counts. Therefore, the surfaces $S_k$ serve merely as labels of the vertices.

The generalisation to the full picture where we consider the complete set of edges $\{\e_k\}$ in $\ct_i$ is straightforward. The corresponding operator, which we again denote $\textfrak{D}_n$, is simply build from the operators associated to each edge $\e_k$ in the initial triangulation $\ct_0$. Further, we denote the limiting operator with $\textfrak{D}_{\smalltriangleup}$. Again, the message is that the operator $\textfrak{D}_{\smalltriangleup}$ corresponds exactly to the Dirac like operator $D_{\smalltriangleup}$. The difference is that $\textfrak{D}_{\smalltriangleup}$ is written in terms of variables from canonical quantum gravity.

This shows that the spectral triples $(\cb_{\smalltriangleup},D_{\smalltriangleup},\ch_\smalltriangleup)$ constructed in section \ref{sectiontrace} give a representation of the Poisson brackets (\ref{com-hol}). The representation is separable and based on a system of graphs dense in $\cm$.

In which sense does the representation of the Poisson algebra (\ref{com-hol}) given by the triple $(\cb_{\smalltriangleup},D_{\smalltriangleup},\ch_\smalltriangleup)$ differ by the representation used in Loop Quantum Gravity? To answer this question we need a precise comparison of the general setup used in this paper with the setup used in Loop Quantum Gravity. This will be the topic of the next section.\\

For later reference we need an expression for the square of the Dirac-like operator (\ref{Dfrak}). We write
\ba 
\textfrak{D}^2_{\smalltriangleup} = \sum_{ij} c_{j} {\bf F}_j^i{\bf F}^i_j\; + \;\mbox{lower order} \;,
\label{SE1}
\ea
where $c_{j}$ are constants depending on the parameters $\{a_i\}$ and where we by lower order refer to terms which are linear in ${\bf F}^i_j$.

\section{Link to Canonical Quantum Gravity II. Diffeomorphisms.}
\label{section2}

%This section is concerned with a detailed comparison between the mathematical machinery used in this paper and the approach used in Loop Quantum Gravity. The purpose is to obtain a better understanding of the spectral triple $(\cb_{\smalltriangleup},D_{\smalltriangleup},\ch_\smalltriangleup)$ and its connection to canonical quantum gravity.

In this section we first introduce the inductive system of piecewise analytic graphs applied in Loop Quantum Gravity. This system entail a space of generalised connections much alike the space $\overline{\ca}^{\smalltriangleup}$ constructed in this paper. We find that the essential difference between the two spaces is separability. This issue is closely related to the manner in which the diffeomorphism group is treated in Loop Quantum Gravity and this paper respectively.

These insights lead us to consider the role of diffeomorphism invariance in Quantum Gravity, in particular the diffeomorphism constraints found in Loop Quantum Gravity. We find that the Hilbert space $\ch_\smalltriangleup$ is, up to a discrete symmetry group, identical to the Hilbert space of diffeomorphism invariant states in Loop Quantum Gravity.

\subsection{Functional spaces of connections in Loop Quantum Gravity}

%In section \ref{section1} - \ref{sectiontrace} we introduced a mathematical machinery designed for the study of functional spaces of connections. The construction is based on an inductive system of simplicial complexes and is strongly motivated by a similar construction found in Loop Quantum Gravity. This construction is also based on an inductive system of graphs, but in this case the system involves piecewise analytic loops and piecewise analytic graphs. Again the underlying space is a space of generalised connections. In contrast to the previous section, this functional space will have an action of the diffeomorphism group.

%The fundamental idea in Loop Quantum Gravity is to consider the space of loops as the dual to the space of connections. With this idea it is possible to equip the space of connections with a projective structure which entail measure and Hilbert space structures.

The following is a review of material known in Loop Quantum Gravity. In particular, we refer to the publications \cite{Ashtekar:1993wf,Ashtekar:1994wa,Marolf:1994cj}.

Let $\cm$ be a real analytic manifold and let $\cp$ be the space of piecewise analytic directed paths on $\cm$. We will consider two paths in $\cp$ to be the same if they differs by trivial backtracking. $\cp$ has a product simply by composing paths. We define the space of generalised $G$-connections by
\ba
\overline{\ca}^{\small a}=\hbox{Hom} (\cp,G)\;,
\label{klo}
\ea
where Hom means maps $\nabla$ from $\cp$ to $G$ satisfying $$\nabla (P_1P_2)=\nabla (P_1)\nabla (P_2)\;.$$ The space $\overline{\ca}^{\small a}$ was studied in \cite{Ashtekar:1994wa}.

Again, we consider the space $\ca$ of smooth $G$-connections. We repeat the argument of section \ref{funcspace}: There is a natural map $\chi_{\small a}:\ca \to \overline{\ca}^{\small a}$ given by 
$$\chi_{\small a} ( \nabla)(P)= Hol(\nabla ,P)\;,$$
where $Hol(\nabla,P)$ denoted the holonomy of $\nabla$ along $P$. 
If we are given two different smooth connections $\nabla_1,\nabla_2\in \ca$ they are going to differ in a point, let us say $m$, and hence also in a neighbourhood of $m$. We can therefore choose a small directed analytic path $P$ in a neighbourhood of $m$ such that $Hol(\nabla_1,P)\not= Hol (\nabla_2,P)$. In other words, $\chi_{\small a}$ is an embedding, and hence the terminology generalised connection is justified.

\subsubsection{Projective structure on $\overline{\ca}^{\small a}$.}

 On $\cm$ we will consider the system  of connected piecewise analytic graphs. This system is directed under inclusions of graphs. Given a piecewise connected analytic graph $\Gamma$ in $\cm$ we denote by $\{ \e_1,\ldots , \e_{n(\Gamma)}\}$ the edges and by $\{ v_1,\ldots, v_{m(\Gamma)}\}$ the vertices.  Let $\cp_\Gamma$ be the set of paths in $\Gamma$ and define 
$$\ca_\Gamma=\hbox{Hom}(\cp_\Gamma, G)\;.$$
An inclusion of graphs $\Gamma' \subset \Gamma$ induces a projection 
\ba
P_{\Gamma' \Gamma}:\ca_\Gamma \to \ca_{\Gamma'}\;.
\label{gogo}
\ea
Since every analytic path is a path in a piecewise analytic graph and vice versa we get
$$\overline{\ca}^{\small a}=\lim_{\stackrel{\Gamma}{\longleftarrow}}\ca_\Gamma ,$$
where the projective limit is taken over all piecewise analytic graphs.

If we choose an orientation of the edges $\{ \e_1,\ldots, \e_{n(\Gamma)}\}$ we obtain again an identification 
$$\ca_{\Gamma}\simeq G^{n(\Gamma)}$$
via the map 
$$\ca_{\Gamma}\ni \nabla \to (\nabla (\e_1),\ldots , \nabla (\e_{n({\Gamma})}))\in G^{n(\Gamma)}.$$
This identification gives rise to various structures on $\ca_\Gamma$ and therefore also on $\overline{\ca}^{\small a}$. For example the topological structure given by the topological structure of $G^{n(\Gamma)}$. Also the projections $P_{\Gamma ' \Gamma}$ are continuous and hence give a topological structure on the projective limit, i.e. $\overline{\ca}^{\small a}$. We note that $\ca$ is dense in $\overline{\ca}^{\small a}$, see \cite{Aastrup} and references therein.

Another structure arising from this identification is the measure or more relevant the Hilbert space structure. Since $G$ is a compact group we can equip $G^{n(\Gamma)}$ uniquely with product Haar measure and define 
$$L^2(\ca_\Gamma)=L^2(G^{n(\Gamma)}).$$
The projections $P_{\Gamma' \Gamma}$ induce embeddings of Hilbert spaces 
$$P_{\Gamma' \Gamma}^*:L^2(\ca_\Gamma')\to L^2(\ca_\Gamma).$$
With this we then define
$$L^2(\overline{\ca}^{\small a})=\lim_{\stackrel{\Gamma}{\longrightarrow}}L^2(\ca_\Gamma).$$
which is non-separable.

The Hilbert space $L^2(\overline{\ca}^{\small a})$ is the Hilbert space that carries the representation of the Poisson algebra (\ref{opbra}) used in Loop Quantum Gravity. It is known as the kinematical Hilbert space.

\subsubsection{Actions of the diffeomorphism group}

Since an analytic diffeomorphism of $\cm$ maps $\cp$ to $\cp$ we get an action of the group of analytic diffeomorphisms $Diff_a(\cm)$ on $\overline{\ca}^a$. 
On the other hand the group of all diffeomorphisms $Diff (\cm)$ acts on $\ca$ but does not extend to an action on $\overline{\ca}^a$. There are several different way of completing $\ca$ depending on the choice of "lattice", for example piecewise analytic paths like in the case described above, various kinds of  smooth paths, where one gets an action of the full diffeomorphism group. See \cite{Fleischhack:2000ij,Fleischhack:2000am} for results and a thorough discussion. The different completions all contain $\ca$ but the crucial difference seems to be the size of the symmetry group.

\subsection{Comparing the spaces $\overline{\ca}^{\smalltriangleup}$ and $\overline{\ca}^{\small a}$}
\label{link}

So far we have introduced the three spaces $\ca$, $\overline{\ca}^{\smalltriangleup}$ and $\overline{\ca}^{\small a}$. $\ca$ is the space of smooth $G$-connections and we demonstrated that
\[
\ca\hookrightarrow \overline{\ca}^{\smalltriangleup}\;,\quad \ca\hookrightarrow \overline{\ca}^{\small a}\;,
\]
which means that both $\overline{\ca}^{\smalltriangleup}$ and $\overline{\ca}^{\small a}$ are spaces of generalised connections. Furthermore, since it is known that $\ca$ is dense in $\overline{\ca}^{\small a}$, and since there exist a natural surjection
\[
\s:\overline{\ca}^{\small a}\rightarrow \overline{\ca}^{\smalltriangleup}
\]
we know that $\ca$, too, is dense in $\overline{\ca}^{\smalltriangleup}$. 
To see how the surjection $\s$ works recall that both $\overline{\ca}^{\smalltriangleup}$ and $\overline{\ca}^{\small a}$ are spaces of homomorphisms
\ba
\overline{\ca}^{\smalltriangleup}=\Hom(\cp_{\smalltriangleup},G)\;,\quad \overline{\ca}^{\small a}=\Hom(\cp_{\small a},G)\;,
\label{xxx}
\ea
and since\footnote{we assume here that the edges in the triangulations $\ct_i$ are analytic. This depends on the embedding $\phi:\ck_i\rightarrow\ct_i$ but is no real restriction.} $\{\cp_{\smalltriangleup}\}\hookrightarrow\{\cp_{\small a}\}$ then $\s$ is just the corresponding surjection between the spaces (\ref{xxx}) of homomorphisms.

To be precise we should state that we now think of the space $\overline{\ca}^{\smalltriangleup}$ in terms of an embedding
\[
\phi:\ck_i\rightarrow \ct_i
\]
of the projective system $\{\ck_i\}$,
and that we therefore identify
\[
\overline{\ca}^{\smalltriangleup}=\lim_{\stackrel{\ct}{\leftarrow}}\ca_\ct\;.
\]

In the following we will demonstrate the relationship between and role played by these three spaces of connections.

First, it is important to see that the shift from $\ca$ to a larger space of generalised connections is necessary to equip $\ca$ with topological and Hilbert space structures. The identification of $\overline{\ca}^{\smalltriangleup}$ and $\overline{\ca}^{\small a}$ as pro-manifolds is the key step to obtain this. The space $\ca$ itself is not a pro-manifold.

The key difference between the two spaces $\overline{\ca}^{\smalltriangleup}$ and $\overline{\ca}^{\small a}$ is that $\overline{\ca}^{\small a}$ has an action of the (analytic) diffeomorphism group; $\overline{\ca}^{\smalltriangleup}$ does not. Further, the Hilbert space structure associated to the two spaces are respectively non-separable for $\overline{\ca}^{\small a}$ and separable for $\overline{\ca}^{\smalltriangleup}$. That is, the way $\ca$ is completed is decisive for how large a symmetry group remains. This shows that there is a direct link between separability of the Hilbert space structure and the action of the diffeomorphism group.

The following diagram illustrate the relationship between the three spaces of connections:
\[
\begin{array}{ccccc}
&&\overline{\ca}^{\small a}&\leftarrow& \mbox{diff}_{\small a}(\cm)\nn\\
&\nearrow&&& \nn\\
\ca &&\hspace{-3mm}\s\downarrow && \nn\\
&\searrow &&& \nn\\
&&\overline{\ca}^{\smalltriangleup}&\leftarrow& \mbox{diff}(\triangle)
\end{array}
\]
where we by diff$(\triangle)$ refer to the discrete symmetries which remain on the infinite triangulation underlying $\overline{\ca}^{\smalltriangleup}$.

It is interesting that the construction of the Dirac-like operator (\ref{didi}) requires the shift from $\overline{\ca}^{\small a}$ to $\overline{\ca}^{\smalltriangleup}$: In \cite{Aastrup:2005yk} the authors attempted to construct such an operator on $\overline{\ca}^{\small a}$ and a corresponding Hilbert space. The attempt was unsuccessful exactly because the number of possible projections (\ref{gogo}) is too large to permit a Hilbert space structure carrying a Dirac-like operator similar to (\ref{didi}). The shift from $\overline{\ca}^{\small a}$ to $\overline{\ca}^{\smalltriangleup}$ reduces the type of embeddings between graphs sufficiently.
We appear to find the following hierarchy:
\[
\begin{array}{cl}
\ca: & \mbox{No Hilbert space structure}\nn\\
     & \mbox{No Dirac-like operator}\nn\\
     & \mbox{Action of diff($\cm$)}\nn\\
     &\nn\\
\overline{\ca}^{\small a}: & \mbox{Hilbert space structure, non-separable}\nn\\
& \mbox{No Dirac-like operator}\nn\\
& \mbox{Action of analytic diffeomorphisms}\nn\\
&\nn\\
\overline{\ca}^{\smalltriangleup}:& \mbox{Hilbert space structure, separable}\nn\\
& \mbox{Dirac-like operator}\nn\\
& \mbox{Action of diff($\triangle$)}\nn
\end{array}
\]

One way to think of the space $\overline{\ca}^{\smalltriangleup}$ is to see it as the space of (smooth) connections subjected to a sort of gauge fixing of the diffeomorphism group\footnote{In fact, space $\overline{\ca}^{\small a}$ is, in this line of thinking, subjected to another partial gauge fixing by reducing the symmetry group from the group of diffeomorphisms to the .group of analytical diffeomorphisms.}. Here gauge fixing is meant in the sense that a symmetry group is (partly) removed while the integration at hand still involves the entire space, in this case the space of smooth connections $\ca$. In this sense the inner product (\ref{funcint}) resembles a functional integral over $\ca$ "up to diffeomorphisms".

Another way to think of the space $\overline{\ca}^{\smalltriangleup}$ is to relate it to the construction of the Hilbert space $\ch_{diff}$ of diffeomorphism invariant states in Loop Quantum Gravity, see \cite{Fairbairn:2004qe}. 
To do this we first notice that each triangulation in particular is a graph, and we hence get an embedding
$$\lim_{\stackrel{\ct}{\longrightarrow}} L^2(\ca_{\ct}) \stackrel{\iota}{\hookrightarrow} \lim_{\stackrel{\G}{\longrightarrow}}L^2(\ca_{\G}).$$
On the other hand consider the Hilbert space formally defined by 
$$\ch_{diff}= \hbox{diffeomorphism invariant states in }\lim_{\stackrel{\G}{\longrightarrow}}L^2(\ca_{\G}).$$
The elaborate definition is written in \cite{Fairbairn:2004qe}. There is a surjection 
$$\lim_{\stackrel{\G}{\longrightarrow}}L^2(\ca_{\G})\stackrel{q}{\to} \ch_{diff}$$
given by meaning vectors over the action of the diffeomorphism group. We thus get a map 
$$\lim_{\stackrel{\ct}{\longrightarrow}} L^2(\ca_{\ct})\stackrel{\Xi}{\longrightarrow} \ch_{diff}$$
by composing. This map is also going to be a surjection since each graph can, via a diffeomorphism, be mapped into a triangulated graph. The map is not injective because of the symmetries of the triangulated graph. 

We have the following commutative diagram
\ba
\begin{array}{ccc}
&  & \parbox{8mm}{$\stackrel{\displaystyle{\lim}} {\longrightarrow}$}  L^2(\ca_{\G})\\
&\stackrel{\iota}{\nearrow} &\downarrow q\\
\parbox{8mm}{$\stackrel{\displaystyle{\lim}} {\longrightarrow}$} L^2(\ca_{\ct}) &\stackrel{\Xi}{\longrightarrow} &\ch_{diff}
\end{array}
\ea
The amount by which the map $\Xi$ fails to be injective is exactly the symmetry group diff$(\triangle)$ of discrete diffeomorphisms of triangulations.
Therefore, we can think of the Hilbert space $L^2(\ca_{\ct})$, and thereby also $\ch_\smalltriangleup$, as Hilbert spaces of diffeomorphism invariant states up to the discrete group diff$(\triangle)$.
In this picture one should therefore think of a loop associated to a simplicial complex as a kind of equivalence class of smooth loops where the equivalence is with respect to smooth diffeomorphisms.

It is interesting that the construction of $\ch_{\smalltriangleup}$ relies on a partition of the diffeomorphism group in a 'countable' and a 'over countable' part. Here, the countable part is diff$(\triangle)$. This partition is a compromise between two opposite considerations:
\begin{itemize}
\item
For the Hilbert space $\ch_{\smalltriangleup}$ to carry a representation of the loop algebra it must have an action of at least parts of the diffeomorphism group since the loop algebra itself is not diffeomorphism invariant. This requirement works to maximise the size of the remaining symmetry group diff$(\triangle)$.
\item
The construction of a Dirac type operator acting on $\ch_{\smalltriangleup}$ requires a strict control over permissible embeddings of the type (\ref{proj}). This requirement works to minimise the size of diff$(\triangle)$.
\end{itemize}
Therefore, the identification of the symmetry group diff$(\triangle)$ is essential for the construction of the spectral triple $(\cb_{\smalltriangleup},D_{\smalltriangleup},\ch_\smalltriangleup)$.

Let us now return to the question regarding the different representations of the Poisson algebras (\ref{equaltime}) and (\ref{com-hol}), the representation used in Loop Quantum Gravity and the representation contained in the triple $(\cb_{\smalltriangleup},D_{\smalltriangleup},\ch_\smalltriangleup)$. We find that the different representations correspond to different choices of holonomy loops in (\ref{com-hol}). The representation used in Loop Quantum Gravity involves the non separable Hilbert space $L^2(\overline{\ca}^{\small a})$ whereas the representation presented in this paper involves the separable Hilbert space $L^2(\overline{\ca}^{\smalltriangleup})$. The central difference between the two representations lies therefore in the corresponding symmetry groups. The larger the class of loops is, the larger is the symmetry group.

%To do this notice first that each triangulation in particular is a graph, and we hence get an embedding
%$$\lim_{\stackrel{\ct}{\longrightarrow}} L^2(\ca_{\ct}) \stackrel{\iota}{\hookrightarrow} \lim_{\stackrel{\G}{\longrightarrow}}L^2(\ca_{\G}).$$
%where we by $\ca_\ct$ mean the space $\ca_\ck$, now with the simplex $\ck$ mapped into a triangulation $\ct$. This embedding makes only sense when $\ct$ is piecewise analytic. On the other hand consider the Hilbert space formally defined by
%$$\ch_{diff}= \hbox{diffeomorphism invariant states in }\lim_{\stackrel{\G}{\longrightarrow}}L^2(\ca_{\G}).$$
%The elaborate definition is written in \cite{Fairbairn:2004qe}, see also below. There is a surjection 
%$$\lim_{\stackrel{\G}{\longrightarrow}}L^2(\ca_{\G})\stackrel{q}{\to} \ch_{diff}$$
%given by meaning vectors over the action of the diffeomorphism group. We thus get a map 
%$$\lim_{\stackrel{\ct}{\longrightarrow}} L^2(\ca_{\ct})\stackrel{\Xi}{\longrightarrow} \ch_{diff}$$
%by composing. This map is also going to be a surjection since each graph can, via a diffeomorphism, be mapped into a triangulated graph. The map is not injective because of the symmetries of the triangulated graph. 

%We have the following commutative diagram
%\ba
%\end{array}
%\ea
%To understand the exact relationship between the Hilbert spaces $L^2(\overline{\ca}^{\smalltriangleup})$ and $\ch_{diff}$ we need some details on the construction of the Hilbert space $\ch_{diff}$ from Loop Quantum Gravity. For further details we refer to \cite{Aastrup:2005yk}. 

\section{Area operators}
\label{sectionarea}

The area operators are an important set of operators in Loop Quantum Gravity \cite{Ashtekar:1996eg}. It turns out that the area operator also exist within the framework described by the spectral triple $(\cb_{\smalltriangleup},D_{\smalltriangleup},\ch_\smalltriangleup)$. These operators are, however, not very natural objects to consider within this framework. However, we find that the square of the Dirac-like operator $D_{\smalltriangleup}$ has a natural interpretation in terms of a global area-squared operator.

\subsection{Area operators in Loop Quantum Gravity}

Classically, the area of a 2-dimensional surface $S$ in $\S$ is given by
\[
A(S)=\int_S \sqrt{dF^a\cdot dF^a}\;.
\]
where the area element $dF^a$ was defined in (\ref{areaelement}). To convert this expression into a form suitable for quantization
consider a partition of $S$ into $N$ smaller surfaces $S_n$ such that for any N we have $\bigcup_n S_n=S$. Then the area of $S$ can be written
\ba 
A(S)=\lim_{N\rightarrow \infty}\sum_{n=1}^N \sqrt{F^i_{S_n}F^j_{S_n}\d_{ij}}\;.
\label{clasarea}
\ea
In Loop Quantum Gravity, the area operator is constructed by substituting the classical flux variable $F^i_{S_n}$ with the corresponding operators ${\bf F}^i_{S_n}$
\ba 
{\bf A}(S)=  \lim_{N\rightarrow \infty}\sum_{n=1}^N \sqrt{{\bf F}^i_{S_n}{\bf F}^j_{S_n}\d_{ij}}\;.
\label{areaI}
\ea
The point here is that this operator, at the level of a given graph, is well defined. This is due to the fact that the number of intersections of any graph with the surface $S$ is finite. Continued subdivisions of $S$ are obsolete once the resolution is reached where each $S_n$ contains one intersection point. If the area element $S_n$ does not involve an intersection with the graph $\G$ then the corresponding operator ${\bf F}^i_{S_k}$ vanishes. This means that the surface $S$ has a minimal subdivision into elementary cells $S_n$ each containing one intersection point with the graph. The area operator obtains the form
\[
{\bf A}(S)=  \sum_i {\bf A}(p_i)\;,
\]
where the sum runs over intersection points $p_n$ and where
\ba 
{\bf A}(p_n)= \sqrt{{\bf F}^j_{S_n}{\bf F}^k_{S_n}\d_{jk}}\;.
\label{SE2}
\ea
The operator ${\bf A}(p_n)$ basically assigns an area to the intersection point $p_n$.
The spectrum of the area operator ${\bf A}(S)$, at the level of a given graph $\G$, can be computed and reads \cite{Ashtekar:2004eh}
\[
\mbox{Spec}({\bf A}(S)) = \left\{4\p \gamma l^2_{Pl}\sum_{p}\sqrt{j_p(j_p + 1)}\right\}\;,
\]
where $p$ are intersection points between $S$ and the graph $\G$ and $j_p$ are positive half integers (for details and subtleties we refer to \cite{Ashtekar:2004eh}). $l_{Pl}$ is the Planck length.

%This also provides new insight to the relationship between the flux operator (\ref{mmm}) and its corresponding triad operator. Basically, when shrinking the surface $S_i$ to a point $v_i$ the flux operator ${\bf F}_{S_i}$ becomes the Hodge dual to the triad operator
%\[
%{\bf F}^a_{S_i}= \int_{S_i} \e_{klm}{\bf E}^{ma} dx^k \wedge dx^l \sim \ast{\bf E}^a (v_i)
%\]
%where a constant with dimension length squared should be added for the dimensions to match.

\subsection{A global area operator}

The spectral triple $(\cb_{\smalltriangleup},D_{\smalltriangleup},\ch_\smalltriangleup)$ involves another representation of the operator algebra (\ref{opbra}), now based on curves in the inductive system of triangulations. Therefore, it is possible to repeat the line of reasoning from Loop Quantum Gravity and define a second area operator, now based on the Dirac-like operator $D_{\smalltriangleup}$ and the vectors $\hat{\ce}^j_i$.

Recall the interpretation (\ref{coRRe}) which relates the vectors $\hat{\ce}^j_i$ to the flux operators $F_i^j$. This leads to an area operator 
%We start by reversing the expression (\ref{Dfrak}) to extract an expression for ${\bf F}$ in terms of the Dirac-like operator $\textfrak{D}_{\smalltriangleup}$. The exact expression is complicated due to the adjoint actions involved in the sections $\ce^i_j$ on the $i$'th copy of $G$ and because of the parameters $\{a_i\}$. Therefore, for simplicity we give here the general form
%\ba
%{\bf F}^i(\textfrak{D}_{\smalltriangleup}) =  \sum \pm c(a_i)tr\left( \bar{\ce}\cdot \textfrak{D}_{\smalltriangleup}\right) 
%\label{revers}
%\ea
%where the trace is over the Clifford algebra and $c(a_i)$ are constants depending on the parameters $a_i$.
%Next we obtain an expression for ${\bf F}^i$ in terms of $D_{\smalltriangleup}$ by substituting [skrives om, drop dette]
%\[
%{\bf F}^i(\textfrak{D}_{\smalltriangleup})\rightarrow{\bf F}^i( D_{\smalltriangleup})\;.
%\]
%in equation (\ref{revers}). With this expression 
%we obtain again an area operator corresponding to (\ref{clasarea}), now expressed in terms of the vectors $\ce^j_i$
\ba 
{\bf A}_{{\smalltriangleup}}(S)=\lim_{N\rightarrow \infty}\sum_n \sqrt{{\bf F}^i_{S_n}{\bf F}^j_{S_n}\d_{ij}}\;.
\label{areaII}
\ea
where ${\bf F}^i_{S_n}$ are as explained in section \ref{canqua}.

The difference between the operators (\ref{areaI}) and (\ref{areaII}) lies not in the expression of the operators themselves but rather in the Hilbert spaces on which they act. The former acts on a non-separable Hilbert space whereas the latter acts on the separable Hilbert space $\ch_\smalltriangleup$.

The area operator (\ref{areaII}), however, does not appear to be a natural object in the construction presented here. It involves a  surface $S$ which has no natural place in this setting. 

We can nevertheless apply the line of reasoning from Loop Quantum Gravity to obtain a better understanding of the operator $D_{\smalltriangleup}$. 
Consider again the interpretation of $D_{\smalltriangleup}$ in terms of conjugate variables of canonical gravity. If we combine equation (\ref{SE1}) with equation (\ref{SE2}) we obtain
\ba 
\textfrak{D}^2 = \sum_{i} c_{i} A(v_i)^2 + \ldots
\label{bibi}
\ea
where, in the limit, the sum runs over {\it all} vertices in the inductive system of triangulations $\{\ct_i\}$. It is clear that the set of all vertices is a dense set $\cm$. Therefore, the sum over all vertices, weighted with the sequence $\{a_i\}$ and the a multiplicity factor given by the valency of the vertex, defines an integral over $\cm$
\[
\sum_{v_i}... \rightarrow \int_{\cm} d\mbox{Vol}\;.
\]
This provides us with a surprising interpretation of $\textfrak{D}^2$ and thereby of $D_{\smalltriangleup}^2$. According to (\ref{bibi}) it is an operator which is related to the area of {\it the entire manifold} $\cm$
\ba 
D_{\smalltriangleup}^2 \sim \int_\cm \Big({\bf A}(x)\Big)^2 + ...
\label{areAl}
\ea
Clearly, the spectrum of $\Delta_{\smalltriangleup}$ is discrete.

\section{The spectral action}
\label{EINstein}

The identification of the operator $D^2_{\smalltriangleup}$ as a global operator related to the area of the underlying manifold leads us to a new interpretation of the spectral triple $(\cb_{\smalltriangleup},D_{\smalltriangleup},\ch_\smalltriangleup)$. In this section we suggest that the operator $D^2_{\smalltriangleup}$ should be thought of as an action. Subsequently, we point out that the spectral action of $D_{\smalltriangleup}$ resembles a partition function. The argumentation presented in the following is tentative.

The first indication that $(D_{\smalltriangleup})^2$ may be understood in terms of an action is, as already mentioned, that is has the form of an integral over an underlying manifold. 
 
The second indication is directly related to the classical Einstein-Hilbert action. Within the language of Noncommutative Geometry the Einstein-Hilbert action has a natural interpretation as an area of the underlying manifold. With the previous section in mind this suggest that the operator $(D_{\smalltriangleup})^2$ is somehow related to the Einstein-Hilbert action. 

Let us go into some details. Consider a 4-dimensional, Riemannian spin-geometry described by the real, even spectral triple $(A,D,H)$ where $A$ is a commutative $C^\ast$-algebra. 
%According to the reconstruction theorem \cite{Rennie:2006pi} the state space is a manifold $\cm$ and the Dirac operator has the usual form
%\ba 
%D=\sum_i \gamma^A E_A^\m\nabla_\m
%\label{dadada}
%\ea
%where $\gamma_i$ denote gamma matrices and $\nabla_\m$ is the Levi-Civita connection.
The Eucledian Einstein-Hilbert action on $\cm$ can be computed directly from $D$. It is proportional to the Wodzicki residue of the inverse square of the Dirac operator $D$ \cite{Kastler:1993zj}. If we denote by $Tr^+$ the Dixmier trace and define the noncommutative integral $\fint$ by
\[
\fint f := Tr^+ f |D|^{-4}
\]
one has
\[
\fint D^{2}=\frac{-1}{48\p^2}\int_\cm R\sqrt{g}d^4x\;,
\]
where $dv=\sqrt{g}d^4x$ is the volume form and $R$ is the scalar curvature. It is in this sense that one may interpret the Einstein-Hilbert action as the "two-dimensional measure of a four manifold", the "area" of $\cm$ \cite{Connes:2000ti}. 

This resembles the findings of the previous section where we saw that the operator $D^2_{\smalltriangleup}$ has a natural interpretation in terms of a global area-squared operator over an underlying manifold. %The spectrum of $D^2_{\smalltriangleup}$ is discrete and clearly the nature of $D^2_{\smalltriangleup}$ is 'quantum'. Thus, the idea emerges that $D^2_{\smalltriangleup}$ might be thought of as a quantized version of the Einstein-Hilbert action.
If we ignore the obvious issue of 'area' vs. 'area-squared' one may speculate whether there is some deeper relation between the operator $D^2_{\smalltriangleup}$ and the Einstein-Hilbert action.

These considerations entail an interesting interpretation of the spectral action of $D_{\smalltriangleup}$. Consider the quantity
\ba 
Tr \exp(-s D^2_{\smalltriangleup})\;,
\label{initial}
\ea
where $s$ is a real parameter, and let us perform a formal calculation to obtain a better understanding of this quantity. We write
\ba 
Tr \exp(-s D^2_{\smalltriangleup})&=& 
\sum_{\psi_n}\langle \psi_n|\exp{\left(-s D^2_{\smalltriangleup}\right)}|\psi_n\rangle
\nn\\
&=& \int_{\overline{\ca}^{\smalltriangleup}}[d\nabla]\langle\d_\nabla|\exp\left(-s D^2_{\smalltriangleup}\right)|\d_\nabla\rangle
\nn\\
&=&\int_{\overline{\ca}^{\smalltriangleup}}[d\nabla]\left(\int_{\overline{\ca}^{\smalltriangleup}}[d\nabla_1]\d_{\nabla}(\nabla_1)\exp\left(-s D^2_{\smalltriangleup}\d_{\nabla}\right)(\nabla_1) \right)
\nn\\
&=& \int_{\overline{\ca}^{\smalltriangleup}}[d\nabla]\exp\left(-s D^2_{\smalltriangleup}\d_{\nabla}\right)(\nabla)\;,
\label{formalcalc}
\ea
where $\{\d_{\nabla}\}$ is the orthogonal set of delta functions on the space $\overline{\ca}^{\smalltriangleup}$ and where $[d\nabla]$ denotes the measure on $\overline{\ca}^{\smalltriangleup}$ introduced in section \ref{projlim}. The final expression in (\ref{formalcalc}) may, strictly speaking, not make perfect mathematical sense. However, we know that the initial quantity (\ref{initial}) is well defined. This formal calculations shows that the spectral action (\ref{initial}) has the form of a formal Feynman integral where the expression
\[
D^2_{\smalltriangleup}\d_{\nabla}
\] 
plays the role of a classical action.

\section{Including the sequences $\{a_n\}$ as dynamical variables}
\label{includea}

Up till this point we have constructed the spectral triple $(\cb_{\smalltriangleup},D_{\smalltriangleup},\ch_\smalltriangleup)$ and related it to Quantum Gravity. We have shown that it contains information of the Poisson brackets of canonical gravity and that it partly incorporates diffeomorphism invariance. Two questions remains to be addressed:
\begin{enumerate}
\item
The spectral triple relies on the divergent sequence $\{a_i\}$. What structure does this sequence represent and how do we deal with it?
\item
How do we incorporate the remaining diffeomorphisms contained in the group diff$(\triangle)$.
\end{enumerate}

One solution to the first question would be to fix the sequence $\{a_i\}$ in a manner which counts a partition of an edge with a factor $\frac{1}{2}$. However, we will argue that there may be another solution which is more natural. This section is concerned with this issue.

In section \ref{cangrav} we demonstrated that the triple $(\cb_{\smalltriangleup},D_{\smalltriangleup},\ch_\smalltriangleup)$ has a natural interpretation in terms of Poisson brackets of canonical gravity. In this picture the Dirac type operator $D_{\smalltriangleup}$, including the sequence $\{a_i\}$, is a linear sum of the flux operators involving the {\it densitised} vielbein operator, see equation (\ref{mmm}). This suggest an interpretation of the parameters $a_i$ in terms of the determinant of the vielbein 
\[
%a_{m(s)}  \nabla^{\mbox{\tiny\bf n}}_{\hat{\ce}^{n,s}_i}\sim {\bf E}^m_a \;\leftrightarrow\;
  a_i \sim det(e_m^a)\;.
\]
That the sequence $\{a_i\}$ carries metric information also comes out of the discussion in subsection \ref{rolE} where we relate the sequence $\{a_i\}$ to the scaling behaviour of the construction.

Thirdly, in section \ref{sectionarea} we found that the square of $D_{\smalltriangleup}$ has the form of an integral over the underlying manifold. Here, the measure involved in this integral is given by the sequence $\{a_i\}$.

This suggests, as a possible solution to the first question, that we should include the sequence $\{a_i\}$ in the construction as dynamical variables. If the sequence $\{a_i\}$ has a metric origin, then it seems natural to try to integrate over these degrees of freedom.

Therefore we extend the spectral triple $(\cb_{\smalltriangleup},D_{\smalltriangleup},\ch_\smalltriangleup)$ to include the sequence $\{a_k\}$ as dynamical variables. By doing this we obtain a new triple, denoted $(\cb_t,D_t,\ch_t)$, which is a fibration of triples $(\cb_{\smalltriangleup},D_{\smalltriangleup},\ch_\smalltriangleup)$ over the space of permissible sequences $\{a_k\}$. It turns out that there is a way to obtain this which leaves the spectral triple $(\cb_t,D_t,\ch_t)$ with very few free parameters.

This section gives a presentation of ideas and methods used to construct the triple $(\cb_t,D_t,\ch_t)$. A detailed account of the construction will be presented elsewhere.

To emphasise the dependency on the sequence $\{a_i\}$ we shall in the following write $D_{\smalltriangleup(a_i)}$ instead of $D_{\smalltriangleup}$.

\subsection{The space of permissible sequences $\{a_i\}$}

We would like to think of the parameters $a_i$ as a coordinates in a space $H$ of permissible sequences $\{a_i\}$ (that is, sequences that satisfy the condition given by (\ref{suff})) and to define a Dirac operator on this space. Clearly $H\subset \mathbb{R}^\infty$. The sequence $\{a_i\}$ is characterised by the condition that it diverges sufficiently fast. This means that the inverse sequence $\{a_i^{-1}\}$ converges towards zero:
\ba 
\lim_i a_i^{-1}=0 \;\;\mbox{sufficiently fast}\;.
\label{Con1}
\ea
This convergency condition is easier to work with than condition (\ref{suff}) since convergent sequences can be understood in terms of a Hilbert space structure. Therefore, in a first step we consider $L^2$-sequences. That is, sequences 
\[
\{x_k\}=(x_1,\ldots,x_n,\ldots)\in \mathbb{R^\infty}
\]
which satisfy
\ba 
\|\{x_k\}\|^2 := \sum_k x_k^2 <\infty\;.
\label{L2}
\ea
The space of sequences satisfying (\ref{L2}) is an infinite dimensional, real, separable Hilbert space which we also denote $H$.

The exact relationship between vectors $\{x_k\}$ in $H$ and the sequences $\{a_k\}$ satisfying condition (\ref{suff}) will be determined in the following.

In order to construct a Dirac operator on $H$ we first need a Hilbert space structure {\it over} $H$. The goal is to construct a spectral triple over $H$. 

The techniques used to construct $L^2(H)$ and the corresponding spectral triple are essentially the same techniques we used to construct the space $\overline{\ca}^{\smalltriangleup}$ and the triple $(\cb_{\smalltriangleup},D_{\smalltriangleup},\ch_\smalltriangleup)$. The following is based on ideas by Higson and Kasparov, see \cite{Higson}. 

The strategy is to consider first finite dimensional subspaces $H_n$ of $H$, corresponding Hilbert spaces $L^2(H_n)$ and structure maps
\[
P^\ast_{m,n}:L^2(H_n)\rightarrow L^2(H_m)\;,\quad n\leq m 
\]
between Hilbert spaces. Next step is to construct a spectral triple over each space $H_n$, ensure compatibility with the structure maps $P^\ast_{m,n}$ and finally obtain a triple over $H$ by taking the limit $n\rightarrow\infty$.

There are two potential difficulties with this strategy:
\begin{itemize}
\item
The spaces $H_n$ are non-compact. The construction of the spectral triple $(\cb_{\smalltriangleup},D_{\smalltriangleup},\ch_\smalltriangleup)$ relies strongly on the fact that $G$ is compact. This is a necessary condition to construct the structure maps $P^\ast$ (\ref{PrOj}) between Hilbert spaces. Specifically, we used the fact that the constant function is square integrable. This is no-longer the case when the underlying space is non-compact.
\item
The Dirac type operator, which we aim to construct, will have a non-compact resolvent. This is the same problem we encountered when we constructed the operator $D_{\smalltriangleup}$ and a problem which will always arise when one constructs a Dirac like operator on an infinite dimensional space (see also \cite{Higson}). This problem was solved in section \ref{sec14} by introducing the parameters $\{a_i\}$, whose presence is the very reason why we now wish to construct the extended triple $(\cb_t,D_t,\ch_t)$. Therefore, it appears that we are in danger of a circular argument: Certainly, we do not wish to introduce yet another infinite sequence to ensure a well behaved Dirac operator acting on $L^2(H)$ entailing yet another extension of $(\cb_t,D_t,\ch_t)$ and so forth.
\end{itemize}

It turns out that there exist a way to construct the triple $(\cb_t,D_t,\ch_t)$ which avoids these two technical pitfalls. 

The first problem is related to the fact the canonical map from $\ch_n$ to $\ch_m$ is given by $\xi\rightarrow \xi\otimes 1_{n,m}$, where $1_{n,m}$ is the identity function on the $m-n$ dimensional orthogonal complement to $\ch_n$ in $\ch_m$. However, the function $1_{n,m}$ is not a $L^2$-function and therefore the canonical map is not a map between Hilbert spaces. We can remedy this defect by finding a suitable $L^2$-function $\phi_{n,m}$ to replace the constant function $1_{n,m}$. The choice of function $\phi_{n,m}$ is restricted by the requirement that it must lie in the kernel of the Dirac type operator, which we wish to construct, when this is restricted to the orthogonal complement to $\ch_n$ in $\ch_m$.

The second problem is related to the fact that we are dealing with an infinite dimensional space. 
To solve this problem we modify the canonical algebra of functions on $\ch_n$ to include operators which project onto finite dimensional subspaces.

\subsection{A spectral triple over $\mathbb{R}^\infty$}
\label{gausssection}

We first consider finite dimensional subspaces $H_n=\mathbb{R}^n$ embedded in $H$ by
\[
H_n\ni(x_1,\ldots,x_n)\rightarrow (x_1,\ldots,x_n,0,\ldots)\in H\;.
\]
Before we proceed with the construction we introduce some notation and definitions.
The Gaussian functions on $\mathbb{R}$
\[
\phi_\a(x) =(\p\a)^{1/4} \exp\left(\frac{-x^2}{2\a}\right)\;,
\]
where $\a$ is a real positive number, is a square integrable normalised function on $\mathbb{R}$ and the composition 
\[
\phi_\a(x_1)\cdot\ldots\cdot\phi_\a(x_n) =(\p\a)^{n/4} \exp\left(-\sum_{i+1}^n \frac{x_i^2}{2\a}\right)
\]
is square integrable and normalised on $\mathbb{R}^n$. Define also
\[
\phi_{n,m} (x_{n+1},\ldots,x_m)=\phi_\a(x_{n+1})\cdot\ldots\cdot \phi_\a(x_m)\;.
\]
This is a square integrable and normalised function on the orthogonal complement $H_{n,m}$ of $H_n$ in $H_m$.

Given a function $\xi\in L^2(H_n)$ define a function $P^\ast_{n,m}(f)$, $n<m$, in $L^2(H_m)$ by
\ba 
P^\ast_{n,m}(f)(x_1,\ldots,x_m)= f(x_1,\ldots,x_n)\cdot\phi_{n,m}\;.
\label{Hem}
\ea
Note that $P^\ast_{n,m}$ defines an embedding of Hilbert spaces. We can therefore define the limit
\[
L^2(H):=\lim_{\rightarrow}L^2(H_n)\;,
\]
which is a real, separable Hilbert space over $H$.

Note that, as mentioned in the previous section, the Gauss distributions play the same role in the construction as the constant function did in the construction of $\ch_{\smalltriangleup}$. This solves the first pitfall mentioned above.

Since we wish to construct a Dirac type operator we need the Clifford bundle. We define
\[
L^2(H,Cl(H)):=\lim_{\rightarrow}L^2(H_n,Cl(H_n))\;,
\]
where we have used the unital embedding $Cl(H_n)\rightarrow Cl(H_m)$, $n<m$.

We identify $Cl(H_n)$ with the exterior product $\Lambda^\ast H_n$ and define the operators
\ba
\mbox{ext}(e)&=& \mbox{exterior multiplication by $e$ on $\Lambda^\ast H_n$}\;,\nn\\
\mbox{int}(e)&=& \mbox{interior multiplication by $e$ on $\Lambda^\ast H_n$}\;,\nn\\
c_\pm (e)&=&  \mbox{ext}(e)\pm \mbox{int}(e)\;.\nn
\ea
which means that $c_-(e)$ is Clifford multiplication with $e$ and $c_+(e)$ is Clifford multiplication with $e$ except that $e^2=1$ and not $e^2=-1$. Furthermore, one easily checks that $\{c_+(e),c_-(e)\}=0$.
 
Define $e_i$ to be the vector which is $1$ and the $i$'th place and zero elsewhere. Define the operator $D_n$ on $L^2(H_n,Cl(H_n))$ by
\ba 
D_n = \sum_{i=1}^n \left( c_-(e_i)\frac{\pa}{\pa x_i} + c_+(e_i)\frac{x_i}{\a} \right)\;.
\label{BDirac}
\ea
This operator is known as the Bott-Dirac operator and its construction is due to Higson and Kasparov \cite{Higson}. It is proven in \cite{Higson} that it is self adjoint with compact resolvent. It satisfies
\[
D_n \phi_{1,n}(x_1,\ldots,x_n)=0\;.
\]
In fact, the kernel of $D_n$ is one dimensional and given by the function $\phi_{1,n}(x_1,\ldots,x_n)$. From this we get that
\[
P^\ast_{n,m}(D_n(\xi))=D_m(\xi)
\]
and we obtain a self adjoint unbounded operator $D$ on $L^2(H,Cl(H))$.

The square of $D_n$ is
\ba 
D_n^2 = \sum_i \left(-\frac{\pa}{\pa x_i}\frac{\pa}{\pa x_i} + \frac{x_i x_i}{\a^2}\right)  -\frac{n}{\a}+ \frac{2}{\a}N\;,
\label{harmonic}
\ea
where $N$ is the operator which assigns a differential form its degree. The first part of (\ref{harmonic}) is the harmonic oscillator Hamiltonian which has the spectrum $0,2/\a,4/\a,\ldots$. 

As anticipated, the operator $D$ does not have compact resolvent since the group of all finite perturbations of $\mathbb{N}$ acts on its spectrum. A possible solution to this problem is to modify the canonical algebra of functions acting on $L^2(H,Cl(H))$. In stead of functions on $L^2(H,Cl(H))$ we consider instead functions on finite subspaces of $L^2(H,Cl(H))$ and extend these to the full Hilbert space by tensoring them with projection onto Gauss functions. Thus, effectively we include cut-off's in the algebra. This will ensure that the resolvent of $D$, understood in terms of an interaction with the algebra, is compact.

Define
\[
\Phi_{n,m}:= \mathds{1}_{H_n}\otimes P_{\phi_{n,m}}\in \mathbb{B}(L^2(H_m\\
),Cl(m))\;,
\]
where $P_{\phi_{n,m}}$ is the orthogonal projection onto $\phi_{n,m}$ in $H_{n,m}$. Let $A_n$ be the algebra of operators in $L^2(H_n,Cl(H_n))$ generated by tensor products of elements in $C_0(\mathbb{R})\otimes Cl(1)$ and $\Phi_{k,l}$. Note that $A_n$ embeds in $A_m$, $n<m$, by
\[
A_n\ni a\rightarrow a\otimes \mathds{1}_{H_{n,m}}\cdot \Phi_{n,m}\;.
\]
Define
\[
A = \lim_{\rightarrow} A_n\;.
\]
In particular, $A$ consist of operators on $L^2(H,Cl(H))$. Since the operators in $A$ contain a cut-down to the Gauss distribution from a certain point on, we see that $(A,D,L^2(H,Cl(H)))$ is a spectral triple since
\[
a (\l-D)^{-1} \;,\quad a\in A\;,\l\not\in\mathbb{R}
\]
is a compact operator.

Note that the semi-finite trick with the Clifford bundle, which worked for the spaces of connections does not seem to work in this case, since the kernel of $D$ is one dimensional. Therefore there is no symmetry which can be discarded of.

\subsection{$\theta$-summability}

We would like to interpret condition (\ref{L2}) in terms of $\theta$-summability of the operator $D_{\smalltriangleup(a_i)}$. That is, we wish to establish a relation between sequences $\{x_i\}$ satisfying (\ref{L2}) and sequences $\{a_i\}$ which leave
\ba 
Tr\exp\left( -D_{\smalltriangleup(a_i)}^2 \right)
\label{ARgue}
\ea
finite.
Thus, we seek a function 
\[
f:\mathbb{R}^\infty\rightarrow \mathbb{R}^\infty
\]
which satisfies
\ba 
Tr\exp\left( -D_{\smalltriangleup(a_i)}^2 \right)<\infty \Leftrightarrow f(\{a_i\})\in l^2(\mathbb{N})\;.
\label{JJJ}
\ea

This emphasis on $\theta$-summability has a clear physical motivation. In section \ref{EINstein} we noted that the quantity (\ref{ARgue}) has the form of a path integral. In this light $\theta$-summability simply means that this path integral is finite.

%\subsubsection{A modified Dirac type operator on $\ch_{\smalltriangleup}$}

\subsection{The $U(1)$-case}

As a toy example let us consider the $U(1)$-case and the simplified operator $D'_{\smalltriangleup(a_i)}$ on $U(1)^\infty$ which is the limit of operators of the form 
$$
D'_n := D_1 + a_2 D_2 + \ldots + a_n D_n\;,
$$ 
where $D_i$ is the Dirac operator on the $i$'th copy of $G$.
With this operator we consider expression (\ref{ARgue}) and calculate
\ba 
Tr\exp\left( -(D')_{\smalltriangleup(a_i)}^2 \right) 
&=& 
\left(\sum_{n\in\mathbb{N}^\infty}\exp\left( -\sum a_i^2 n_i^2 \right)\right)  
\nn\\
&=& 
\P_{n} Tr \exp\left( -a_n^2 D_1^2 \right)\;,
\nn 
\ea
where $D_1$ is the Dirac operator on $U(1)$. Taking the logarithm we obtain
\ba 
\ln Tr\exp\left( -(D'_{\smalltriangleup(a_i)})^2 \right) 
&=& 
\sum_n \ln Tr \exp\left( - a_n^2 D_1  \right)
\nn\\
&=& 
\sum_n \ln \sum_k\exp\left( -a_n^2 k^2 \right)\;.
\ea
Thus $$Tr\exp\left( -(D'_{\smalltriangleup(a_i)})^2 \right)<\infty $$ if and only if
\[
\sum_n\sum_{k>1} \exp\left( - a_n^2 k^2 \right) <\infty\;.
\]
If we define the function
\[
f(a)= sgn(a)\sqrt{\sum_{k>1}\exp(-a^2 k^2)  }\;,
\]
we see that, for $G=U(1)$ and for the modified operator, $\theta$-summability of $D'_{\smalltriangleup(a_i)}$ is directly related to the Hilbert space condition
\[
\sum_n f(a_n)^2 <\infty\;.
\]
This means that the sequence $\{a_i\}$ is related to the Hilbert state $\{x_i\}\in H$ through the relation
\ba 
x_i = f(a_i)\;.
\label{RELa}
\ea

\subsection{The triple $(\cb_t,D_t,\ch_t)$}
\label{tHE}

We are now ready to combine the two spectral triples $(\cb_{\smalltriangleup},D_{\smalltriangleup},\ch_\smalltriangleup)$ and 
$(A,D,L^2(H,Cl(H)))$. A point $\{ x_n \}$ in $H$ gives a $\theta$-summable spectral triple on $\cb_{\smalltriangleup}$ via the Dirac operator on $\ch_\smalltriangleup$ defined by the sequence $f^{-1}(\{x_n\})$. Let us denote this Dirac operator by $D_{f^{-1}(\{x_n\})}$. We define the operator $D_t$ acting on 
$$ \ch_t:=  L^2(H,Cl(H)) \otimes  \ch_\smalltriangleup $$ 
by
\ba
D_t(\xi \otimes \eta )(\{x_n\})&=& \nn\\
&&\hspace{-2cm} D (\xi)\otimes \eta (\{x_n\})+(-1)^{\hbox{deg}(\xi)}\xi \otimes D_{f^{-1}(\{x_n\})} (\eta)(\{ x_n\})\;,
\label{DT}
\ea 
where $\hbox{deg}(\xi)$ means the degree of $\xi$ with respect to the degree in $CL(H)$.

One of the requirement that this defines a semi-finite spectral triple over 
$ A \otimes \cb_{\smalltriangleup} :=\cb_t$
is that 
$$\frac{a}{1+D_t^2}, \hbox{ for all } a\in \cb_t,$$
is $\t$-compact, where $\t$ is a trace defined in (\ref{TraCE}) tensored with the trace on $B(L^2(H,Cl(H)))$ where $B$ denotes the bounded operators.

We see that this requirement collide with the following symmetry property: All $a\in \cb_t$ are from a certain step of the form 
$$\mathds{1}\otimes P_{\phi_{n,\infty}}$$
and therefore the operator  
$$\frac{a}{1+D_t^2}$$
is invariant under symmetries permuting all coordinates bigger than $n$.

In order to remedy this we modify the construction of the spectral triple $(A,D,L^2(H,Cl(H)))$ slightly. This modification is based on the observation that the parameter $\a$ in the Gauss distribution $\phi_\a(x)$ is in fact a free parameter. Further, there are an infinite number of these free parameters in the construction of the triple $(A,D,L^2(H,Cl(H)))$, one for each dimension in $H$. Let us denote these with $\{\a_i\}$. By choosing these parameters carefully we can solve the problem mentioned above.

For $D_t$ to have a $Tr$-compact resolvent its eigenvalues must approach infinity. However, for each copy of $G$ we know that its eigenvalues scale with $(\a_i)^{-1}$. Thus, if we require the sequence $\{\a_i\}$ to satisfy 
\[
\lim_{i\rightarrow\infty} \a_i =0\;,
\]
then the triple $(\cb_t,D_t,\ch_t)$ may be spectral.

Let us pause to comment on this idea. It appears that we have simply exchanged one infinite sequence of free parameters $\{a_i\}$ with another infinite sequence $\{\a_i\}$. However, we suggest that the sequence $\{\a_i\}$ is not arbitrary but should be determined completely from symmetry considerations. The idea is that the $\a_i$'s represent an ``average background'' over which the parameters $a_i$ are allowed to fluctuate. Therefore, we suggest to choose a specific sequence $\{\a_i\}$ satisfying the following requirements:
\begin{itemize}
\item 
Copies of $G$ which correspond to the same level in the projective system of graphs should be assigned the same $\a_i$'s.
\item
Given two copies of $G$ where one copy $G_2$ corresponds to a subdivision of the edge associated to the other copy $G_1$, then the corresponding parameters $\a_1,\a_2$ should satisfy
\[
\a_2=\frac{\a_1}{2} \;.
\]
\end{itemize}

The first point reflects a rotational symmetry within each graph. The second point is related to the scaling properties of the construction: a division of a line segment corresponds to a factor $\frac{1}{2}$ (see section \ref{rolE} and the discussion of the role of the parameters $\{a_i\}$).

To implement this we enumerate the barycentric subdivisions with $k$ and introduce the modified Gauss distributions
\[
\phi_k(x) =(\p\a_k)^{1/4} \exp\left(\frac{-x^2}{2\a_k}\right)\;,
\]
where
\ba 
\a_k = 2^{-k}\a\;.
\label{tHe}
\ea
This means that we associate to each edge in the simplicial complexes a parameter $\a_k$. In the initial complex each edge is associated the parameter $\a_0=\a$. Edges arising through the first barycentric subdivision are then associated the parameter $\a_1=\a/2 $ and so forth.

Consider now the triple $(A,D,L^2(H,Cl(H)))$ constructed with with these new Gauss distributions. This means that the Hilbert space embeddings (\ref{Hem}) are modified together with the Bott-Dirac operator (\ref{BDirac}) which is now of the form
\[ 
D_n = \sum_{k=0}^N\sum_{i=1}^{n_k} \left( c_-(e_i)\frac{\pa}{\pa x_i} + c_+(e_i)\frac{x_i}{\a_k} \right)\;,
\]
where $n=\sum_k n_k$ and where the first sum runs over the number of barycentric subdivisions involved in the graph to which $D_n$ is associated, $N$ being the total number of barycentric subdivisions. The second sum runs over the number of edges within a barycentric subdivision. 

One now repeats the above construction leading to the operator (\ref{DT}), now with the modified Bott-Dirac operator.

The details concerning the spectral properties of the triple $(\cb_t,D_t,\ch_t)$ will appear elsewhere.

%\subsection{Diffeomorphism invariance revisited}

\section{Distances on $\overline{\ca}^{\smalltriangleup}$}
\label{DISTa}

On a Riemannian spin-geometry the Dirac operator $D$ contains the geometrical information of the manifold $\cm$. In particular, distances can be formulated purely algebraically due to Connes \cite{ConnesBook}. Given two points $x,y\in\cm$ their distance is given by
\ba
d(x,y)=\sup_{f\in C^{\infty}(\cm)}\big\{ |f(x)-f(y)|\; \| [D,f] \| \leq 1  \big\}\;.
\label{distance}
\ea
On a Noncommutative Geometry the state space replaces the notion of points. It is possible to extend the notion of distance to the state space by generalising (\ref{distance}).

In the present case it is natural to ask whether the Dirac-like operator $D_{\smalltriangleup}$ introduces a metric structure over the space $\overline{\ca}^{\smalltriangleup}$ and if so, how distances should be interpreted in this setting.

The possibility for a distance between smooth connections in $\ca$ and some closure hereof was first considered in \cite{Aastrup:2005yk}. There, however, the distance between two smooth connections was found to be infinite, since two smooth connections will differ on infinitely many different loops. This entails a finite contribution to the distance from infinitely many copies of $G$.

In the present setting the situation is different due to the sequence $\{a_i\}$. The role of the $a_i$'s is to assign a weight to each copy of $G$ and therefore to scale the corresponding distance on $G$. Therefore, given two points in $\overline{\ca}^{\smalltriangleup}$ their distance  
\ba
d(\nabla_1,\nabla_2)=\sup_{a\in\cb_{\smalltriangleup}}\big\{  \|\nabla_1(a)-\nabla_2(a)\|\; \| [D_{\smalltriangleup},a]\|\leq 1 \big\}\;,\quad\nabla_1,\nabla_2\in\overline{\ca}^{\smalltriangleup}\;,
\label{diSt}
\ea
where we choose $\|\cdot\|$ as the operator norm on matrices in $M_l(\mathbb{C})$, is well defined and finite, even for $\nabla_1,\nabla_2\in\ca$. This is related to the fact that although our geometry on $\overline{\ca}^{\smalltriangleup}$ is infinite dimensional it assigns $\overline{\ca}^{\smalltriangleup}$ zero volume.

The distance (\ref{diSt}), when it is applied to smooth connections, is not independent of the embedding $h$ of simplicial complexes into triangulations.

With a metric structure on the space $\overline{\ca}^{\smalltriangleup}$ it is natural to ask which connections are 'close' to each other. Clearly, this depends on the choice of the sequence $\{a_i\}$. However, it is possible to give some general remarks independently of this choice.

Recall that the space $\overline{\ca}^{\smalltriangleup}$ is the limit space $\lim_n G^n$ in an appropriate sense. The distance between $\nabla_1,\nabla_2\in\ca_{\ck_i}$ is simply the sum of geodesic distances on each copy of $G$, weighted with the appropriate parameters $a_i$. In the previous section we found that the role of the parameters $\{a_i\}$ is to scale the different copies of $G$ according to their location in the projective system of simplicial complexes. The larger $a_i$'s corresponds to short distances, now with respect to some manifold.
This means that the the distance between $\nabla_1$ and $\nabla_2$ is larger if the connections differ mostly on those copies of $G$ which are assigned small $a_i$'s. Correspondingly, the distance between $\nabla_1$ and $\nabla_2$ is smaller if $\nabla_1$ and $\nabla_2$ differ mostly on those copies of $G$ which are assigned large $a_i$'s. Therefore, in general, the two generalised connections will be 'close' to each other if they differ mostly at short scales. 

Via Levi-Civita connections we can interpret equation (\ref{diSt}) as a distance between geometries\footnote{In this case, however, the distance formula will be degenerate since different geometries may have identical Levi-Civita connections.}. Again we can say that the distance between two geometries depends on the scale on which they differ.

Let us consider the Abelian case $G=U(1)$ and two points $\nabla_1,\nabla_2$ in $\overline{\ca}^{\smalltriangleup}$. In this case $\nabla_1$ and $\nabla_2$ are given by sequences of angles $\{\theta^1_i\}$ and $\{\theta^2_i\}$ where each angle $\theta^j_i$ corresponds to points $\exp(2\pi\rm{i}\theta^j_i)$ on the $i$'th copies of $U(1)$. Let us for simplicity choose a coordinate system on $G^n$ so that $\theta^j_i$ corresponds to the parameter $a_i$. Thus, for example, if we consider $G\times G$ parametrised by angles $\phi_1$ and $\phi_2$ we choose
\[
\theta_1 = \phi_1 + \phi_2 \;,\quad \theta_2 = \phi_1 - \phi_2
\]
and so forth. Then the distance between $\nabla_1$ and $\nabla_2$ reads
\[
d(\nabla_1,\nabla_2)= \sum_k \frac{2^k}{a_k 2\p}|\theta_k^1-\theta_k^2 |\;,
\]
which should be read together with the condition (\ref{Con1}).

The notion of a distance on a space of connections is not a new one but was discussed already by Feynman \cite{Feynman:1981ss} and Singer \cite{Singer:1981xw}. See also \cite{Orland:1996hm} and references herein. However, these papers all deal with distances between gauge equivalence classes of connections. The construction presented in this paper is quite different from these previous works since two connections which differ by a gauge transformation will, in general, have a non vanishing distance.

\section{Discussion}
\label{DISCu}

Before we end this paper we give a few remarks concerning the physical interpretation of the constructions presented.

\subsection{The quantization scheme}
\label{link2}

 As already mentioned, the shift to Ashtekar variables permits a formulation of General Relativity which is close to ordinary Yang-Mills theory. In the Hamilton formulation the theory involves a configuration space of connections corresponding to ordinary $SU(2)$ Yang-Mills theory. The essential difference between General Relativity and Yang-Mills theory lies in the algebra of constraints which encode the symmetries of the theory. For General Relativity the constraints encode diffeomorphism invariance. These constraints are closely related to the foliation of the spacetime $\cm$ according to
\ba 
\Sigma\times \mathbb{R}\rightarrow \cm\;,
\label{foliation}
\ea
where $\Sigma$ is a 3-dimensional hypersurface. Loosely stated, the constraints encode information about diffeomorphisms within and perpendicular to $\Sigma$. Therefore, the key question in any quantization procedure of gravity is the implementation of diffeomorphism invariance in the construction.

To quantize a constrained theory there are in general two standard approaches:
\begin{itemize}
\item[{\bf a})]
To eleminate first, at a classical level, the constraints and thereby find the reduced phase space of the theory. The quantization procedure is applied to the reduced phase space.
\item[{\bf b})] 
To construct first quantum kinematics for the full phase space by ignoring the constraints, then construct operators corresponding to the classical constraints and finally solve the quantum constraints to obtain the physical Hilbert space.
\end{itemize}
Loop Quantum Gravity follows the second approach which is due to Dirac (see \cite{Henneaux:1992ig} for a detailed exposition). In section \ref{canqua} we demonstrated that the spectral triple $(\cb_{\smalltriangleup},D_{\smalltriangleup},\ch_\smalltriangleup)$ involves a representation of the Poisson algebra (\ref{com-hol}) of General Relativity. In our interpretation, the construction should be understood as a quantization scheme which lies somewhere between the approaches {\bf a}) and {\bf b}).

%In section \ref{canqua} we demonstrated that the spectral triple $(\cb_{\smalltriangleup},D_{\smalltriangleup},\ch_\smalltriangleup)$ involves a representation of the Poisson algebra (\ref{com-hol}) of General Relativity. Our interpretation of the construction differs, however, from a standard 'Dirac type'  canonical quantization of General Relativity on several points, mainly concerning the constraints and the foliation. 

First of all, there is nothing in the geometrical construction that corresponds directly to the constraint algebra of General Relativity. However, in section \ref{section2} we related the Hilbert space $\ch_{\smalltriangleup}$ to the diffeomorphism invariant Hilbert space $\ch_{diff}$ of Loop Quantum Gravity. The difference between $\ch_{\smalltriangleup}$ and $\ch_{diff}$ is the group diff($\triangle$) of discrete diffeomorphisms on the system of simplicial complexes. Therefore, we think of the spectral triple $(\cb_{\smalltriangleup},D_{\smalltriangleup},\ch_\smalltriangleup)$ as a geometrical construction describing a quantization procedure of gravity on a Hilbert space which corresponds to a {\it partial} solution to the (spatial) diffeomorphism constraint. What remains then, in this picture, is the implementation of the remaining diffeomorphisms encoded in the discrete group diff($\triangle$).

The second point in which our construction differs from canonical quantum gravity is the issue of time. Clearly, a foliation of the manifold is essential for canonical gravity since the Hamilton formulation is based on a choice of an explicit time direction.
In contrast to this, the construction presented here does not a priori involve a foliation of space-time. Further, the dimension of the underlying space is completely free. The only restriction is the choice of symmetry group, which, for now, has to be compact\footnote{The case $G=SO(1,3)$, for example, is not permitted.}. In particular, we can consider triangulations of a 4-dimensional manifold.

These observations suggest two possible interpretations of the construction:
First, the construction may be interpreted as a completely covariant construction in four dimensions, involving the full, four-dimensional group of diffeomorphisms. This interpretation leaves no room for a foliation and thus no Hamiltonian constraint. Therefore, a dynamical principle replacing the Hamiltonian constraint must be sought elsewhere. 
This interpretation implies that the group $G$ corresponds to four-dimensional local Lorentz rotations.
Next, if the loop algebra lives on a four-dimensional manifold, then the interpretation of the Dirac type operator $D_{\smalltriangleup}$ in terms of canonical quantum gravity must involve a flux operator ${\bf F}^a_S$ where $S$ is now a three-dimensional hypersurface. Further, the interaction between $D_{\smalltriangleup}$ and the algebra of loops should be understood as a four-dimensional operator algebra containing a representation of the Poisson algebra of General Relativity as a kind of subalgebra.

Alternatively, we may interpret the construction as fundamentally three-dimensional, with $G$ equal to $SU(2)$. This interpretation does not necessarily mean that we consider a foliation of space-time according to (\ref{foliation}). Rather, we would like time to emerge naturally from the construction.

Indeed, let us end this discussion with the remark that the philosophy of Noncommutative Geometry is to seek a time {\it within} the algebraic construction rather than imposing it a priori. Here we think of the Tomita-Takesaki theory \cite{tt} which identifies a one-parameter automorphism group uniquely up to inner automorphisms. It is an interesting question whether this group of modular automorphisms is nontrivial in this or perhaps some similar construction. If the answer is in the affirmative, then this might provide us with an alternative to the foliation and thereby with a new dynamical principle.

\subsection{The group diff$(\triangle)$ and the question of constraints}

Regardless of how one interprets the spectral triple $(\cb_{\smalltriangleup},D_{\smalltriangleup},\ch_\smalltriangleup)$ it is clear that one should seek to obtain invariance under the discrete group diff$(\triangle)$ of graph preserving diffeomorphisms.

Traditionally, given some space $M$, invariance under a symmetry group $\cg$ is obtained by taking the the quotient 
\[
M/\cg\;.
\]
However, Noncommutative Geometry gives an alternative approach to quotient spaces. A simple example is the space of two points identified. Within Noncommutative Geometry this setup is described via two by two matrices where the off-diagonal entries represent the identification of the points. This entails a noncommutative algebra and a structure which, ultimately, leads to the Higgs mechanism in the noncommutative formulation of the Standard Model.

In the case of gravity the relevant symmetry group is the diffeomorphism group. This group is probably too large for an application of the noncommutative approach. However, in the present setup we have instead the much smaller group diff$(\triangle)$ and one could speculate if the machinery of Noncommutative Geometry of quotient spaces could be successfully applied. 
This idea differs fundamentally from a Dirac-type quantization procedure.

Thus, what we suggest is to obtain a formal diffeomorphism invariance by considering the semi-direct product of the loop algebra with the group diff$(\triangle)$
\[
\cb_{\smalltriangleup} \rtimes \mbox{diff}(\triangle)\;,
\]
and thereafter building a spectral triple with an associated Hilbert space $L^2(\ca)\otimes L^2(\mbox{diff}(\triangle))$. Presumably, such a construction will give rise to additional degrees of freedom through fluctuations around the Dirac type operator.

\subsection{Background independence}
\label{background}

In section \ref{link} we argued that the space $\overline{\ca}^{\smalltriangleup}$ should be viewed as a space of connections subjected to a gauge fixing of the diffeomorphism group. Naively, it appears that this gauge fixing does not involve any choice of metric or background 'field' on the manifold. Let us comment on this.

The construction in section \ref{section1} depends primarily on the initial simplicial complex $\ck_o$. The idea is that the basic data entering the construction is the topology of the corresponding manifold. The initial complex gives rise to an initial triangulation $\ct_o$ via a homeomorphism $$h:\ck\rightarrow\cm\;.$$ 
 This initial triangulation introduces a metric structure on $\cm$. The embedding $\ca\hookrightarrow\overline{\ca}^{\smalltriangleup}$ will, of course, depend crucially on the triangulation and in particular on the homeomorphism $h$. However, the construction and spectrum of the Dirac-type operator is independent of $h$ and so is its interaction with the algebra and the construction of the Hilbert space. In fact, the notion of a manifold may be left out altogether. It is only the identification of $\overline{\ca}^{\smalltriangleup}$ as a space of generalised connections that requires a manifold.

Another question is the dependency on the sequence $\{a_i\}$. The Dirac type operator $D_\smalltriangleup$ depends crucially on this sequence as does its spectrum. Thus, one may argue that some degree of background dependency enters the construction with the sequence $\{a_i\}$. In section \ref{includea} we attempted to free the construction of this dependency by making the $a_i$'s dynamical. The price paid, however, is the introduction of a new sequence $\{\a_i\}$ determining the Gauss distributions. Therefore, it seems that the construction, whether we consider the triple $(\cb_{\smalltriangleup},D_{\smalltriangleup},\ch_\smalltriangleup)$ or the triple $(\cb_t,D_t,\ch_t)$, does posses some degree of dependency on a parameter space which one may interpret in terms of a background. The exact nature and implication of this dependency is to be clarified.

\subsection{Additional degrees of freedom}

The construction of the triples $(\cb_{\smalltriangleup},D_{\smalltriangleup},\ch_\smalltriangleup)$ and $(\cb_t,D_t,\ch_t)$ takes canonical quantum gravity as its point of departure. Thus, both spectral triples are a priori of purely gravitational origin. However, there are several indications that the framework presented in this paper contains additional degrees of freedom, both bosonic and fermionic.

First, recall that the framework of Noncommutative Geometry generally involves fermionic degrees of freedom since it involves a Dirac type operator acting on a Hilbert space\footnote{For a commutative algebra the underlying space is a Riemannian geometry and the Dirac operator acts on a Hilbert space of spinors. In the noncommutative formulation of the Standard Model the Hilbert space is labelled by the fermions in the Standard Model.}. The 'fermions' involved in the spectral triple $(\cb_{\smalltriangleup},D_{\smalltriangleup},\ch_\smalltriangleup)$ are clearly very different from the fermions of the Standard Model since they live on a space of connections. However, this space of connections is of course linked to an underlying manifold. A classical limit will presumable involve some kind of delta function on the space of connections (see next section for a discussion hereof) and therefore leave only space-time degrees of freedom for the fermions. Therefore the interesting question is what structures may emerge from these 'fermions' in a classical limit.

Furthermore, during the construction of the the triple $(\cb_{\smalltriangleup},D_{\smalltriangleup},\ch_\smalltriangleup)$ we found that the CAR algebra emerged in an almost canonical fashion. It seems plausible that almost any construction of a Dirac type operator on an infinite dimensional space will naturally entail an infinite dimensional Clifford bundle which, in turn, leads to the CAR algebra. Thus, it seems that Noncommutative Geometry provides us with a mechanism which equips a purely gravitational setting, the construction of a Dirac type operator on an infinite dimensional space of field configurations, with basic elements of fermionic Quantum Field Theory.
In section \ref{sectiontrace} we found that the Hilbert space $\ch_\smalltriangleup$ factorizes into
\[
\ch_\smalltriangleup = \ch_{\smalltriangleup,b}\otimes \ch_{\smalltriangleup,f}\;,
\]
where the CAR algebra acts on $\ch_{\smalltriangleup,f}$. This shows that $\ch_\smalltriangleup$ naturally involves both bosonic and fermionic sectors.

Finally, let us consider the possibility that the noncommutativity of the loop algebra will generate an additional bosonic sector
through the inner automorphisms of the algebra. To explain this consider first a classical finite-dimensional, real spectral triple $(A,D,H)$. 
%That the triple is real means that it is equipped with a real structure $J$ satisfying certain axiomatic requirements \cite{Connes:1996gi}. 
A noncommutative algebra $A$ contains inner automorphisms of the form
\[
\a_u(x)= u x u^\ast\quad \forall x\in A \;,
\]
where $u$ is an arbitrary element of the unitary group, $uu^\ast=u^\ast u =1$. A change of representation of the algebra $A$ from
$\pi$ to  $\pi\circ \a_u^{-1}$ is equivalent to the replacement of the Dirac operator $D$ by
\ba 
\tilde{D}= D + A' + J A' J^{-1}\;,
\label{asdf}
\ea
where $A'= u[D,u^\ast]$ is a noncommutative one-form, and where $J$ is a real structure. In the noncommutative formulation of the Standard Model coupled to gravity \cite{Connes:1996gi} the entire bosonic sector of the Standard Model, including the Higgs boson, is generated by this type of fluctuations of the Dirac operator by a general one-form $A'=\sum a_i[D,b_i]$ with $a_i,b_i\in A$.

This mechanism is general. The interesting question is what kind of bosonic sector the inner automorphisms of the spectral triple $(\cb_{\smalltriangleup},D_{\smalltriangleup},\ch_\smalltriangleup)$ will generate.  \\

%We find the prospect that pure Quantum Gravity may, as a "free" spin-off, contain additional matter degrees of freedom, fascinating. Where the message coming from finite-dimensional Riemannian spin geometry (commutative spectral triples) is that fermions carry classical geometrical data, then the message coming from this paper seems to be that quantized fermions carry data whose origin is the quantum theory of gravity.

The general idea behind these remarks is that pure Quantum Gravity may, as a "free" spin-off, contain matter degrees of freedom. Thus, we believe that one should {\it not} attempt to couple matter degrees of freedom to the construction
presented in this paper but rather hope to see matter emerge dynamically.

\section{Conclusion and outlook}
\label{section4}

In this paper we establish a link between the mathematics of Noncommutative Geometry and the field of canonical quantum gravity. We construct a spectral triple $(\cb_{\smalltriangleup},D_{\smalltriangleup},\ch_\smalltriangleup)$ over a space of connections and show that the Poisson structure of General Relativity, formulated in terms of loop variables, is encoded in the interaction between the Dirac type operator $D_{\smalltriangleup}$ and the loop algebra $\cb_{\smalltriangleup}$. The Hilbert space $\ch_\smalltriangleup$ corresponds to a partial solution of the diffeomorphism constraint of canonical gravity. The inner product of $\ch_\smalltriangleup$ involves a functional integral over a space of connections and the Dirac type operator $D_{\smalltriangleup}$ has the form of a global functional derivation. Consequently, we interpret the triple in terms of a non-perturbative, background independent, quantum field theory. \\

The construction is based on a projective system of simplicial complexes. The simplicial complexes are related through repeated barycentric subdivisions. The triple $(\cb_{\smalltriangleup},D_{\smalltriangleup},\ch_\smalltriangleup)$ is the limit of spectral triples formulated at the level of finite simplicial complexes. Since the operation of barycentric subdivision is countable the limit triple is separable and spectral.\\

The square of the Dirac type operator has, in terms of canonical quantum gravity, a natural interpretation as global  operator related to the area operators known in Loop Quantum Gravity. We interpret the operator $(D_{\smalltriangleup})^2$ in terms of an action and show that the spectral action of $D_{\smalltriangleup}$ has the form of a Feynman path integral. Thus, at the core of the construction we find an object which resembles a partition function related to Quantum Gravity.\\

The construction of the spectral triple $(\cb_{\smalltriangleup},D_{\smalltriangleup},\ch_\smalltriangleup)$ differs from a traditional canonical quantization procedure of General Relativity in the way the group of diffeomorphisms is treated. Rather than encoding the symmetries of the classical theory in a set of constraints the construction works directly on a Hilbert space $\ch_\smalltriangleup$ which corresponds to a partial solution of the (spatial) diffeomorphism constraint. The existence of the spectral triple relies on a particular split of the diffeomorphism group into a countable and an over-countable part. The over-countable part is discarded of; the group of countable diffeomorphisms is the group diff$(\triangle)$ of diffeomorphisms between simplicial complexes. This means that the Poisson algebra of General Relativity is represented on the separable Hilbert space $\ch_\smalltriangleup$.\\

The Dirac type operator $D_{\smalltriangleup}$ depends on an infinite sequence of parameters $\{a_i\}$. These parameters determine the scaling behaviour of the construction. We believe that a correct understanding and treatment of these parameters is essential. In this paper we propose a possible way to treat the sequence $\{a_i\}$. Since the sequence is seen to have metric origin we propose to include it as a dynamical variable in the construction. This leads to a new triple, denoted $(\cb_t,D_t,\ch_t)$, which includes the spectral triple $(\cb_{\smalltriangleup},D_{\smalltriangleup},\ch_\smalltriangleup)$ as well as a sector permitting the sequence $\{a_i\}$ to vary. 
The triple $(\cb_t,D_t,\ch_t)$ is constructed to ensure that the operator $D_{\smalltriangleup}$ is $\theta$-summable. Thus, we permit only sequences $\{a_i\}$ which leaves the spectral action of $D_{\smalltriangleup}$ finite.\\ %The spectral properties of the triple $(\cb_t,D_t,\ch_t)$ will be analysed elsewhere.\\

Furthermore, the Dirac type operator $D_\smalltriangleup$ defines a distance on the underlying space of connections. Clearly, this distance depends strongly on the sequence $\{a_i\}$. However, we find that a general feature of this distance function is that two connections are "close" if they differ mostly at short scales. \\

%The square of the Dirac type operator $D_{\smalltriangleup}$ resembles an integral over an underlying manifold. We interpret this as a kind of action and find that the spectral action of $D_{\smalltriangleup}$ strongly resembles a Feynman path integral. Thus, at the core of the construction we find a background independent, nonperturbatively defined quantum field theory related to gravity.\\

It is possible to read this paper in a more conservative way, discarding the role of Noncommutative Geometry and reading it as a reformulation of Loop Quantum Gravity. If we ignore the construction of the Dirac type operator $D_{\smalltriangleup}$ and focus instead on the algebra $\cb_{\smalltriangleup}$ and the Hilbert space $\ch_{\smalltriangleup}$, without the Clifford bundle, and consider the algebra of the vectors $\hat{\ce}^i_j$, then, as already mentioned, we obtain a representation of the Poisson algebra of General Relativity on a separable Hilbert space. Therefore, this Hilbert space, let us denote it $\ch_{\smalltriangleup}'$, replaces the otherwise non-separable Hilbert space $L^2(\overline{\ca}^{\small a})$ known as the kinematical Hilbert space in Loop Quantum Gravity. In Loop Quantum Gravity the kinematical Hilbert space is the Hilbert space on which the constraints of General Relativity are defined. Thus, one could formulate the complete set of constraints of General Relativity in terms of operators acting on the Hilbert space $\ch_{\smalltriangleup}'$. First, as explained in section \ref{section2}, the Hilbert space $\ch_{\smalltriangleup}'$ has an action of the reduced set of diffeomorphisms diff$(\triangle)$. Therefore the spatial diffeomorphism constraint should be formulated in terms of the group diff$(\triangle)$. Next, one may likewise formulate the Gauss and Hamiltonian constraints of Loop Quantum Gravity on $\ch_{\smalltriangleup}'$. The central message here is that it is possible to formulate Loop Quantum Gravity in terms of a separable kinematical Hilbert space. It is an interesting question whether this observation will have an impact on any of the challenges facing Loop Quantum Gravity.\\

Let us finally remind the reader that the focus of this paper is the physical significance of the spectral triple $(\cb_{\smalltriangleup},D_{\smalltriangleup},\ch_\smalltriangleup)$. The detailed mathematical analysis of the triple is given in \cite{Aastrup}.\\

\noindent{\bf\large Outlook}\\

More analysis is needed to understand the physical and mathematical significance of the spectral triple $(\cb_{\smalltriangleup},D_{\smalltriangleup},\ch_\smalltriangleup)$. First of all it is imperative to understand the role and proper treatment of the sequence $\{a_i\}$ since this sequence is central to the existence of the triple. Let us here just say that the sequence $\{a_i\}$ should be understood in connection with the group of diffeomorphisms diff$(\triangle)$ since elements hereof are given by rearrangement of the parameters $a_i$. Having said this, let us list other issues which we think deserves attention.
\begin{itemize}
\item[-] 
First, we have seen that the spectral action resembles a Feynman path integral. We believe that the computation and analysis of the spectral action is the most interesting task to address at the present stage of the project. Here one should consider the Dirac operator which involves the inner fluctuations described in section \ref{DISCu}.
\item[-] 
A prime issue for any theory or framework for non-perturbative Quantum Gravity is the formulation of a semi-classical limit. In the present case the aim is to obtain a classical limit which involves not only a smooth geometry - characterised by a commutative $\ast$-algebra - but to obtain a limit which includes an additional matrix factor of the type that characterises the almost commutative algebra in Connes' formulation of the Standard Model. Indeed, since Connes' geometrical realization of the Standard Model is so attractive and powerful as it stands, it remains to understand why the algebra which lies at the heart of this formulation should have this particular noncommutativity. We suggest that the source of this noncommutativity lies in pure Quantum Gravity. Recall that the loop algebra $\cb_{\smalltriangleup}$ is essentially an almost commutative algebra over the space $\ca$ of connections. That is, it is a product of smooth functions over $\ca$ and a matrix factor $M_l(\mathbb{C})$. If we think of a classical limit as the emergence of a single geometry it seems reasonable to expect something close to a delta function peaked around a connection $\nabla$. However, when we apply the loop algebra on a delta function it reduces to the matrix factor $M_l(\mathbb{C})$ or some subalgebra hereof. Furthermore, if we keep in mind that the construction of the spectral triple $(\cb_{\smalltriangleup},D_{\smalltriangleup},\ch_\smalltriangleup)$ involves a choice of a basepoint\footnote{See appendix \ref{Appendix} for an extension of the triple $(\cb_{\smalltriangleup},D_{\smalltriangleup},\ch_\smalltriangleup)$ which avoids this choice of basepoint.}, then it seems possible that a similar construction which does not involve this basepoint will entail a smearing of the matrix factor over the manifold.
Thus, we speculate that the matrix factor behind the noncommutative formulation of the Standard Model emerges in the classical limit from the noncommutative algebra of holonomy loops.

We suspect this issue to be related to the calculation of the spectral action. However, one should also investigate whether ideas from Loop Quantum Gravity concerning coherent states \cite{Thiemann:2002vj} can be applied. 
\item[-]
The question about {\it time} is fundamental to any general covariant theory since such theories have no preferred time flow. This situation is even more complicated when one attempts to include quantum theory since this will presumable lead to a theory which involves some notion of superpositions of geometries and thus does not permit any notion of a predetermined time. A possible solution to this problem has been proposed by Connes (see for example \cite{Connes:2000ti}. See also \cite{Connes:1994hv} for similar ideas developed by Connes and Rovelli). The idea is that the concept of time is intimately linked to the noncommutativity of the algebra of observables of Quantum Gravity. Specifically, it is a fundamental property of von Neumann algebras that they possess a 1-parameter family of automorphisms which is unique up to inner automorphisms. Thus, the idea is that this group of automorphisms, the modular group, should be understood in terms of a time. It is therefore natural to ask whether the noncommutative ${\ast}$-algebras introduced in this paper give rise to a nontrivial modular group and whether this can be interpreted as a time flow. 

Alternatively, one might try to exploit the fact that the construction presented in this paper is basically quantum mechanics on the group $G$ taken infinitely many times. Here, each copy of $G$ corresponds to a degree of freedom originating somewhere on the underlying manifold. Thus, one could consider the time evolution, with respect to one copy of $G$, given by the operator $\exp(i\Delta t)$ where $\Delta$ is the Laplace operator. This suggest that the unitary operator
\[
U(t):=\exp(\rm{i}(D_{\smalltriangleup})^2t)
\]
may be thought of as a time evolution operator.
%\item[-]
%In the noncommutative formulation of the Standard Model it is, as previously mentioned, the noncommutativity of the algebra which generates the bosonic sector through so-called inner fluctuations of the Dirac operator. Although the situation in this paper is fundamentally different - the nature of the Dirac type operator $D_{\smalltriangleup}$ is quite different to the Dirac operator of Connes formulation of the Standard Model - it is nevertheless an interesting question to ask what kind of bosonic sector the inner fluctuations of the operator $D_{\smalltriangleup}$ will generate. Since the spectral triple $(\cb_{\smalltriangleup},D_{\smalltriangleup},\ch_\smalltriangleup)$ does involve inner automorphisms it seems clear that additional degrees of freedom will arise. Again, it seems plausible that this question should be analysed in combination with a classical limit. 
\item[-]
The construction of the spectral triple $(\cb_{\smalltriangleup},D_{\smalltriangleup},\ch_\smalltriangleup)$ starts with a simplicial complex. It is an important question to determine the exact dependency of the triple on the choice of initial complex. For instance, consider two simplicial complexes $\ck_1$ and $\ck_2$ chosen so that their union is the barycentric subdivision of yet another simplicial complex $\ck_3$. Does the construction depend on whether one chooses $\ck_3$ or the union of $\ck_1$ and $\ck_2$ as the initial complex? The answer is, a priori, yes, since the two choices will come with different sequences of parameters $\{a_i\}$. The exact nature of this dependency needs to be clarified. Clearly, the idea is that the spectral triple should depend only on topological data coming from the underlying manifold. A related issue is the merging of different spectral triples based on different simplicial complexes. By gluing complexes it should be possible to move from one topological setting to another. These issues are all connected with the sequence $\{a_i\}$ and the group of diffeomorphisms in diff$(\triangle)$. 
\item[-]
A related issue is the possibility to obtain a similar construction based on a different projective system of graphs. The choice of simplicial complexes (or, triangulations) and barycentric subdivisions seems natural but is not compulsory. In \cite{Aastrup} we provide certain necessary requirements for a system of graphs to be suitable for the construction of a spectral triple. These requirements leave room for projective systems of graphs which are not simplicial complexes. Again, more analysis is needed to determine the dependency of the final construction on different choices of graphs. It is clear that a projective system of graphs must be countable in order to permit the construction of a spectral triple. 
\item[-]
It is desirable to be able to deal also with a non-compact gauge group such as $SO(3,1)$. At present, the compactness of the gauge group is essential for the construction of the Hilbert space to work. Basically, we need the identity to be an $L^2$-function with respect to the Haar measure. One could speculate whether the techniques of Higson and Kasparov \cite{Higson} might be applied to resolve this issue. Here, the trick is to use a Dirac operator with a nontrivial square-integrable kernel and to define embeddings between Hilbert spaces via this kernel.
\item[-]
One should further clarify the relation between the construction presented here and Loop Quantum Gravity. In particular, it would be interesting to understand if there is a relation between the operator $D_{\smalltriangleup}$ and the Hamilton constraint. Here one should most likely consider the fluctuated version of $D_{\smalltriangleup}$, with respect to inner automorphisms, since this operator involves also the loop algebra. 
\item[-]
It is an interesting question whether the construction presented in this paper has anything to say about nonperturbative gauge theory. Consider a single line segment and the sequence $a_n=a_0 2^n$, where $n$ corresponds to the number of subdivisions of the segment. This is a natural sequence to consider but we know that the asymptotic behaviour $a_n\sim 2^n$ is too weak for the Dirac type operator to have a compact resolvent. If we consider instead the sequence $a_n=a_0 (2+|\e|)^n$ then $D_{\smalltriangleup}$ will have a compact resolvent as long as $\e$ is non-zero. This setup is then extended to all line segments in the projective system of graphs. As long as $\e$ is non-zero the corresponding spectral action of $D_{\smalltriangleup}$ is well defined and resembles a Feynman path integral over a space of connections. The interesting question is what theory this object represents. It seems clear that it should be understood as a non-perturbative Quantum Field Theory involving a gauge field. The setup resembles lattice gauge theory with the crucial difference that a lattice spacing is absent and that one does not have the freedom to choose an action. It would be interesting to calculate the spectral action and, subsequently, to take the limit $\e\rightarrow 0$. 
Presumably, this limit will lead to divergences since the operator $D_{\smalltriangleup}$ ceases to have compact resolvent when $\e=0$. One can speculate whether these divergences might correspond to the divergences found in perturbative Quantum Field Theory.
\item[-]
Of course, the exact construction and analysis of the triple $(\cb_t,D_t,\ch_t)$ should be carried out. A publication with these details is under preparation.\\[3ex]
\end{itemize}

\noindent{\bf Acknowledgements}\\

We thank the following colleagues for fruitful discussions: Christian Fleischhack, Victor Gayral, Gerd Grubb, Troels Harmark, Thordur Jonsson, Mario Paschke, Adam Rennie, Carlo Rovelli, Thomas Schucker, Christian Voigh and Raimar Wulkenhaar. Furthermore, we are grateful to the following institutes for hospitality during visits: the Institute of Mathematics in Reykjavik, Iceland; The Max Planck Institute for Mathematics in the Sciences, Leipzig, Germany; the Isaac Newton Institute for Mathematical Sciences, Cambridge, UK; the Institute of Theoretical Physics in Marseilles, France.

Johannes Aastrup was  funded by the German Research Foundation (DFG) within the research projects  {\it Deformation Theory for Boundary Value Problems} and {\it Geometrische Strukturen in der Mathematik} (SFB 478).

\appendix

\begin{appendix}

\section{A spectral triple without the basepoint}
\label{Appendix}

The diffeomorphism invariance obtained so far only includes diffeomorphisms which preserve the basepoint introduced in section \ref{section1}.
The role of the basepoint is to equip the algebra of loops with a product. Because the choice of basepoint partly breaks diffeomorphism invariance we would like to obtain a structure which avoids the basepoint.
 It turns out that such a construction does exist. Instead of the group structure of loops this more general construction is based on a groupoid structure of path.

Once again we start with an abstract simplicial complex $\ck$ with vertices $\{v_i\}$ and edges $\{\e_j\}$. Consider the Hilbert space
\[
\ch_{\ck} = L^2(\{v_i\}\times \ca_{\ck},M_l(\mathbb{C}))\;,
\]
where we recall that $\ca_{\ck}=G^{n(\ck)}$. For now we omit the Clifford bundle which is not necessary for the construction and representation of the algebra.

In fact, there are two natural algebras to consider. Denote by $\OO$ the set of loops in $\ck$ with arbitrary basepoint and consider first the algebra generated by these loops equipped with the product
\[
f_{L_1}\cdot f_{L_2}= \left\{ \begin{array}{ll}
0 & \mbox{if basepoints differ} \\
f_{L_1\cdot L_2} & \mbox{if basepoints coincide.}  
                              \end{array}\right.
\]
Again we can construct a norm via the matrix norm on $G$ and we obtain a $C^\ast$-algebra which we denote $\cb_{\OO}$.

The second option, which is perhaps more natural, is to consider paths in $\ck$. Denote by $\cp$ the set of paths in $\ck$ and denote by $\cb_{\cp}$ the algebra generated by such paths with a natural product. Clearly, $\cb_{\OO}\hookrightarrow \cb_{\cp}$. Concretely, we let the algebra $\cb_\cp$ be given by its representation
which is as follows. Given a path $p$ that starts in $v_1$ and ends in $v_2$ we write
\[
(f_p \cdot \xi) (v_j,\nabla) =\left\{ \begin{array}{ll}
0 & \mbox{if}\; v_j\not= v_2  \\
\nabla(p)\cdot \xi(v_1,\nabla) & \mbox{if}\; v_j=v_2\;,  
                              \end{array}\right.
\]
where $\nabla(p)$ is defined as in (\ref{actionon}), just with a path instead of a loop.

In this setup a path can now be seen as an operator which combines the holonomy along the path with a matrix structure
\[
 | v_m\rangle\langle v_n |
\]
and we notice that loops with arbitrary basepoint are found on the diagonal of this matrix structure. Thus, a natural trace will pick out all the loops and thereby, in terms of holonomy loops, the gauge covariant elements.

The inner product on the Hilbert space $\ch_\ck$ also involves a sum over vertices. The rest of the construction can be carried out in a similar manner to the construction of the triple $(\cb_{\smalltriangleup},D_{\smalltriangleup},\ch_\smalltriangleup)$.

This formulation seems better suitable for a semi-classical limit since it involves the points of the manifold, which, presumable, should emerge in such a limit.

\section{On diffeomorphism invariance}
\label{DIFFinva}

Although the Hilbert space $L^2(\overline{\ca}^{\smalltriangleup})$ does not have an action of the full diffeomorphism group it is possible to introduce a notion of analytic diffeomorphisms on $L^2(\overline{\ca}^{\smalltriangleup})$ via the space $L^2(\overline{\ca}^{\small a})$. This, in turn, allows us to extend the action of certain operators on $L^2(\overline{\ca}^{\smalltriangleup})$ to the larger Hilbert space $L^2(\overline{\ca}^{\small a})$ and thereby introduce a notion of (analytic) diffeomorphism invariance for these operators. This setup involves an embedding of general piecewise analytic graphs $\G$ into triangulations $\ct_i$ coming from the projective system $\{\ck_i\}$. Therefore, the action of these operators on $L^2(\overline{\ca}^{\smalltriangleup})$ will depend on a choice of embedding. It is important to realize that this construction does not work for the operator $D_{\smalltriangleup}$ since we do not have a Clifford bundle over the space $L^2(\overline{\ca}^{\small a})$. 

The following should be read as a rough idea or strategy as to how one introduces a notion of diffeomorphism invariance on $L^2(\overline{\ca}^{\smalltriangleup})$, rather than a complete analysis.

Let us go into details. Given an element $\xi_\G\in L^2(\overline{\ca}^{\small a})$ associated to a piece-wise analytic graph $\G$ we choose an embedding of a suitable simplicial complex $\ck_i$
\[
\phi:\ck_i\rightarrow\ct_i\;,
\]
where $\ct_i$ is a triangulation in $\cm$,
so that $\G$ lies in $\ct_i$. Consider an operator $\co$ on $L^2(\overline{\ca}^{\smalltriangleup})$. This could for example be the Laplace operator. The action of $\co$ on $\xi_\G$ is defined as 
\[
\co(\xi_\G):= \phi(\co(\phi^{-1}\xi_\G))\;.
\]
If we map $d:\xi_\G\rightarrow\xi_{d(\G)}$ with a diffeomorphism $d$
then the action of $\co$ changes accordingly
\[
\co(\xi_{d(\G)}):= \phi^{\prime}(\co((\phi^{\prime})^{-1}\xi_\G))\;,
\]
where $\phi^{\prime}$ is a new embedding. We obtain the diagram 
\ba
\begin{array}{ccccccc}
\xi_\G &\stackrel{\phi^{-1}}{\longrightarrow} & \xi_{\smalltriangleup}&  \longrightarrow  &   \co\xi_{\smalltriangleup} & \stackrel{\phi}{\longrightarrow} & (\co\xi_{\smalltriangleup})_{\G}\\
d\downarrow && d_{\smalltriangleup}\downarrow && d_{\smalltriangleup}\downarrow && d\downarrow  \\
\xi_{d(\G)} &\stackrel{(\phi^{\prime})^{-1}}{\longrightarrow} & \xi^{\prime}_{\smalltriangleup}&  \longrightarrow  &   \co\xi^{\prime}_{\smalltriangleup} & \stackrel{\phi^{\prime}}{\longrightarrow} & (\co\xi^{\prime}_{\smalltriangleup})_{d(\G)}
\end{array}
\label{diagram}
\ea
where $d_{\smalltriangleup}$ and $d^\prime_{\smalltriangleup}$ belong to diff$(\triangle)$. To have a diffeomorphism invariant state means that the state is invariant under diff$(\triangle)$. This is exactly the space $\ch_{diff}$ of diffeomorphism invariant states mentioned in the previous subsection. A diffeomorphism invariant operator $\co$ is an operator for which the centre part of the diagram (\ref{diagram}) commute
\[
\begin{array}{ccc}
 \xi_{\smalltriangleup}&  \longrightarrow  &   \co ( \xi_{\smalltriangleup}) \\
 d_{\smalltriangleup}\downarrow & \circlearrowleft   & d_{\smalltriangleup}\downarrow  \\
 \xi^{\prime}_{\smalltriangleup}&  \longrightarrow  &   \co (\xi^{\prime}_{\smalltriangleup}) 
\end{array}
\]
This reflects the simple observation that a loop is, by itself, not an observable since it is not self-adjoint. Only the combination $L + L^{-1}$ is self-adjoint, exactly because it is invariant with respect to the symmetry group of the loop.

\section{Symmetric states}

In Quantum Field Theory the vacuum state is defined as the unique translational invariant state. In the present setting there are certain states which display a high degree of symmetry and which may be thought in terms of a ground state.

Consider first the spectral triple $(\cb_{\smalltriangleup},D_{\smalltriangleup},\ch_\smalltriangleup)$ and the two states
\[
\psi_0(\nabla)=\psi_0(g_1,g_2,\ldots,g_n,\ldots)= \d_{G_1}(id)\cdot \d_{G_2}(id)\cdot\ldots\cdot \d_{G_n}(id)\cdot\ldots\;, 
\]
and
\[
\phi_0(\nabla)= 1\;.
\]
Here $\d_{G_n}$ is the delta-function on the $n$'th copy of $G$ and $c$ is a constant. The action of a loop $L$ on the first state reads
\[
f_L\cdot\psi_0(\nabla) = \psi_0(\nabla)
\]
for any $L$ in $\cb_{\smalltriangleup}$. This means that the entire algebra $\cb_{\smalltriangleup}$ collapses into the identity on the state $\psi_0$. This corresponds to flat space. This state, however, does not lie in the Hilbert space, nor in the domain of the operator $D_{\smalltriangleup}$. In contrast to this the state $\phi_0$ lies in the kernel of $D_{\smalltriangleup}$. 
Both states $\psi_0$ are invariant under diff$(\triangle)$ since they are invariant under any permutation of the argument $(g_1,\ldots,g_n,\ldots)$.

If we consider instead the spectral triple $(\cb_t,D_t,\ch_t)$ there are again two states which are highly symmetric. Consider first
\[
\ch_t\ni\Psi_0(\nabla,\bar{x}) = \psi_0(\nabla)\otimes \xi_o(\bar{x})\;,
\]
where
\[
\xi_0(\bar{x})=\xi_0(x_1,x_2,\ldots,x_n,\ldots)=\d_{x_1}(\a_1)\cdot \d_{x_2}(\a_2)\cdot\ldots\cdot \d_{x_n}(\a_n)\cdot\ldots 
\]
This state fixes the sequence $\{a_i\}$ on the 'background' sequence $\{\a_i\}$ which was, in section \ref{tHE}, fixed according to relation (\ref{tHe}). Again, this states does not lie in the domain of the Dirac operator $D_t$. Notice however, that by choosing the parameters $a_i$'s according to relation (\ref{tHe}) one obtains a Dirac type operator $D_{\smalltriangleup}$ which is, up to an overall factor, invariant under a shift in the parameters $a_i$. Such a shift corresponds to a change of scale. 

Next, we may also consider the state
\[
\ch_t\ni\Phi_0(\nabla,\bar{x}) = \phi_0(\nabla)\otimes \eta_o(\bar{x})\;,
\]
where
\[
\eta_0(\bar{x})=\phi_1(x_1)\cdot \phi_{2}(x_2)\cdot\ldots\cdot \phi_{n}(x_n)\cdot\ldots 
\]
Here $\phi_{n}(x_n)$ is the Gauss distribution which we used to construct the triple $(\cb_t,D_t,\ch_t)$. Since the state $\eta_0$ lies in the kernel of $D_a$ the state $\Phi_0$ lies in the kernel of $D_t$. 
%Again, the state $\Phi_0$ is symmetric under shifts of the parameters $a_i$ (or, equivalently, $x_i$).

\end{appendix}

\end{document}